\documentclass[reprint,amsmath,amssymb,aps]{revtex4-2}

\usepackage[english]{babel}
\usepackage[utf8]{inputenc}
\usepackage{graphicx}
\usepackage{floatflt}
\usepackage{blindtext}
\usepackage{titlesec}
\usepackage[shortlabels]{enumitem}
\usepackage{amsthm}
\usepackage{subfig}
\usepackage{listings}
\usepackage{listingsutf8}
\usepackage{framed}
\usepackage{minibox}

\usepackage{float}
\usepackage{wrapfig}
\usepackage[strict]{changepage}
\usepackage{pgfplots}
\usepackage{tikz}
\usepackage{caption}
\usepackage{subcaption}
\usetikzlibrary{matrix}
\pgfplotsset{width=11cm,compat=1.9}
\usepgfplotslibrary{external}
\usepackage{adjustbox}
\usepackage{stackengine}
\makeatletter
  \def\my@tag@font{\normalsize}
  \def\maketag@@@#1{\hbox{\m@th\normalfont\my@tag@font#1}}
  \let\amsmath@eqref\eqref
  \renewcommand\eqref[1]{{\let\my@tag@font\relax\amsmath@eqref{#1}}}
\makeatother

\usepackage{braket}
\usepackage{slashed}
\usepackage{cases}
\usepackage{physics}
\usepackage{dsfont}
\usepackage{caption}
\usepackage{epsfig}
\usepackage{array}
\usepackage{hhline}
\usepackage{verbatim}
\usepackage{empheq}
\usepackage{graphicx}
\usepackage{mathrsfs}
\usepackage{mathtools}
\usepackage{xcolor}
\usepackage{layout} 
\usepackage[utf8]{inputenc}
\usepackage{xcolor}
\usepackage{tcolorbox}
\tcbuselibrary{breakable}
\usepackage{tcolorbox}
\usepackage{amstext} 
\usepackage{array}
\usepackage[vcentermath]{youngtab}
\usepackage{nicematrix}
\usepackage{hyperref}

\usepackage{tabularx}
\usepackage{natbib}
 
\newcommand{\beq}{\begin{eqnarray}}
\newcommand{\eeq}{\end{eqnarray}}
\newcommand{\bea}{\begin{eqnarray}}
\newcommand{\bqa}{\begin{eqnarray}}
\newcommand{\eea}{\end{eqnarray}}
\newcommand{\be}{\begin{equation}}
\newcommand{\ee}{\end{equation}}

\begin{document} 

\title{Baryons, Skyrmions and $\theta$-periodicity anomaly in chiral and vector-like gauge theories}
\author{Stefano Bolognesi}
\email{stefano.bolognesi@unipi.it}
\affiliation{Department of Physics E. Fermi, University of Pisa, and INFN Sezione di Pisa,
Largo Pontecorvo, 3, Ed. C, 56127 Pisa, Italy}
\author{Andrea Luzio}
\email{andrea.luzio@sns.it}
\affiliation{Theoretical Particle Physics Laboratory, EPFL, 1015 Lausanne, Switzerland}
\author{Giacomo Santoni}
\email{giacomo.santoni@alumni.sns.it}
\affiliation{Department of Physics, University of Rome "La Sapienza" and INFN Sezione di Roma1, Piazzale A. Moro 2, Roma, I-00185, Italy}

\begin{abstract}

In this paper, we study the baryons and solitons of chiral and vector-like $SU(N)$ gauge theories with matter in mixed one and two-index representations. Focusing on the Color-flavor locked  (CFL) phase, we compute the topology of the coset of their low-energy EFT. We find that in the chiral models under consideration, Skyrmions are always absent. We also show, however, that some of these models admit heavy baryons that are expected to be stable, because their decay into the lighter degrees of freedom of the EFT is forbidden by the unbroken symmetry group. This mismatch suggests that some deeper dynamical mechanism must be responsible with either the instability of the seemingly stable heavy baryons or the unreliability of the Skyrme model in the low-energy EFT. In the vector-like models all the expected baryons are mirrored by Skyrmions. 

Then we turn to the study of domain walls. We determine some aspects of their dynamics by matching the $\theta$-periodicity anomaly. We find that, for complete CFL, the $\theta$-periodicity anomaly is always matched without introducing new dynamical degrees of freedom in the low-energy EFT. If part of the color group is unbroken, new dynamical degrees of freedom must be added to the low-energy EFT in the domain-wall background with few exceptions.
\end{abstract}

\maketitle

\section{Introduction}

In recent years, new and old techniques have been used to understand the dynamics of strongly coupled gauge theories: the 't Hooft anomaly matching is a standard tool to constrain the possible IR phases of a certain UV theory, and generalized anomaly matching has been recently deployed to shine some light on the IR behavior of chiral gauge theories \cite{Bolognesi:2020mpe,Bolognesi:2021yni}. 

However, it is often simple to construct IR phases that automatically reproduce all matching constraints by considering an Higgs regime, where a quark bilinear plays the role of the elementary Higgs scalar. In such a weakly coupled theory, which often reduces to a Color-flavor locking (CFL) mechanism\footnote{Being the quark bilinear charged both under color and flavor symmetries, its condensation breaks the two groups to a diagonal subgroup. Such a mechanism is known as Color-flavor locking.}, one can work out the symmetry-breaking pattern and the massless spectrum. Then, by assuming that the same phase is present also in the strong-interacting theory, one obtains a perfectly valid phase that passes all the anomaly matching checks. 

This construction, dubbed natural anomaly matching, is reviewed in the context of chiral gauge theories in \cite{Bolognesi:2024bnm}, and makes it necessary to consider other consistency checks to elucidate whether these phases are actually realized. 

The study of topological solitons in the IR EFT, particularly when combined with large-$N$ reasoning, has been insightful in understanding fundamental QCD, orbifold QCD, and their supersymmetric counterparts. In this paper, we extend these ideas for several phases of chiral and vector-like $SU(N)$ gauge theories with quarks in mixed one- and two-index representations\footnote{The hadronic spectrum in chiral gauge theories has recently been considered in \cite{Kristensen:2024vmi,Girmohanta:2019cth}.}.

In the first part of the paper, we focus on the Skyrmion-baryon correspondence. 

A baryonic operator is a gauge-invariant operator charged under some of the $U(1)$ conserved charges. In general, there are two types of baryonic operators, which have different $N$ dependence:  ``light-baryons'', made of a $\mathcal{O}(N^0)$ number of fermions, and ``heavy-baryons'', built with the completely antisymmetric $\epsilon$ tensor of $SU(N)$. 

Fundamental QCD and orbifold QCD have only heavy baryons, which therefore interpolate stable particles (the baryons) that can be modeled as solitons (Skyrmions) of the low-energy effective action \cite{Skyrme:1962vh,Witten:1979kh,Adkins:1983ya,Bolognesi:2006ws,Cherman:2006iy}. 

Conversely, in the theories considered, light and heavy baryons are both present, and whether they form stable particles depends on the structure of the conserved charges. The naive expectation is that in those theories where stable heavy baryons are known to exist, a corresponding Skyrmion must be present in the low-energy EFT. We find that in the chiral gauge theories considered in this work this expectation is not always met. In particular in the Georgi-Glashow (GG) models, there are some heavy baryons that are forbidden to decay into light baryons and NGBs by the selection rules, while the topology of the coset in the low-energy EFT is trivial.

Recently, the soliton-baryon correspondence has been extended to theories without Nambu-Goldstone bosons (NGBs), but rather pseudo-Nambu-Goldstone (pseudo-NGBs), by introducing another possible mechanism by which the low-energy effective action can detect a massive spectrum of stable particles: the baryons correspond to pancake-like objects, where the disk is made of a metastable domain wall of the low-energy theory. This idea is at work, for example, in $N_{\rm f}=1$ fundamental QCD \cite{Komargodski:2018odf} (see also \cite{Bigazzi:2022luo,Karasik:2022tmd,Lin:2023qya}).

Because of this, it is important to study the domain walls exhibited by these theories. In particular, to stabilize the pancake, it is fundamental that the worldsheet of the domain wall hosts a TQFT\footnote{Such a TQFT provides degrees of freedom to the edge of the disk.}. Some features of such TQFT, in particular its anomalies, can be inferred by studying the so-called $\theta$ periodicity anomaly \cite{Cordova:2019jnf, Cordova:2019uob}. 

A gauge theory possesses a $\theta$-periodicity anomaly if, in the presence of some background gauge fields with fractional topological charge, the Euclidean partition function acquires a nontrivial phase when the $\theta$-parameter is shifted by $2\pi$. Due to the renormalization-group invariance of the partition function, the low-energy effective field theory should have the same anomaly. By requiring the $\theta$-periodicity anomalies in the UV and the IR to match, one can obtain nontrivial information about the matter content of the low-energy effective field theory \cite{Anber:2019nze,Cordova:2019jnf, Cordova:2019uob, Kitano:2020evx}. 

In the second part of this paper, we will show that for theories in a complete CFL phase, there is no such anomaly. In case of partial CFL, i.e., when part of all of the $SU(N)$ group remains un-Higgsed, a $\theta$ periodicity anomaly is present, and so a related effective action on the world volume of the pseudo-NGB domain wall, which is a certain Chern-Simons theory.

Here is a brief outline of the paper:
\begin{itemize}
    \item In section \ref{sec:cfl} we illustrate general aspects of the CFL phase, and the corresponding IR EFT.
    \item In section \ref{sec:baryons} and \ref{sec:introvect}, we will describe the chiral theories we are focusing on, their baryonic operators, and the stability of corresponding particles. 
    \item In section \ref{sec:skwzw}, we compute the homotopy groups of the low-energy effective actions to show whether it is possible to construct Skyrmions, and write Wess-Zumino-Witten (WZW) terms. The result is consistent with the expectation from the UV analysis of baryonic operators and the conserved $U(1)$ charges.
    \item In section \ref{sec:abj} we compute the $\theta$-periodicity anomaly, and in section \ref{sec:dw} we discuss the effective action on the domain wall required to match it. This allows us to briefly discuss the possibility of having pancake-like solitons in such a class of models.  
\end{itemize}
We conclude in section \ref{sec:conclusion}, where some general lessons are discussed.

\section{The color-flavor locking phase}
\label{sec:cfl}
In this section, we collect some results on the color-flavor locking (CFL) phase that will be used in the rest of the work.

\subsection{Color-flavor locking mechanism}
\label{sec:cfldef}
Consider a quantum field theory with an internal symmetry group
\begin{align}
    G=\frac{G_{\rm c}\times G_{\rm f}}{\Gamma_G}
\end{align}
where $\Gamma_G$ is a discrete abelian subgroup of $G_{\rm c}\times G_{\rm f}$. Let $H_{\rm c}\subset G_{\rm c}$ and $H_{\rm f}\subset G_{\rm f}$, and suppose that there are two isomorphic subgroups $H_{\rm cf}^{({\rm c})}\subset G_{\rm c} $ and $H_{\rm cf}^{({\rm f})}\subset G_{\rm f} $ and let
\begin{align}
    H_{\rm cf}=\left(H_{\rm cf}^{({\rm c})}\times H_{\rm cf}^{({\rm f})} \right)_{\text{diag}}
\end{align}
We define
\begin{align}
    H=\frac{H_{\rm c}\times H_{\rm f}\times H_{\rm cf}}{\Gamma_H}
\end{align}
where $\Gamma_H$ is a discrete abelian subgroup of $H_{\rm c}\times H_{\rm f}\times H_{\rm cf}$. Color-flavor locking (CFL) occurs if
\begin{align}
\label{eq:sb}
    G \overset{\text{SSB}}{\longrightarrow} H
\end{align}
We adopt the following terminology:
\begin{itemize}
    \item When $H_{\rm c}=\{1\}$ we say that there is \emph{complete} CFL.
    \item When $H_{\rm c}\neq \{1\}$ we say that there is \emph{partial} CFL.
\end{itemize}
Although devoted to the Higgs mechanism, Refs. \cite{Frohlich:1980gj, Frohlich:1981yi} provides a useful framework to properly define the CFL mechanism. The ``order parameter'' for the CFL must be some scalar operator $\phi$ that transforms nontrivially under both $G_{\rm c}$ and $G_{\rm f}$. The symmetry group $G$ breaks down to $H$ when the minimum of the effective potential $V(\phi)$ is a gauge-orbit 
\begin{equation}
\label{eq:orbit}
    V^{-1}(0)=\{\mathcal{R}(g_{\rm c}(x))\cdot \phi_0\ |\ g_{\rm c}(x)\in G_{\rm c}\}
\end{equation}
where $\phi_0$ is a constant representative of the orbit whose stability subgroup is $H$
\begin{align}
    \mathcal{R}(h)\cdot \phi_0 = \phi_0 \qquad \forall\ h\in H
\end{align}
and $\mathcal{R}$ is a nontrivial representation of $G$. One can use $\phi$ to construct gauge-invariant operators that interpolate between the vacuum and the states of the theory, and compute their correlation functions in perturbation theory by expanding around some point of the orbit \eqref{eq:orbit}.

It is by no means necessary for $\phi$ to have a nonzero vacuum expectation value in every gauge for CFL to occur. In particular, the authors of Refs. \cite{Frohlich:1980gj, Frohlich:1981yi} explicitly show that in temporal gauge this is not the case even without invoking Elitzur's theorem \cite{Elitzur:1975im}.

The description above is precise if there is a separation of scales between the CFL scale and the strong coupling scale, $\Lambda\ll |\phi_0|$. Let us proceed with this assumption, which we will relax later.

\subsection{Low-energy spectrum}
In this subsection we describe the low-energy spectrum arising from the symmetry-breaking pattern of Eq. \eqref{eq:sb} by means of the Callan-Coleman-Wess-Zumino (CCWZ) construction \cite{Coleman:1969sm, Callan:1969sn, Weinberg:1975gm}. Since they are irrelevant in the counting of the degrees of freedom, we will take $\Gamma_G=\Gamma_H=\{1\}$.
\par We describe the physical and unphysical NGBs with a field
\begin{align}
    \left(U_{\rm c}, U_{\rm f}\right)\in G_{\rm c}\times G_{\rm f}
\end{align}
Its transformation laws under the ordinary color gauge symmetry and the hidden gauge symmetry \cite{Leutwyler:1993iq} are
\begin{align}
    &\text{color:}\ && \left(U_{\rm c}, U_{\rm f}\right)\longmapsto \left(g_{\rm c}U_{\rm c}, U_{\rm f}\right) \nonumber \\
    &\text{hidden:}\ && \left(U_{\rm c}, U_{\rm f}\right)\longmapsto \left(U_{\rm c}h_{\rm cf}^{-1}h_{\rm c}^{-1}, U_{\rm f}h_{\rm cf}^{-1}h_{f}^{-1}\right)
\end{align}
where $g_{\rm c}\in G_{\rm c}$, $h_{\rm c}\in H_{\rm c}$, $h_{\rm cf}\in H_{\rm cf}$ and $h_{\rm cf}\in H_{\rm cf}^{({\rm c})}, H_{\rm cf}^{({\rm f})}$. We use a color gauge transformation to choose the unitary gauge $U_{\rm c}=1$. The residual transformations are those with $g_{\rm c}=h_{\rm c}h_{\rm cf}$. This tells us that the physical NGBs live on the coset space
\begin{align}
    \frac{G_{\rm f}}{H_{\rm f}\times H_{\rm cf}^{({\rm f})}}
\end{align}
Then, the number of physical NGBs is equal to the number of broken flavor generators. The gauge field for the color symmetry is $a$. In the low-energy effective action, $a$ appears through the combination
\begin{align}
    a_U=U_{\rm c}^{-1}aU_{\rm c}+iU_{\rm c}^{-1}dU_{\rm c}
\end{align}
Let us decompose the color Lie algebra $\mathfrak{g}_c=\mathfrak{h}_{\rm c}\oplus \mathfrak{f}_c$ where $\mathfrak{h}_{\rm c}$ is the Lie algebra of $H_{\rm c}$ and $\mathfrak{f}_c$ is its orthogonal complement. If we decompose
\begin{align}
\label{eq:au}
    a_U=a_U^{\mathfrak{h}_{\rm c}}+a_U^{\mathfrak{f}_c}\ , \qquad a_U^{\mathfrak{h}_{\rm c}}\in \mathfrak{h}_{\rm c}\ , \ a_U^{\mathfrak{f}_c}\in \mathfrak{f}_c
\end{align}
Under the hidden gauge group $a_U^{\mathfrak{h}_{\rm c}}$ transforms as a connection, while $a_U^{\mathfrak{f}_c}$ transforms homogeneously
\begin{align}
    &a_U^{\mathfrak{h}_{\rm c}}\longmapsto h_{\rm c}a_U^{\mathfrak{h}_{\rm c}}h_{\rm c}^{-1}-i(dh_{\rm c})h_{\rm c}^{-1}\nonumber \\
    &a_U^{\mathfrak{f}_c}\longmapsto h_{\rm c}a_U^{\mathfrak{f}_c}h_{\rm c}^{-1}
\end{align}
This tells us that in the effective Lagrangian we can construct a mass term for $a_U^{\mathfrak{f}_{\rm c}}$. Hence, in the CFL phase, the number of massive gauge bosons is equal to the number of broken color generators.
\par We summarize here what theories in the CFL mechanism look like at different energy scales:
\begin{itemize}
    \item At the symmetry-breaking scale $E\sim v$ the effective field theory contains massless NGBs and spin-$\frac{1}{2}$ baryons, the massive gauge bosons, the Higgs field, and a massive pseudo-NGBs arising from the anomalous $U(1)_{\rm an}$ symmetry. If there is a residual confining gauge group $H_{\rm c}$, other massive hadrons are present at this energy scale.
    \item In the deep IR region $E \ll v$, one expects the effective field theory to contain only the massless degrees of freedom: the NGBs and the spin-$\frac{1}{2}$ baryons.
\end{itemize}

\subsection{Color-flavor locking and strong coupling} 
\label{sec:cflstrong}

CFL is a particular realization of a phase with the   
\begin{equation}
    \frac{G_{\rm f}}{\Gamma_G} \to  \frac{H_{\rm f}\times H_{\rm cf}}{\Gamma_H}\; \label{eq:gsb}
\end{equation}
global symmetry-breaking pattern. In particular, the language above is precise if there is a separation of scales between the symmetry-breaking scale and the strong-coupling scale i.e. $|\phi_0| \gg \Lambda$. If the separation of scales is not present, then the CFL mechanism cannot provide quantitative predictions, but it might still be useful for qualitative ones. 

More precisely, suppose that it is possible to add an elementary scalar that transforms as $\phi_0$ under the internal symmetries. By tuning its potential to be of the symmetry-breaking type, one obtains a weakly coupled scenario. By tuning the potential to give a large positive mass to the scalar, it can be integrated away at very high energy, bringing us back to the gauge theory we are aiming to describe. If we suppose that the potential can be smoothly deformed from one case into the other, in such a way that no phase transition occurs, then some of the qualitative predictions one obtains in the weakly coupled model persist in the strongly coupled theory\footnote{A familiar setting where there is continuity between the CFL regime and a strongly coupled regime is finite density QCD with three flavors. In that case, depending on the chemical potential, it is more convenient to use the CFL description \cite{Alford:1998mk} or a chiral Lagrangian description. However, the two regimes are believed to be connected, and the qualitative features match \cite{Schafer:1998ef}.}.

For example, if there is no separation of scales, one can expect the vector boson resonances to become less sharp and to be at the same energy scale as the other vector mesons. The lack of separation of scales affects also the baryon spectrum of a theory with quarks in higher representation, as discussed at the end of subsection \eqref{sec:ggintro}.

In any case, even the fact that the low-energy effective action is a nonlinear sigma model with $G_{\rm f}/(H_{\rm f}\times H_{\rm cf})$ remains unchanged.  

\subsection{Background gauge fields}
\label{sec:background}
A common way to study a theory is to couple its current to background gauge fields and compute how the partition function depends on these external probes. In the following, this idea will be central in defining the $\theta$-periodicity anomaly. In this section, we present some subtleties that arise when one considers the CFL phase.

In general, as the internal symmetry group is $G=\frac{G_{\rm c}\times G_{\rm f}}{\Gamma_G}$, to couple the UV theory to a background gauge field, one provides a connection on an $\frac{G_{\rm f}}{\Gamma_G}$ bundle. Then the partition function of the theory reads\footnote{In the notation of subsection \ref{app:1formquant}, $\protect\widetilde{A}$ corresponds to the $\protect\widetilde{A}_i^G$, $\protect\widetilde{A}_i^{U(1)}$ collectively and $B$ corresponds to the $B_i$ and $b$ collectively.}
\begin{align}
\label{eq:zuvir}
    &\mathcal{Z}[\widetilde{A}, B]= \nonumber \\
    &\int\limits_{\substack{\text{compatible} \\ G_{\rm c}/\Gamma_G\text{-bundles}}} [d\chi_{UV}][d\widetilde{a}]\ \mathrm{exp}\Big(-S_{UV}[\chi_{UV}, \widetilde{a}, \widetilde{A}, B]\Big)  
\end{align}
where $S_{UV}$ is the microscopic action and $\chi_{UV}$ denotes the matter degrees of freedom of the UV theory collectively. The path integral is performed only on those $\frac{G_{\rm c}}{\Gamma_G}$-bundles that combine with the $\frac{G_{\rm f}}{\Gamma_G}$-bundle to form an $\frac{G_{\rm c}\times G_{\rm f}}{\Gamma_G}$-bundle. More explicitly, the 't Hooft fluxes of the gauge bundle are fixed by the external field $B$ (e.g. see \cite{Bolognesi:2020mpe}).

As we explained in subsection \ref{sec:cfldef}, CFL occurs when the minimum of the effective potential $V(\phi)$ for the ``order parameter'' $\phi$ is a gauge-orbit with a constant representative $\phi_0$ and $H$ as a stability subgroup. This constant representative can exist only on $H$-bundles. In the other sectors, the integrand $e^{-S}$ is exponentially suppressed because the deviations of $\phi$ from the value $\phi_0$ are necessarily large. The necessity of this restriction was previously noted in Ref. \cite{Yonekura:2020upo}, where it was applied to global symmetry-breaking. As shown in Refs. \cite{Brower:2003yx, Aoki:2007ka} the restriction of the path-integral domain to some topological sectors is not harmful as long as the volume of the spacetime is sufficiently large.

Therefore, if CFL occurs, if $\widetilde{A}, B$ are connections on a $\frac{H_{\rm f}\times H_{\rm cf}}{\Gamma_H}$ bundles, the partition function is effectively reduced to
\begin{align}
    &\mathcal{Z}[\widetilde{A}, B]\sim \int\limits_{H\text{-bundles}} [d\chi_{UV}][d\widetilde{a}]\ \mathrm{exp}\Big(-S_{UV}[\chi_{UV}, \widetilde{a}, A, B]\Big)\;.  
\end{align}
This observation is crucial for the study of the $\theta$-periodicity anomaly in the CFL phase performed in section \ref{sec:abj}.

\section{Chiral gauge theories: symmetries and baryons}
\label{sec:baryons}
In this section, we discuss the symmetries and classify the baryons of some chiral gauge theories previously studied in Refs. \cite{Bolognesi:2020mpe, Bolognesi:2021yni, Bolognesi:2023xxv, Bolognesi:2024bnm}, namely the $\psi\eta$ model, the $\chi\eta$ models, the Bars-Yanckielowicz (BY) models and the Georgi-Glashow (GG) models.

\subsection{$\psi\eta$ model}
\label{sec:intropsieta}

\paragraph{Lagrangian and symmetries}

The $\psi\eta$ model consists of the Weyl fermions
\begin{equation}
    \psi^{\{ij\}}  \qquad \eta_i^A
\end{equation}
in the direct-sum representation
\begin{align}
    \yng(2)\oplus (N+4)\, \bar{\yng(1)}\;.
\end{align}
where $i,j$ are color indices and $A$ is $a$ flavor index. The continuous symmetries of the classical action are shown in Table \ref{tab:psietaTab}. The symmetries $U(1)_{\psi}\times U(1)_{\eta}$ are broken by the ABJ anomaly to
\begin{align}
    & U(1)_{\psi}\times U(1)_{\eta} \longrightarrow U(1)_{\psi\eta} \times \mathbb{Z}_2^F  && \text{for even $N$} \nonumber \\
    & U(1)_{\psi}\times U(1)_{\eta} \longrightarrow U(1)_{\psi\eta}  && \text{for odd $N$}
\end{align}
where $U(1)_{\psi\eta}$ is the non-anomalous combination of $U(1)_{\psi}$ and $U(1)_{\eta}$ and in table \ref{tab:psietaTab} we denoted as $U(1)_{\rm an}$ the anomalous continuous subgroup. The global symmetry group of the model, including discrete factors, is
\begin{align}
    \frac{SU(N)\times SU(N+4) \times U(1)_{\psi\eta}\times \mathbb{Z}_2^F}{\mathbb{Z}_{N}\times \mathbb{Z}_{N+4}} \qquad & \text{for even }N\;, \\
    \frac{SU(N)\times SU(N+4) \times U(1)_{\psi\eta}}{\mathbb{Z}_{N}\times \mathbb{Z}_{N+4}}  \qquad & \text{for odd }N\;,  \label{eq:sympsieta}
\end{align}
where the denominator is present because a transformation in the center of color (flavor) can be erased by an $U(1)_{\psi\eta}$ transformation (combined with a fermion parity, for even $N$).

\begin{table}
    \renewcommand*{\arraystretch}{1.3}
\centering 
    \begin{tabular}{|c|c|c|c|c|c|}
    \hline
    fields  &  $SU(N)_{\rm c}  $    &  $ SU(N+4)$     &   $ {U}(1)_{\psi\eta}   $ & $U(1)_{\rm an} $  \\
     \hline 
        $\psi^{\{ij\}}$   &   $ { \yng(2)} $  &    $  \frac{N(N+1)}{2} \cdot (\cdot) $    & $   +\frac{N+4}{N^*}$  & $+\frac{N^*(N+3)}{2}$ \\
      $ \eta^{A}_i$      &   $  (N+4)  \cdot   {\bar  {\yng(1)}}   $     & $N \, \cdot  \, {\yng(1)}  $     &   $  - \frac{N+2}{N^*} $ & $-\frac{N^*(N+1)}{2}$ \\
    \hline
    \end{tabular}
\caption{\footnotesize Classical continuous symmetries of the $\psi\eta$ lagrangian. We decomposed $U(1)_{\psi}\times U(1)_{\eta}$ into its anomaly-free subgroup $U(1)_{\psi\eta}$ and its anomalous subgroup $U(1)_{\rm an}$. Notice that $N^*=\mathrm{gcd}(N,2)$.}
\label{tab:psietaTab}
\end{table}

\paragraph{The CFL phase}

In \cite{Bolognesi:2020mpe}, it has been shown that the $\psi\eta$ model is not fully compatible with a phase where, in the IR, the entire symmetry group is unbroken, and all the `t Hooft anomalies are reproduced by massless composite fermions\footnote{To prove this point, it is necessary to consider the entire global structure of the symmetry group. In other words, to correctly background gauge the $\mathbb{Z}^{(1)}_N$ symmetry.}. It is compatible with the color-flavor locking phase with condensates\footnote{This is one of the simplest realizations of the NAM paradigm \cite{Bolognesi:2024bnm}.}
\begin{widetext}
\begin{align}
    & \expval{\psi^{\{ij\}}\eta_j^A}=c_{\psi\eta}\Lambda^3\delta^{iA}\ , && i,A=1,\dots,N  \nonumber \\
    &\expval{\psi^{\{ij\}}\eta_i^{\ell_1}\eta_j^{\ell_2}\psi^{\{i'j'\}}\eta_{i'}^{\ell_3}\eta_{j'}^{\ell_4}}=c_{\psi\eta\eta}\Lambda^{9}\epsilon^{\ell_1\ell_2\ell_3\ell_4}\ , && \ell_i=N+1,\dots N+4.
\end{align}
\end{widetext}
The unbroken symmetry group is then
\begin{align}
    H=\frac{SU(N)_{\rm cf}\times SU(4)_{\rm f}\times U(1)'\times \mathbb{Z}^F_2}{\mathbb{Z}_N\times \mathbb{Z}_4}
\end{align}
The denominator of $H$ consists of the transformations
\begin{align}
    & \mathbb{Z}_{N}:\ \left(e^{i\frac{2\pi k}{N}},\ 1,\ e^{-i\frac{2\pi k}{N}},\ 1\right) && k=0,\dots, N-1\nonumber \\
    & \mathbb{Z}_4:\  \left(1,\ e^{i\frac{2\pi \ell}{4}},\ e^{-i\frac{2\pi \ell}{4}},\ e^{i\pi\ell} \right) && \ell=0,\dots,3
\end{align}
The massless degrees of freedom consists on the $8N+1$ non-Abelian NGBs relative to the coset $\frac{SU(N+4)\times U(1)_{\psi\eta}}{SU(N)\times SU(4)\times U(1)'}$ and the three massless spin-$\frac{1}{2}$ baryons\footnote{The two-dimensional $\epsilon$-tensor $\epsilon^{\alpha\beta}$ acting on spinor indices is defined through the conventions $\epsilon^{12}=-\epsilon^{21}=-\epsilon_{12}=\epsilon_{21}=1$.}
\begin{widetext}
\begin{align}
\label{eq:masslesspsieta}
    & B^{[A_1,B_1]}_{\alpha}=\epsilon^{\beta\gamma}\psi^{\{ij\}}_{\beta}\eta_{i\gamma}^{[A_1}\eta_{j\alpha}^{B_1]}\ , && A_1,B_1=1,\dots,N \nonumber \\
    & B^{[A_1,B_2]}_{\alpha}=\epsilon^{\beta\gamma}\psi^{\{ij\}}_{\beta}\eta_{i\gamma}^{[A_1}\eta_{j\alpha}^{B_2]}\ , && A_1=1,\dots,N \ , \ B_2=N+1,\dots,N+4 
\end{align}
\end{widetext}
The transformation laws of the UV and IR fields under the unbroken symmetry group are shown in Table \ref{tab:psietacfl}.

\begin{table}[H]
\renewcommand*{\arraystretch}{1.3}
{
    \centering 
    \begin{tabular}{|c|c|c|c|c|  }
    \hline
      & fields   &  $SU(N)_{\rm cf}  $    &  $ SU(4)_{\rm f}$     &   $  U(1)^{\prime}   $    \\
    \hline
    {\rm UV}&  $\psi^{\{ij\}}$   &   $ { \yng(2)} $  &    $  \frac{N(N+1)}{2} \cdot   (\cdot) $    & $   +1  $      \\
    & $ \eta_{i A_1}$      &   $  {\bar  {\yng(2)}} \oplus {\bar  {\yng(1,1)}}  $     & $N^2 \, \cdot  \, (\cdot )  $     &   $ - 1 $   \\
    &  $ \eta_{i}^{A_2}$      &   $ 4  \cdot   {\bar  {\yng(1)}}   $     & $N \, \cdot  \, {\yng(1)}  $     &   $ - \frac{1}{2}  $     \\
    \hline 
    $ \phantom{{\bar{ \bar  {\bar  {\yng(1,1)}}}}}\!  \! \!\! \! \!  \!\!\!$  {\rm IR}&      $ B^{[A_1  B_1]}$      &  $ {\bar  {\yng(1,1)}}   $         &  $  \frac{N(N-1)}{2} \cdot  (\cdot) $        &    $   -1 $      \\
   &   $ B^{[A_1 B_2]}$      &  $   4 \cdot {\bar  {\yng(1)}}   $         &  $N \, \cdot  \, {\yng(1)}  $        &    $ - \frac{1}{2}$      \\
  & $\pi_1$ & $\yng(1)$ & $\yng(1)$ & $1/2$ \\
  & $\pi_2$ & $(\cdot)$ & $(\cdot)$ & $0$ \\
    \hline
    \end{tabular}  
    \caption{\footnotesize Unbroken symmetries in the CFL phase in the $\psi\eta$ model and fermions charges. The indices are $i, A_1=1,\dots , N$ and $A_2= N+1,\dots, N+4$.}
    \label{tab:psietacfl}
}
\end{table}

\paragraph{Light baryons}

In this work, we will refer to the baryons without $\epsilon$ tensor as \emph{light baryons}. The massless baryons in Eq. \eqref{eq:masslesspsieta} are not the only baryons without $\epsilon$-tensor that can be constructed. The complete list is
\begin{align} 
\label{eq:baryons101psieta}
    &B^{[AB]}_{\{\alpha\beta\gamma\}}=\psi^{\{ij\}}_{\{\gamma}  \eta_{i,\alpha}^{[A}  \eta_{j,\beta\}}^{B]}\;, && \nonumber \\
    & B^{[AB]}_{\alpha}=\epsilon^{\beta\gamma}\psi^{\{ij\}}_{\gamma}  \eta_{i,\alpha}^{[A}  \eta_{j,\beta}^{B]}\;, && \nonumber \\
    & B^{\{AB\}}_{\alpha}=\epsilon^{\beta\gamma}\psi^{\{ij\}}_{\alpha}  \eta_{i,\beta}^{\{A}  \eta_{j,\gamma}^{B\}}\;, &&A,B=1,2, \ldots, N+4\;,
\end{align}
The spin-$\frac{3}{2}$ baryons $B^{[AB]}_{\{\alpha\beta\gamma\}}$ cannot be massless due to the Weinberg-Witten theorem \cite{Weinberg:1980kq}. Anomalies \cite{Bolognesi:2020mpe} guarantee that the baryons in Eq. \eqref{eq:masslesspsieta} and table \ref{tab:psietacfl} remain exactly massless, while the rest can acquire a mass.

\paragraph{Heavy baryons and their stability}

In this work, we will refer to the baryons possessing at least one $\epsilon$-tensor as \emph{heavy baryons}. A heavy baryon in the $\psi\eta$ model can be constructed by attaching to some $\epsilon$ tensors the following \emph{words}
\begin{align}
    &\psi^{\{ij\}}\ , \quad (\psi\eta)^{i}_{A_1}\ , \quad (\psi\eta)^{i}_{A_2}\ , \quad \bar{\eta}^{iA_2}\ , \quad \bar{\eta}^{iA_2} \nonumber \\
    & i, A_1 = 1,\dots, N\ , \quad A_2=N+1, \dots, N+4
\end{align}
The charges of the words under the unbroken group can be read from table \ref{tab:psietacfl}. There are many structures that may emerge from this construction. The key question whether they are stable or they are allowed by selection rules to decay into lighter degrees of freedom like massless/light baryons and NGBs. We verified numerically that these decays seems to be always allowed, so heavy baryons are likely to be unstable.

\subsection{$\chi\eta$ model}
\label{sec:introchieta}

\paragraph{Lagrangian and symmetries}
The $\chi\eta$ model consists of the Weyl fermions
\begin{equation}
    \chi^{[ij]} \qquad \eta_i^A
\end{equation}
in the direct-sum representation
\begin{align}
    \yng(1,1)\oplus (N-4)\, \bar{\yng(1)}\;.
\end{align}
where $i,j$ are color indices and $A$ is a flavor index. The continuous symmetries of the classical action are shown in Table \ref{tab:chietaTab}.  The symmetries $U(1)_{\chi}\times U(1)_{\eta}$ are broken by the ABJ anomaly to
\begin{align}
    & U(1)_{\chi}\times U(1)_{\eta} \longrightarrow U(1)_{\chi\eta} \times \mathbb{Z}_2^F  && \text{for even $N$} \nonumber \\
    & U(1)_{\chi}\times U(1)_{\eta} \longrightarrow U(1)_{\chi\eta}  && \text{for odd $N$}
\end{align}
where $U(1)_{\chi\eta}$ is the non-anomalous combination of $U(1)_{\chi}$ and $U(1)_{\eta}$ and in table \ref{tab:chietaTab} we denoted as $U(1)_{\rm an}$ the anomalous continuous subgroup. The global symmetry group of the model is therefore
\begin{align}
\frac{SU(N)\times SU(N-4) \times U(1)_{\chi\eta}\times \mathbb{Z}^F_2}{\mathbb{Z}_{N}\times \mathbb{Z}_{N-4}} \qquad & \text{for even }N \\
\frac{SU(N)\times SU(N-4) \times U(1)_{\chi\eta}}{\mathbb{Z}_{N}\times \mathbb{Z}_{N-4}}  \qquad & \text{for odd }N \label{eq:symmetriesChiEta}
\end{align}

\begin{table}[H]     \renewcommand*{\arraystretch}{1.3}
    \centering 
    \begin{tabular}{|c|c |c|c|c|  }
    \hline
       &  $SU(N)_{\rm c}  $    &  $ SU(N-4)$     &   $ U(1)_{\chi\eta}   $ & $U(1)_{\rm an}$   \\
    \hline 
      $\chi$   &   $ {\bar  { \yng(1,1)}}   $  &    $  \frac{N(N-1)}{2} \cdot (\cdot) $    & $\frac{N-4}{N^*}$ &  $+\frac{N^*(N-3)}{2}$  \\
     $ {\eta}^{A}$      &   $  (N-4)  \cdot   { {\yng(1)}}   $     & $N \, \cdot  \, {\yng (1)}  $     &   $ -\frac{N-2}{N^*}  $ & $-\frac{N^*(N-1)}{2}$  \\
    \hline
    \end{tabular}
    \caption{\footnotesize Classical continuous symmetries of the $\chi\eta$ lagrangian. We decomposed $U(1)_{\chi}\times U(1)_{\eta}$ into its anomaly-free subgroup $U(1)_{\chi\eta}$ and its anomalous subgroup $U(1)_{\rm an}$. We wrote $N^*=\mathrm{gcd}(N,2)$.} 
    \label{tab:chietaTab}
\end{table}

\paragraph{The CFL phase}
In Ref. \cite{Bolognesi:2021yni} it was shown that the 't Hooft anomaly-matching conditions are not fulfilled in the confining phase. They are compatible with the color-flavor locking phase with condensates\footnote{Which cannot be weakly coupled, as shown in \cite{Li:2025tvu} using exact functional RG methods.}
{\small
\begin{align}
    & \expval{\chi_{[ij]}\eta^{A j}}=c_{\chi\eta}\Lambda^3\delta^a_i \ , && i,A=1,\dots,N-4 \nonumber \\
    & \expval{\chi^{[i_1i_2]}\chi^{[i_3i_4]}}=c_{\chi\chi}\epsilon^{i_1i_2i_3i_4}\Lambda^3\ , && i_1,\dots,i_4=N-3,\dots,N
\end{align}}
where $c_{\chi\eta}$ is a gauge-dependent constant. The unbroken symmetry group is then, for $N$ even
\begin{equation}
\label{eq:chietaunbroken}
    H=\frac{SU(N-4)_{\rm cf}\times SU(4)_{\rm c}\times U(1)'\times\mathbb{Z}^F_2}{\mathbb{Z}_{N-4}\times \mathbb{Z}_4}
\end{equation}
$\mathbb{Z}^F_2$ is included in $U(1)'$ if $N$ is odd. The denominator of $H$ consists of the transformations
\begin{align}
\label{eq:chietacenter}
    & \mathbb{Z}_{N}:\ \left(e^{i\frac{2\pi k}{N-4}},\ 1,\ e^{i\frac{2\pi k}{N-4}},\ 1\right) && k=0,\dots, N-5 \nonumber \\
    & \mathbb{Z}_4:\  \left(1,\ e^{i\frac{2\pi \ell}{4}},\ e^{-i\frac{2\pi \ell}{4}},\ e^{i\pi\ell} \right) && \ell=0,\dots,3
\end{align}
The massless degrees of freedom consist of the spin-$\frac{1}{2}$ baryons
\begin{align}
\label{eq:masslesschieta}
    & B^{\{A,B\}}_{\alpha}=\epsilon^{\beta\gamma}\chi^{[ij]}_{\beta}\eta_{i\gamma}^{\{A}\eta_{j\alpha}^{B\}}\ , && A,B=1,\dots,N-4
\end{align}
The transformation laws of the UV and IR fields under the unbroken symmetry group are shown in Table \ref{tab:chietacfl}.

\begin{table}[H]     \renewcommand*{\arraystretch}{1.3}
    \centering 
    \begin{tabular}{|c|c|c |c|c|  }
    \hline
      &      &  $ SU(N-4)_{\rm cf} $     &   $ U(1)^{\prime} $     &  $SU(4)_{\rm c}  $     \\
    \hline
     {\rm UV}&  $\chi^{[i_1A]}$     &    $  {\bar  { \yng(1,1)}}   $    & $+1$   &   $\frac{(N-4)(N-5)}{2}\cdot (\cdot)  $ \\
    &  $\chi_{[i_2j_2]}$   &    $  4   \cdot {\bar  { \yng(1)}} $    & $+\frac{1}{2}$   &   $ (N-4) \cdot {\bar  { \yng(1)}}   $     \\
    & $\chi_{[i_2j_2]}$   &    $  \frac{4 \cdot 3}{2} \cdot (\cdot) $    & $0$    &   ${\bar  { \yng(1,1)}}   $     \\
    & $ \eta^{i_1A}$          & $\yng(2) \oplus \yng(1,1)$     &   $ - 1 $    &   $  (N-4)^2  \cdot  (\cdot)   $  \\
    & $ \eta^{i_2 A}$         & $4\, \cdot  \, {\yng (1)}  $     &   $ - \frac{1}{2} $     &   $  (N-4)  \cdot  \yng(1)  $  \\
    \hline 
      {\rm IR}&     $ B^{\{AB\}}$        &  $ {\yng(2)}$        &    $ - 1 $     &  $  \frac{(N-4)(N-3)}{2} \cdot ( \cdot )    $      \\
    \hline
    \end{tabular}
    \caption{\footnotesize Unbroken symmetries in the CFL phase of the $\chi\eta$ model. The indices are $i_1, A, B=1, \dots, N-4$ and $i_2, j_2=N-3, \dots N$.}
    \label{tab:chietacfl}
\end{table}

\paragraph{Light baryons}

The complete list of light baryons that can be constructed from the elementary fields are
\begin{align} 
\label{eq:baryons101chieta}
    &B^{\{AB\}}_{\{\alpha\beta\gamma\}}=\chi^{[ij]}_{\{\gamma}  \eta_{i,\alpha}^{\{A}  \eta_{j,\beta\}}^{B\}} && \nonumber \\
    & B^{\{AB\}}_{\alpha}=\epsilon^{\beta\gamma}\chi^{[ij]}_{\gamma}  \eta_{i,\alpha}^{\{A}  \eta_{j,\beta}^{B\}} && \nonumber \\
    & B^{[AB]}_{\alpha}=\epsilon^{\beta\gamma}\chi^{[ij]}_{\alpha}  \eta_{i,\beta}^{[A}  \eta_{j,\gamma}^{B]} &&A,B=1,2, \ldots, N-4
\end{align}
The spin-$\frac{3}{2}$ baryons $B^{[AB]}_{\{\alpha\beta\gamma\}}$ cannot be massless due to the Weinberg-Witten theorem \cite{Weinberg:1980kq}. Anomalies \cite{Bolognesi:2020mpe} guarantee that the baryons in Eq. \eqref{eq:masslesschieta} and table \ref{tab:chietacfl} remain exactly massless, while the rest can acquire a mass.

\paragraph{Heavy baryons and their stability}
Heavy baryons can be constructed by attaching to some $\epsilon$-tensors the following words
\begin{align}
    &\chi^{[ij]}\ ,\quad (\chi \eta)^{iA} \ , \quad \bar{\eta}^{iA} \nonumber \\
    & i=1, \dots, N\ , \quad A=1, \dots, N-4
\end{align}
The charges of the words under the unbroken group can be read from table \ref{tab:chietacfl}. There are many structures that may emerge from this construction. We proved numerically that the decay of heavy baryons into massless and light baryons is not always allowed. A discussion on the heavy baryon stability and on what kind of one-particle states the heavy-baryon operators interpolated can be found at the end of subsection \ref{sec:ggintro}.

\subsection{Bars-Yankielowicz (BY) models}
\label{sec:introby}

\paragraph{Lagrangian and symmetries}
The $\psi\eta$ models generalizes to the Bars-Yankielowicz (BY) models, consisting of the Weyl fermions
\begin{align}
    \psi^{\{ij\}} \quad \eta_i^A \quad \xi^{ia}
\end{align}
in the direct-sum representation
\begin{align}
    \yng(2)\oplus (N+4+p)\,\bar{\yng(1)}\oplus p \,\yng(1)
\end{align}
where $i,j$ are color indices while $a, A$ are flavor indices. The continuous symmetries form the group\footnote{It is cubersome to write the global form of the faithful symmetry group for such models \cite{Bolognesi:2021yni}.}

\begin{align}
    SU(N+4+p)\times SU(p)\times U(1)_{\psi\eta}\times U(1)_{\psi\xi}\;.
\end{align}
The charges of the various elementary fermions are summarized in Table  \ref{tab:by}. We omitted the anomalous subgroup $U(1)_{\rm an}$ since it is irrelevant for the rest of this work.

\begin{table*}     
\renewcommand*{\arraystretch}{1.3}
\centering 
{\footnotesize
    \begin{tabular}{|c|c|c|c|c|c|
    }
    \hline
     &  $SU(N)_{\rm c}  $ &  $ SU(N+4+p)$&  $ SU(p)$ & $ {U}(1)_{\psi\eta}$  &   $ {U}(1)_{\psi\xi}  $ 
     \\
    \hline 
    $\psi$   &   $ { \yng(2)} $  &    $  \frac{N(N+1)}{2} \cdot (\cdot) $    & $   \frac{N(N+1)}{2} \cdot (\cdot)  $  & $N +4 +p$ & $p $  \\
    $ \eta$      &   $  (N+4+p)  \cdot   {\bar  {\yng(1)}}   $     & $N  \cdot  {\yng(1)}  $     & $N (N+4+p) \, \cdot  (\cdot)   $   &$-(N+2)$ & $0$   \\ 
    $ \xi$      &   $  p \cdot   {  {\yng(1)}}   $     & $N p \, \cdot  (\cdot)   $     &  $ N  \cdot   { {\yng(1)}} $ &$0$&$-(N+2)$  \\
    \hline   
    \end{tabular}
}
\caption{\footnotesize \small Non-anomalous continuous symmetries of the BY lagrangians. We decomposed $U(1)_{\psi}\times U(1)_{\eta}\times U(1)_{\xi}$ in the anomaly-free subgroups $U(1)_{\psi\eta}$ and $U(1)_{\psi\xi}$.}
\label{tab:by}
\end{table*}

\paragraph{The CFL phase} It has been shown that the 't Hooft anomaly matching conditions are compatible with the color-flavor locking phase, with condensates
\begin{widetext}
\begin{align}
    &\expval{\psi^{\{ij\}}\eta_j^A}=c_{\psi\eta}\Lambda^3\delta^{jA}\ , && j,A=1,\dots,N \nonumber \\
    &\expval{\xi^{ia}\eta_{i}^A}=c_{\xi\eta}\Lambda^3\delta^{N+4+a, A}\ , && a=1,\dots,p\ , \ A=N+5,\dots N+4+p
\end{align}
\end{widetext}
In this phase, the low-energy spectrum consists of $8 N+2N p +(p^2-1)+8 p+2$ NGBs, and of the massless spin-$\frac{1}{2}$ baryons
\begin{align}
\label{eq:masslessBY}
     & {B}^{[A_1B_1]}_{\alpha} = \epsilon^{\beta\gamma}  \psi^{\{ij\}}_{\gamma}   \eta_{i,\alpha}^{[A_1}  \eta_{j,\beta}^{B_1]} \\
    &{B}^{[A_1A_2]}_{\alpha} = \epsilon^{\beta\gamma} \psi^{\{ij\}}_{\gamma}   \eta_{i,\alpha}^{[A_1}  \eta_{j,\beta}^{A_2]}
   \nonumber \\
    &   A_1,B_1 = 1, \dots, N \;,  \quad A_2=N+1, \dots, N+4
\end{align}
The symmetries of the color-flavor locking phase are shown in table \ref{tab:bycfl}.

\begin{table*}
   \renewcommand*{\arraystretch}{1.3}
{\footnotesize
\begin{tabular}{|c|c|c|c|c|c|c|  }
    \hline
    & &   $SU(N)_{{\rm cf}_{\eta}}   $    &  $SU(4)_{\eta}$     &  $ SU(p)_{\eta\xi}$ &   $  U(1)_{\psi \eta}^{\prime}$   &   $ U(1)_{\psi \xi}^{\prime}$ \\
    \hline
    UV & $\psi^{\{ij\}}$   &   $ { \yng(2)} $  &    $  \frac{N(N+1)}{2} \cdot   (\cdot) $    & $ \frac{N(N+1)}{2} \cdot   (\cdot)   $   &  $N +4 +p$ &  $p $  \\
    & $ \eta_{iA_2}$      &   $  {\bar  {\yng(2)}} \oplus {\bar  {\yng(1,1)}}  $     & $N^2  \cdot  (\cdot )  $     &   $ N^2  \cdot  (\cdot ) $    &$-(N +4 +p)$ & $-p $\\
    & $ \eta_{i}^{A_2}$      &   $ 4  \cdot   {\bar  {\yng(1)}}   $     & $N  \cdot  {\yng(1)}  $     &   $ 4 N  \cdot  (\cdot ) $   & $-\frac{N+ p + 4}{2}$  & $-\frac{p}{2}$\\
    & $ \eta_{iA_3}$      &   $ p  \cdot   {\bar  {\yng(1)}}$     & $ N p  \cdot  (\cdot )  $     &   $N  \cdot  \bar{\yng(1)}  $    &$0$ &$N+2$ \\
    & $ \xi^{ia}$      &   $ p  \cdot   { {\yng(1)}}   $     & $N p  \cdot  (\cdot )  $     &   $ N  \cdot   { {\yng(1)}}   $   & $0$  & $-(N+2)$\\
    \hline
    IR & $ {B}^{[A_1B_1]} $ &  $ {\bar  {\yng(1,1)}}$ &  $\frac{N(N-1)}{2} \cdot  (\cdot) $ & $\frac{N(N-1)}{2} \cdot  (\cdot)$ & $-(N +4 +p)$ &  $-p $\\
    & ${B}^{[A_1A_2]} $ &  $4 \cdot {\bar{\yng(1)}}$  &  $N  \cdot  {\yng(1)} $ & $4 N  \cdot  (\cdot ) $   & $-\frac{N+ p + 4}{2}$  & $-\frac{p}{2}$\\
    & $\pi_1$  &   $\yng(1)$  & $(\cdot)$ & $\bar{\yng(1)}$ & $n+p+4$ & $n+p+2$ \\
    & $\pi_2$  &   $\yng(1)$  & $\yng(1)$ & $(\cdot)$ & $\frac{n+p+4}{2}$  & $\frac{p}{2}$ \\
    & $\pi_3$  &   $(\cdot)$  & $(\cdot)$ & $(\text{Adj})$ & $0$ & $0$ \\
    & $\pi_4$  &   $(\cdot)$  & $\yng(1)$ & $\yng(1)$ & $\frac{n+p+4}{2}$  & $\frac{2N +p+4}{2}$ \\
    &$\pi_5,\; \pi_6$ &  $(\cdot)$  & $(\cdot)$ & $(\cdot)$ & $0$ & $0$ \\
    \hline 
    \end{tabular} }
    \caption{\small Unbroken symmetries of the CFL phase of the BY models. The indices are $i, A_1, B_1=1,\dots, N$, $A_2=N+1,\dots, N+4$, $A_3 = N+5,\dots, N+4+p$ and $a=1,\dots, p$.}    
    \label{tab:bycfl}
\end{table*}

\paragraph{Light baryons}
The complete list of light baryons that can be constructed from the elementary fields are
\begin{widetext}
\begin{align}
    &{(B_{1})}^{AB}_{\alpha\beta\gamma}= \psi^{\{ij\}}_{\gamma}   \eta_{i,\alpha}^{A}  \eta_{j,\beta}^{B} && \nonumber \\
    & {(B_{2})}^{a}_{A,\alpha\beta\gamma}= \bar{\psi}_{\{ij\},\gamma}  \bar{\eta}^{i}_{A,\alpha}\xi^{j,a}_{\beta} && \nonumber \\
    & {(B_{3})}_{ab, ,\alpha\beta\gamma}=  \psi^{\{ij\}}_{\gamma}  \bar{\xi}_{i,a,\alpha}  \bar{\xi}_{j,b,\beta} && A=1,\dots, N+4+p\ , a=1,\dots p
\end{align}
\end{widetext}
These operators can be decomposed into a spin-$\frac{3}{2}$ representations and two spin-$\frac{1}{2}$ representations as in Eq. \eqref{eq:baryons101psieta}. Again, the spin-$\frac{3}{2}$ baryons cannot be massless \cite{Weinberg:1980kq}. Anomaly matching \cite{Bolognesi:2021hmg} guarantees that the baryons in Eq. \eqref{eq:masslessBY} and table \ref{tab:bycfl} remain exactly massless, while the rest can acquire a mass.

\paragraph{Heavy baryons and their stability}

Heavy baryons can be constructed by attaching to some $\epsilon$-tensors the following words
\begin{widetext}
\begin{align}
    &\psi^{\{ij\}} \qquad (\psi\bar{\eta})^{i}_{A_1} \qquad (\psi\bar{\eta})^{i}_{A_2} \qquad (\psi\bar{\eta})^{i}_{A_3} \qquad (\psi\xi)^{ia} \nonumber \\
    & \bar{\eta}^i_{A_1} \qquad \bar{\eta}^i_{A_2} \qquad \bar{\eta}^i_{A_3} \qquad \xi^{ia} \nonumber \\
    & i, A_1 = 1, \dots, N\ , \quad A_2= N+1,\dots, N+4\ , \quad A_3 = N+5,\dots, N+4+p\ , \quad a=1, \dots, p
\end{align}
\end{widetext}
The charges of the words under the unbroken subgroup can be read from table \ref{tab:bycfl}. We verified numerically that the decay of heavy baryons into lighter degrees of freedom (massless and light baryons and NGBs) seems to be always allowed by selection rules.

\subsection{Georgi-Glashow (GG) models}
\label{sec:ggintro}

\paragraph{Lagrangian and symmetries}
The $\chi\eta$ models generalizes to the Georgi-Glashow (GG) models, consisting of the Weyl fermions
\begin{align}
    \chi^{[ij]} \qquad \eta_i^A \qquad \xi^{ia}
\end{align}
in the direct-sum representation
\begin{align}
    \yng(1,1)\oplus (N-4+p)\,\bar{\yng(1)}\oplus p \,\yng(1)
\end{align}
where $i,j$ are color indices while $a, A$ are flavor indices. The non-anomalous continuous symmetries form the group
\begin{align}
    SU(N-4+p)\times SU(p)\times U(1)_{\chi\eta}\times U(1)_{\chi\xi}
\end{align}
The charges of the various elementary fermions are summarized in Table \ref{tab:ggTab}.

\begin{table*}
   \renewcommand*{\arraystretch}{1.3}
    \centering
    {\footnotesize
    \begin{tabular}{|c|c|c|c|c|c|}
    \hline
          & $SU(N)_{\text{c}}$ & $SU(N-4+p)$ & $SU(p)_{\xi}$ & $U(1)_{\chi\eta}$ & $U(1)_{\chi\xi}$ \\
         \hline
         $\chi$ & $\yng(1,1)$ & $\frac{N(N-1)}{2}\cdot (\cdot)$ & $\frac{N(N-1)}{2}\cdot (\cdot)$ & $N-4+p$ & $p$ \\
         $\eta$ & $(N-4+p)\cdot \bar{\yng(1)}$ & $N\cdot \yng(1)$ & $N(N-4+p)\cdot(\cdot)$ & $-(N-2)$ & $0$ \\
         $\xi$ & $p\cdot\yng(1)$ & $pN\cdot (\cdot) $ & $N\cdot \yng(1) $ & $0$ & $-(N-2)$  \\
         \hline
    \end{tabular}}
    \caption{\footnotesize \small Non-anomalous symmetries of the GG lagrangians. We decomposed $U(1)_{\chi}\times U(1)_{\eta}\times U(1)_{\xi}$ in the anomaly-free subgroups $U(1)_{\chi\eta}$ and $U(1)_{\chi\xi}$.}
    \label{tab:ggTab}
\end{table*}

\paragraph{The CFL phase} It has been shown that the 't Hooft anomaly matching conditions are compatible with the color-flavor locking phase, with condensates
\begin{widetext}
\begin{align}
\label{eq:ggvevs}
    &\expval{\chi^{[ij]}\eta_j^A}=c_{\psi\eta}\Lambda^3\delta^{jA}\ , && j,A=1,\dots,N-4 \nonumber \\
    &\expval{\xi^{ia}\eta_{i}^A}=c_{\xi\eta}\Lambda^3\delta^{N+4+a, A}\ , && a=1,\dots,p\ , \ A=N-3,\dots N-3+p \nonumber \\
    & \expval{\chi^{[i_1i_2]}\chi^{[i_3i_4]}}=c_{\chi\chi}\epsilon^{i_1i_2i_3i_4}\Lambda^3\ , && j_1,\dots,j_4=N-3,\dots,N
\end{align}
\end{widetext}
In this phase, the low-energy spectrum consists of the $2Np-8N +1$ NGBs and of the massless spin-$\frac{1}{2}$ baryons
\begin{equation}
\label{eq:masslessGG}
    { B}^{\{AB\}}_{\alpha} =  \epsilon^{\beta\gamma} \chi^{[ij]}_{\beta}   \eta_{i\gamma}^{A}  \eta_{j\alpha}^{B} \;,  \qquad  A,B = 1, \dots, N-4
\end{equation}
The symmetries of the color-flavor locking phase are shown in table \ref{tab:ggcfl}. 

\begin{table*}
   \renewcommand*{\arraystretch}{1.3}
{  \centering 
    \footnotesize{\begin{tabular}{|c|c|c |c|c|c|c|  }
    \hline
    & &    $SU(N-4)_{{\rm cf}_{\eta}}   $  &  $SU(4)_{{\rm c}}$ &  $ SU(p)_{\eta\xi}$ &   $  U(1)_{\chi \eta}^{\prime}$   &   $ U(1)_{\chi \xi}^{\prime}$ \\
    \hline
    UV & $\chi^{[i_1j_1]}$     &       $ {\yng(1,1)} $ & $  \frac{(N-4)(N-5)}{2} \cdot   (\cdot) $  & $ \frac{(N-4)(N-5)}{2} \cdot   (\cdot)   $   &  $\frac{(N-4+p) N}{(N-4)}    $  &  $p \frac{N}{N-4}$  \\
    & $\chi^{[i_1j_2]}$ &   $ 4 \cdot { \yng(1)} $  &  $  (N-4)  \cdot    { \yng(1)} $    & $ 4(N-4) \cdot   (\cdot)   $   &$\frac{(N-4+p) N}{2(N-4)}  $  &  $\frac{p N}{2(N-4)}$  \\
    & $\chi^{[i_2j_2]}$  &    $  6  \cdot   (\cdot) $   &   $ { \yng(1,1)} $   & $ 6  \cdot   (\cdot)   $   &  $0$  & $0$  \\
    & $ \eta_{i_1A_1}$ &  $  {\bar  {\yng(2)}} \oplus {\bar  {\yng(1,1)}}  $     &   $(N-4)^2  \cdot   (\cdot )    $   &   $ (N-4)^2  \cdot  (\cdot ) $   & $ -\frac{(N-4+p) N}{(N-4)} $  & $-\frac{p N}{N-4} $\\   
    & $ \eta_{i_1 A_2}$ & $ p  \cdot \bar{\yng(1)} $  &   $p(N-4)  \cdot  (\cdot )  $      &   $ (N-4)  \cdot \bar{\yng(1)} $    &$-2 -2 \frac{p}{N-4}$ & $N-2 -\frac{2 p}{N-4}  $\\
    & $ \eta_{i_2A_1}$ &   $  4   \cdot   {\bar  {\yng(1)}}$     & $ (N-4)  \cdot   {\bar  {\yng(1)}} $  &   $ 4  (N-4)     \cdot   (\cdot )  $    &$-\frac{(N-4+p) N}{2(N-4)} $ &$ -\frac{p N}{2(N-4)} $ \\
    & $ \eta_{i_2 A_2}$ &   $ 4 p \cdot  (\cdot ) $  & $ p \cdot   {\bar  {\yng(1)}}   $     &   $4  \cdot \bar{\yng(1)}   $    &$\frac{N-4+p }{2}$ &$N-2  + \frac{p }{2}$ \\
    & $ \xi^{i_1a}$  &   $  p  \cdot   { {\yng(1)}}    $     & $p(N-4)   \cdot  (\cdot )  $     &   $ (N-4)   \cdot   { {\yng(1)}}   $   & $2 + 2\frac{p}{N-4}$  & $-(N-2)+\frac{2p}{N-4}$\\
    & $ \xi^{i_2a}$  &   $4 p   \cdot  (\cdot )   $     & $  p  \cdot   { {\yng(1)}}    $     &   $ 4  \cdot   { {\yng(1)}}   $   & $-\frac{N-4+p}{2}$  & $-(N-2)-\frac{p}{2}$\\
    \hline 
    IR & $ B^{\{A_1B_2\}}   $ &  $  {\bar  {\yng(2)}}  $  &  $\frac{(N-4)(N-3)}{2}\cdot(\cdot)$  &   $\frac{(N-4)(N-3)}{2}\  \cdot  (\cdot ) $   &  $ -\frac{(N-4+p) N}{(N-4)} $  & $-\frac{p N}{N-4} $\\
    & $\pi_1$ & $\bar{\yng(1)}$ &  $(\cdot)$ & $\yng(1)$ & $-\frac{(N+p-4)(N-2)}{N-4}$ & $-\frac{(N+p-4)(N-2)}{N-4}$ \\
    & $\pi_2$ & $p^2\cdot(\cdot) $ & $p^2\cdot(\cdot) $ & (Adj) & $0$ & $0$ \\
    & $\pi_3$ & $\cdot(\cdot) $ & $\cdot(\cdot) $ & $\cdot(\cdot) $ & $0$ & $0$ \\
    \hline
    \end{tabular}}  
    \caption{\footnotesize \small Unbroken symmetries in the CFL phase of the GG models. The indices are $i_1,j_1, A_1, B_1=1,\dots, N-4$, $i_2,j_2=N-3,\dots N$, $A_2=N-4,\dots N-4+p $, $a=1,\dots, p$}
    \label{tab:ggcfl}
}
\end{table*}

The complete list of light baryons that can be constructed from the elementary fields are
\begin{align}
    &{(B_{1})}^{AB}_{\alpha\beta\gamma}= \chi^{[ij]}_{\gamma}   \eta_{i,\alpha}^{A}  \eta_{j,\beta}^{B} && \nonumber \\
    & {(B_{2})}^{a}_{A,\alpha\beta\gamma}=    \bar{\chi}_{[ij],\gamma}  \bar{\eta}^{i}_{A,\alpha}\xi^{j,a}_{\beta} && \nonumber \\
    & {(B_{3})}_{ab, ,\alpha\beta\gamma}=  \chi^{[ij]}_{\gamma}  \bar{\xi}_{i,a,\alpha}  \bar{\xi}_{j,b,\beta} && A,B=1,\dots, N\ , a, b = 1,\dots, p
\end{align}
These operators can be decomposed into a spin-$\frac{3}{2}$ representations and two spin-$\frac{1}{2}$ representations as in Eq. \eqref{eq:baryons101chieta}. Again, the spin-$\frac{3}{2}$ baryons cannot be massless \cite{Weinberg:1980kq}. Anomaly matching \cite{Bolognesi:2021hmg} guarantees that the baryons in Eq. \eqref{eq:masslessGG} and table \ref{tab:ggcfl} remain exactly massless, while the rest can acquire a mass.

\paragraph{Heavy baryons and their stability}
Heavy baryons can be constructed by attaching to some $\epsilon$-tensors the following words
\begin{widetext}
\begin{align}
    &\chi^{[ij]} \qquad (\chi\eta)^{i}_{A_1} \qquad (\chi\eta)^{i}_{A_2} \qquad (\chi\bar{\xi})^{ia} \nonumber \\
    & \bar{\eta}^{iA_1} \qquad \bar{\eta}^{iA_2} \qquad \xi^{ia} \nonumber \\
    &i=1, \dots, N\ , \quad A_1=1,\dots, N-4\ , \quad A_2=N-3,\dots, N-4+p\ , \quad a=1,\dots, p
\end{align}
\end{widetext}
The charges of the words under the unbroken subgroup can be read from table \ref{tab:ggcfl}. In contrast to the BY models, not all heavy baryons are allowed by the unbroken symmetries to decay into lighter particles. An explicit example of heavy baryon whose decay into light baryons and NGBs is forbidden is
\begin{align}
    &\epsilon_{i_1\dots i_N}\bar{\eta}^{i_1A_1}\dots \bar{\eta}^{i_NA_N}  && A_1,\dots A_N = N-4, \dots N-4+p
\end{align}
As anticipated in subsections \ref{sec:cflstrong} and \ref{sec:introchieta}, there is a class of heavy baryons that is affected by the separation of scales or lack thereof. Its simplest representative
\begin{align}
\label{eq:dressed}
    &\epsilon_{i_1\dots j_4}(\chi\eta)^{i_1A_1}\dots (\chi\eta)^{i_{N-4}A_{N-4}}\bar{\eta}^{j_1B_1}\dots\bar{\eta}^{j_4B_4} \nonumber \\
    & A_1,\dots A_{N-4} = 1 ,\dots, N-4 \nonumber \\
    & B_1,\dots, B_4 = N-3,\dots, N-4+p
\end{align}
At weak coupling, if we are in a perturbative CFL regime, it is tempting to freeze the $(\chi\eta)^{iA}$ to their vacuum expectation value as in Eq. \eqref{eq:ggvevs}. The remaining dynamical degrees of freedom are those of a $SU(4)_{\rm c}$ baryon interpolated by the gauge-variant (with respect to the full gauge group) operator
    \begin{align}
        &\epsilon_{j_1\dots j_4}\bar{\eta}^{j_1B_1}\dots\bar{\eta}^{j_4B_4} \nonumber \\
        & j_1,\dots j_4 = N-3,\dots, N\nonumber \\
        & B_1,\dots, B_4 = N-3,\dots, N-4+p
    \end{align}
However, we are interested in the strongly coupled scenario where there is no separation of scales, therefore, we expect that we cannot disentangle the NGBs from the operator \eqref{eq:dressed} interpolate a heavy baryon with mass $\mathcal{O}(N)$.


The factorization of the operator \eqref{eq:dressed} into $N-4$ NGBs and one $SU(4)_{\rm c}$ baryon occurs only if $f_{\pi}\gg  \Lambda_{SU(4)_{\rm c}}$, where $f_{\pi}$ is the NGB decay constant, that is the CFL symmetry breaking scale, while $\Lambda_{SU(4)_{\rm c}}$ is the $SU(4)_{\rm c}$ renormalization-group invariant scale. However, this happens only if the symmetry breaking occurs at weak coupling like in the Standard Model. When the symmetry breaking occurs at strong coupling, like in the present case, we expect
\begin{align}
    f_{\pi}\simeq \Lambda_{SU(4)_{\rm c}}
\end{align}
Hence, even initially assuming that the operator \eqref{eq:dressed} creates one $SU(4)_{\rm c} $ dressed with $N-4$ NGBs, the binding between them is so strong that they cannot be disentangled and they must effectively be considered a one-particle state with binding energy that scales as $\mathcal{O}(N)$.

\section{Vector-like gauge theories: symmetries and baryons}
\label{sec:introvect}
In this section, we consider two vector-like gauge theories with mixed representations. We discuss the symmetries and classify the baryon operators.

\subsection{$\psi\tilde{\psi}\eta\tilde{\eta}$ models}

\paragraph{Lagrangian and symmetries}
The $\psi\tilde{\psi}\eta\tilde{\eta}$ models contains the Weyl fermions
\begin{align}
    \psi^{\{ij\}A} \qquad \tilde{\psi}^A_{\{ij\}} \qquad \eta_i^a \qquad \tilde{\eta}^{ia}
\end{align}
in the direct-sum representations
\begin{align}
    N_{\psi}\ \yng(2)\oplus N_{\psi}\ \bar{\yng(2)}\oplus N_\eta\ \bar{\yng(1)}\oplus N_{\eta}\ \bar{\yng(1)}
\end{align}
where $i,j$ are color indices while $A,a$ are flavor indices. The global symmetries of this model are shown in table \ref{tab:psietavec}. The ABJ anomaly breaks the axial $U(1)$'s subgroups to 
\begin{equation}
    U(1)_{\psi A}\times U(1)_{\eta A}\longrightarrow U(1)_{\psi\eta}\times \mathbb{Z}_{\widetilde{N}}\;,
\end{equation}
where $\widetilde{N}=2 \cdot {\rm gcd}(N_\psi (N+2), N_\eta)$.

\begin{table*}
   \renewcommand*{\arraystretch}{1.3}
    \centering
    \begin{tabular}{|c|c|c|c|c|c|c|}
    \hline
         & $SU(N)_{\rm c}$ & $SU(N_{\psi})_L$ & $SU(N_{\psi})_{R}$ & $U(1)_{\psi V}$ &  $U(1)_{\psi\eta}$ & $U(1)_{\rm an}$\\
        \hline
        $\psi$ & $N_{\psi}\cdot \yng(2)$ & $\frac{N(N+1)}{2}\cdot \yng(1)$ & $\frac{N(N+1)}{2}\cdot (\cdot)$ & $+1$ & $-N_{\eta}$ & $+1$ \\
        $\tilde{\psi}$ & $N_{\psi}\cdot \bar{\yng(2)}$ & $\frac{N(N+1)}{2}\cdot (\cdot)$ & $\frac{N(N+1)}{2}\cdot \yng(1)$ & $-1$ & $-N_{\eta}$ & $+1$ \\
        \hline
        \hline
        & $SU(N)_{\rm c}$ & $SU(N_{\eta})_L$ & $SU(N_{\eta})_{R}$ & $U(1)_{\eta V}$ & $U(1)_{\psi\eta}$ &  $U(1)_{\rm an}$ \\
        \hline
        $\eta$ & $N_{\eta}\cdot \bar{\yng(1)}$ & $N\cdot \yng(1)$ & $N\cdot (\cdot)$ & $+1$ & $N_{\psi}(N+2)$  & $+1$ \\
        $\tilde{\eta}$ & $N_{\eta}\cdot \yng(1)$ & $N\cdot (\cdot)$ & $N\cdot \yng(1)$ & $-1$ & $N_{\psi}(N+2)$ &  $+1$ \\
        \hline
    \end{tabular}
    \caption{\footnotesize Symmetries of the lagrangian of the $\psi\tilde{\psi}\eta\tilde{\eta}$ vector-like model. We omitted the action of the symmetries $SU(N_{\psi})_{L, R}$ and $U(1)_{\psi V}$ on $\eta, \tilde{\eta}$ and the action of  $SU(N_{\eta})_{L, R}$ and $U(1)_{\eta V}$ on $\psi, \tilde{\psi}$ because they are trivial. The subgroup $U(1)_{\rm an}$ is broken by the axial anomaly.}
    \label{tab:psietavec}
\end{table*}

\paragraph{Chiral symmetry breaking}
The fermion bilinears form the gauge-invariant condensates
\begin{align}
\label{eq:veccond}
    &\langle\tilde{\psi}^A\psi^B\rangle=c_{\tilde{\psi}\psi}\Lambda^3\delta^{AB} 
    &&\expval{\tilde{\eta}^a\eta^b}=c_{\tilde{\eta}\eta}\Lambda^3\delta^{ab}
\end{align}
where $c_{\tilde{\psi}\psi}$ and $c_{\tilde{\eta}\eta}$ are gauge-invariant constants. The unbroken symmetry group is then
\begin{align}
\label{eq:psietavecunbrk}
    H=\frac{SU(N)_{\rm c}\times SU(N_{\psi})_V\times SU(N_{\eta})_V\times U(1)_{\psi V}\times U(1)_{\eta V}}{\mathbb{Z}_N\times \mathbb{Z}_{N_\psi}\times \mathbb{Z}_{N_\eta}}\;,
\end{align}
where
\begin{align}
\label{eq:psietavectcenter}
    & \mathbb{Z}_N:\ \left(e^{i\frac{2\pi k}{N}}, 1,1, e^{-i\frac{4\pi k}{N}}, e^{i\frac{2\pi k}{N}}\right)\;, && k = 0,\dots N-1\;,\nonumber \\
    & \mathbb{Z}_{N_\psi}:\ \left(1, e^{i\frac{2\pi \ell}{N_{\psi}}}, 1, e^{-i\frac{2\pi \ell}{N_{\psi}}}, 1\right)\;, && \ell = 0,\dots, N_{\psi}-1\;, \nonumber \\
    & \mathbb{Z}_{N_\eta}:\ \left(1, 1, e^{i\frac{2\pi \ell}{N_{\eta}}},1 , e^{-i\frac{2\pi \ell}{N_{\eta}}}\right)\;, && m = 0,\dots, N_{\eta}-1\;,
\end{align}
In this phase, the massless degrees of freedom consist of $N_{\psi}^2+N_{\eta}^2-1$ NGBs created by the operators
\begin{align}
    &U^{AB}=\tilde{\chi}^A\chi^B && V^{ab}=\tilde{\eta}^a\eta^b
\end{align}

\paragraph{Light baryons}
In the phase characterized by the condensates \eqref{eq:veccond} the low-energy theory consists of the $N_{\psi}^2+N_{\eta}^2-1$ NGBs. The theory admits the light baryons 
\begin{align}
\label{eq:lightpsieta}
    B^{A,bc}_{\alpha\beta\gamma}=\psi^{\{ij\}A}_{\gamma}\eta_{i,\alpha}^b\eta_{j,\beta}^c \quad \text{and} \quad \tilde{B}^{A\ bc}=\tilde{\psi}_{\{ij\},\gamma}^A\tilde{\eta}^{i,b}_{\beta}\tilde{\eta}^{j,c}_{\beta}\;,
\end{align} 
that can be decomposed into a spin-$\frac{3}{2}$ baryon and two spin-$\frac{1}{2}$ baryons as in Eq. \eqref{eq:baryons101psieta}. The spin-$\frac{3}{2}$ baryon cannot be massless \cite{Weinberg:1980kq}, while the spin-$\frac{1}{2}$ baryons acquire a mass through the coupling with the mesons. This mass is not suppressed in the large-$N$ limit, as we now show. The large-$N$ scaling of the simplest meson and baryon correlators is
\begin{align}
    & \expval{\bar{B}U B} \sim N^2\;, && \expval{\bar{B}V B} \sim N^2\;,  && \expval{B\bar{B}}\sim N^2\;,\nonumber \\
    & \expval{UU}\sim N^2\;, && \expval{VV}\sim N\;, \nonumber \\
    & \expval{U}\sim N^2\;, && \expval{V}\sim N\;,
\end{align}
and similarly for the two- and three-point correlators involving $\tilde{B}$. It follows that the low-energy effective largangian for the mesons and light-baryons looks like
\begin{align}
    \mathcal{L}\sim& \frac{1}{N^2}\left(\bar{B}\slashed{\partial}B+\bar{\tilde{B}}\slashed{\partial}\tilde{B}\right)+\frac{1}{N^2}\bar{U}(-\partial^2)U+\frac{1}{N}\bar{V}(-\partial^2)V \nonumber \\
    +&\frac{g_{\psi}}{N^4}\left( \bar{B}UB+\bar{\tilde{B}}U\tilde{B}\right)+ \frac{g_{\eta}}{N^3}\left( \bar{B}VB+\bar{\tilde{B}}V\tilde{B}\right)\;.
\end{align}
The Yukawa couplings $g_{\psi}$ and $g_{\eta} $ have been chosen to satisfy parity conservation, that in vector-like gauge theories cannot be spontaneously broken \cite{Vafa:1984xg}. Then, the mass $m_B$ of the light-baryons scale as
\begin{equation}
    m_B\sim N^2\left(\frac{g_{\psi}}{N^4}\expval{U}+\frac{g_{\eta}}{N^3}\expval{V}\right)\sim N^0\;.
\end{equation}
The transformation laws of the UV and IR fields under the unbroken symmetry group are shown in table \ref{tab:psietavecunbrk}.

\begin{table*}
   \renewcommand*{\arraystretch}{1.3}
    \centering
    \begin{tabular}{|c|c|c|c|c|c|c|}
    \hline
        & & $SU(N)_{\rm c}$ & $SU(N_{\psi})_V$ & $SU(N_{\eta})_{V}$ & $U(1)_{\psi V}$ &  $U(1)_{\eta V}$\\
        \hline
        UV &$\psi$ & $N_{\psi}\cdot \yng(2)$ & $\frac{N(N+1)}{2}\cdot \yng(1)$ & $\frac{N(N+1)}{2}\cdot (\cdot)$ & $+1$ & $0$ \\
        &$\tilde{\psi}$ & $N_{\psi}\cdot \bar{\yng(2)}$ & $\frac{N(N+1)}{2}\cdot \bar{\yng(1)}$ & $\frac{N(N+1)}{2}\cdot (\cdot)$ & $-1$  & $0$ \\
        &$\eta$ & $N_{\eta}\cdot \bar{\yng(1)}$ & $N\cdot (\cdot)$ & $N\cdot \yng(1)$ & $0$    & $+1$ \\
        &$\tilde{\eta}$ & $N_{\eta}\cdot \yng(1)$ & $N\cdot (\cdot)$ & $N\cdot \bar{\yng(1)}$ & $0$  & $-1$  \\
        \hline
        IR$^*$ & $ B $ & $\frac{N_{\psi}N_{\eta}(N_{\eta}-1)}{2}\cdot (\cdot) $ & $ \frac{N_{\eta}(N_{\eta}-1)}{2}\cdot \yng(1)$  & $N_{\psi}\cdot \yng(1,1)$ & $+1$ & $+2$ \\
         & $\tilde{B}$ & $\frac{N_{\psi}N_{\eta}(N_{\eta}-1)}{2}\cdot (\cdot) $ & $ \frac{N_{\eta}(N_{\eta}-1)}{2}\cdot \bar{\yng(1)}$ & $N_{\psi}\cdot \bar{\yng(1,1)}$ & $-1$ & $-2$ \\
         & $\pi_1$ & $(\cdot) $ & (Adj) & $N_{\psi}^2\cdot(\cdot)$ & $0$ & $0$ \\
         & $\pi_2$ & $(\cdot)$ & $N_{\eta}^2\cdot(\cdot)$ & (Adj) & $0$ & $0$ \\
        \hline
    \end{tabular}
    \caption{\footnotesize Unbroken symmetries of the $\psi\tilde{\psi}\eta\tilde{\eta}$ vector-like model. }
    \label{tab:psietavecunbrk}
\end{table*}

\paragraph{Heavy baryons}
We now turn to heavy baryons. We indicate a generic heavy baryon as
\begin{equation}
\label{eq:genericbaryonpsi}
    {\cal B}_{(l,n,m,p,\bar{n},\bar{m},\bar{p})}
\end{equation}
where $l$ is the number of $\epsilon_{i_1\dots i_N}$ if $l>0$ (or $\epsilon^{i_1\dots i_N}$ if $l<0$ ). We have $N l$ indices than can be saturated with $n$ $\psi^{\{ij\}}$ or $\bar{n}$ $\bar{\tilde{\psi}}^{\{ij\}}$ , $m$ $\psi^{\{i k\}}\eta_{k}$ or $\bar{m}$ $\bar{\tilde{\psi}}^{\{i k\}}\bar{\tilde{\eta}}_{k}$ , $p$ $\bar{\eta}^i $ or $\bar{p}$ ${\tilde{\eta}}^{i}$. $N l = 2n + m +p+2\bar{n} + \bar{m} +\bar{p}$.  For example the with $l=1$ $n=\bar{n}=0$, and limiting to only left-handed fermions  we have 
{\small
\begin{equation}
  {\cal B}_{(1,0,m,p,0,0,0)}   = \epsilon_{i_1\dots i_N} \psi^{\{i_1 k_1\}}\eta_{k_1} \dots \psi^{\{i_{m} k_{m}\}}\eta_{k_{m}} \tilde{\eta}^{m+1} \dots \tilde{\eta}^{N}\;.
\end{equation}}
The baryons of QCD with the fundamental quarks are, in this notation
\begin{align}
  &{\cal B}_{(1,0,0,N,0,0,0)}   = \epsilon_{i_1\dots i_N} \  \tilde{\eta}^{1} \dots \tilde{\eta}^{N} \nonumber \\
  &{\cal B}_{(1,0,0,0,0,0,N)}   = \epsilon_{i_1\dots i_N} \  \bar{\eta}^{1} \dots \bar{\eta}^{N}\;.
\end{align}
\sloppy The fully antisymetric baryons with only $\psi$ or $\tilde{\psi}$ are ${\cal B}_{(N+1,\frac{N (N+1)}{2},0,0,0,0,0)}  $, ${\cal B}_{(N+1,0,0,0,0,0,\frac{N (N+1)}{2})}  $. These are the ones discussed in \cite{Bolognesi:2006ws}. The set of numbers $(l,n,m,p,\bar{n},\bar{m},\bar{p})$ allows to determine the transformation laws of a given baryon only under the abelian and discrete symmetry subgroups. The information concerning the transformation properties under the non-abelian subgroups cannot be encoded in the integers $(l,n,m,p,\bar{n},\bar{m},\bar{p})$ since it depends nontrivially on the index structure of each baryon. The $U(1)_{\eta V}$, $U(1)_{\psi V}$ charges and fermionic number of the baryons are given in table \ref{tablebaryonsvectorialpsi}.

\begin{table}[H]
   \renewcommand*{\arraystretch}{1.3}
\centering 
\begin{tabular}{ |c| c|c|c| }
\hline
baryons  &      $U(1)_{\psi V}$  &    $U(1)_{\eta V}$   &  ${\mathbb{Z}}_{2}^F$   \\
\hline  
 $ B = \psi\eta\eta $      & $1$        & $2$   & $-1$ \\
 \hline  
 $ \tilde{B} = \tilde{\psi}\tilde{\eta}\tilde{\eta}  $      & $-1$        & $-2$   & $-1$ \\
   \hline  
 $  {\cal B}_{(1,0,0,N,0,0,0)} $    &$0$& $N$   & $(-1)^{N}$  \\
  \hline  
 $ {\cal B}_{(N+1,\frac{N (N+1)}{2},0,0,0,0,0)}  $    &$\frac{N (N+1)}{2}$& $0$  & $(-1)^{\frac{N (N+1)}{2}}$  \\
  \hline  
 ${\cal B}_{(l,n,m,p,\bar{n},\bar{m},\bar{p})} $    &$n-\bar{n}$& $m+p-\bar{m}-\bar{p}$   & $(-1)^{n + \bar{n} +p+\bar{p}}$  \\
	\hline
	\end{tabular}
\caption{\footnotesize Baryons and the charges under the unbroken $U(1)$ symmetries in $\psi\tilde{\psi}\eta\tilde{\eta} $ the vector-like model. 
}\label{tablebaryonsvectorialpsi}
\end{table}

\begin{figure*}
	\begin{center}
		\includegraphics[scale=0.70]{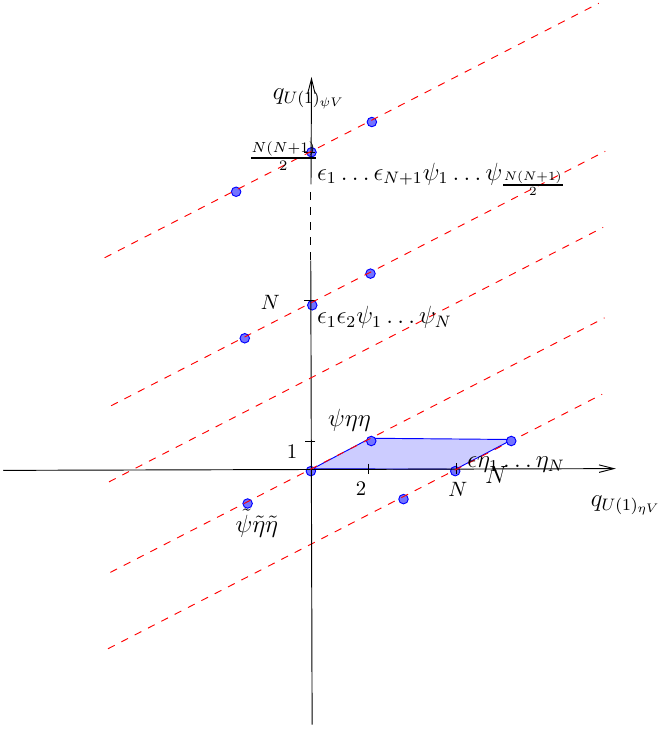} 
		\caption{Lattice of possible baryonic charges. The area in purple is the fundamental cell. The light baryons do not span the entire lattice.}
		\label{baryoncharges}
	\end{center}
\end{figure*}

The lattice of possible $U(1)_{V\psi}\times U(1)_{V\eta}$ charges is the subset of $\mathbb{Z}\times \mathbb{Z}$ satisfying
\begin{equation}
    2 q_{U(1)_{\eta V}}-q_{U(1)_{\psi V}}=0 \mod N
\end{equation}
A basis for such sub-lattice is given by the vectors $(2, 1)$ and $(0, N_{\rm f})$, corresponding to the light baryon, $\psi\eta\eta$, and the hevay baryon, $\epsilon_{i_1\dots i_N} \  \tilde{\eta}^{1} \dots \tilde{\eta}^{N}$, respectively (see Fig.~\ref{baryoncharges} for a cartoon).  This implies the stability of some heavy baryon.

\subsection{$\chi\tilde{\chi}\eta\tilde{\eta}$ models}

\paragraph{Lagrangian and symmetries}
The $\chi\tilde{\chi}\eta\tilde{\eta}$ models contains the Weyl fermions
\begin{align}
    \chi^{[ij]A}\ , \quad \tilde{\chi}^A_{[ij]}\ , \quad \eta_i^a\ , \quad \tilde{\eta}^{ia}
\end{align}
in the direct-sum representations
\begin{align}
    N_{\chi}\ \yng(1,1)\oplus N_{\chi}\ \bar{\yng(1,1)}\oplus N_\eta\ \bar{\yng(1)}\oplus N_{\eta}\ \bar{\yng(1)}
\end{align}
where $i,j$ are color indices while $A,a$ are flavor indices. This theory has been considered before in \cite{Armoni:2005wt}.\footnote{With $N=3$ and $N_{\chi}=1$, the main purpose of \cite{Armoni:2005wt,Armoni:2014ywa} was to have a large $N$ planar equivalence of multi-flavor QCD with supersymmetric YM.} The ABJ anomaly breaks the axial symmetries
\begin{equation}
    U(1)_{\chi A}\times U(1)_{\eta A}\longrightarrow \mathbb{Z}_{\tilde{N}}\;, 
\end{equation}
where $\widetilde{N}={\rm gcd}(2N_{\chi} (N-2),2N_{\eta})$. The global symmetries of this model are shown in table \ref{tab:chietavec}.

\begin{table*}
   \renewcommand*{\arraystretch}{1.3}
    \centering
    \begin{tabular}{|c|c|c|c|c|c|c|}
    \hline
         & $SU(N)_{\rm c}$ & $SU(N_{\chi})_L$ & $SU(N_{\chi})_{R}$ & $U(1)_{\chi V}$ &  $\widetilde{U}(1)_{\chi\eta}$  & $U(1)_{\rm an}$ \\
        \hline
        $\chi$ & $N_{\chi}\cdot \yng(2)$ & $\frac{N(N-1)}{2}\cdot \yng(1)$ & $\frac{N(N-1)}{2}\cdot (\cdot)$ & $+1$ & $-N_{\eta}$ & $+1$  \\
        $\tilde{\chi}$ & $N_{\chi}\cdot \bar{\yng(2)}$ & $\frac{N(N-1)}{2}\cdot (\cdot)$ & $\frac{N(N-1)}{2}\cdot \yng(1)$ & $-1$ & $-N_{\eta}$ & $+1$  \\
        \hline
        \hline
        & $SU(N)_{\rm c}$ & $SU(N_{\eta})_L$ & $SU(N_{\eta})_{R}$ & $U(1)_{\eta V}$ & $\widetilde{U}(1)_{\chi\eta}$ & $U(1)_{\rm an}$  \\
        \hline
        $\eta$ & $N_{\eta}\cdot \bar{\yng(1)}$ & $N\cdot \yng(1)$ & $N\cdot (\cdot)$ & $+1$ & $N_{\chi}(N-2)$ & $+1$  \\
        $\tilde{\eta}$ & $N_{\eta}\cdot \yng(1)$ & $N\cdot (\cdot)$ & $N\cdot \yng(1)$ & $-1$ & $N_{\chi}(N-2)$ & $+1$  \\
        \hline
    \end{tabular}
    \caption{\footnotesize Symmetries of the lagrangian of the $\chi\tilde{\chi}\eta\tilde{\eta}$ vector-like model. We omitted the action of the symmetries $SU(N_{\chi})_{L, R}$ and $U(1)_{\chi V}$ on $\eta, \tilde{\eta}$ and the action of  $SU(N_{\eta})_{L, R}$ and $U(1)_{\eta V}$ on $\chi, \tilde{\chi}$ because they are trivial. The subgroup $U(1)_{\rm an}$ is broken by the axial anomaly.}
    \label{tab:chietavec}
\end{table*}

\paragraph{Chiral symmetry breaking}
The fermions form the gauge-invariant condensates
\begin{align}
    &\expval{\tilde{\chi}^A\chi^B}=c_{\tilde{\chi}\chi}\Lambda^3\delta^{AB} 
    &&\expval{\tilde{\eta}^a\eta^b}=c_{\tilde{\eta}\eta}\Lambda^3\delta^{ab}
\end{align}
where $c_{\chi}$ and $c_{\tilde{\eta}\eta}$ are gauge-invariant constants.
The unbroken symmetry group is then
\begin{align}
\label{eq:chietavecunbrk}
    H=\frac{SU(N)_{\rm c}\times SU(N_{\chi})_V\times SU(N_{\eta})_V\times U(1)_{\chi V}\times U(1)_{\eta V}}{\mathbb{Z}_N\times \mathbb{Z}_{N_\chi}\times \mathbb{Z}_{N_\eta}}
\end{align}
where
\begin{align}
\label{eq:chietaveccenter}
    & \mathbb{Z}_N:\ \left(e^{i\frac{2\pi k}{N}}, 1,1, e^{-i\frac{4\pi k}{N}}, e^{i\frac{2\pi k}{N}}\right) && k = 0,\dots N-1\nonumber \\
    & \mathbb{Z}_{N_\chi}:\ \left(1, e^{i\frac{2\pi \ell}{N_{\chi}}}, 1, e^{-i\frac{2\pi \ell}{N_{\chi}}}, 1\right) && \ell = 0,\dots, N_{\chi}-1 \nonumber \\
    & \mathbb{Z}_{N_\eta}:\ \left(1, 1, e^{i\frac{2\pi \ell}{N_{\eta}}},1 , e^{-i\frac{2\pi \ell}{N_{\eta}}}\right) && m = 0,\dots, N_{\eta}-1
\end{align}
In this phase, the massless degrees of freedom consist of the $N_{\chi}^2+N_{\eta}^2-1$ NGBs created by the operators
\begin{align}
    &U^{AB}=\tilde{\chi}^A\chi^B && V^{ab}=\tilde{\eta}^a\eta^b
\end{align}

\paragraph{Light baryons} One can construct the baryons
\begin{align}
    & B^{A,bc}_{\alpha\beta\gamma}=\chi^{[ij]}_{\gamma}\eta^b_{i,\alpha}\eta^c_{j,\beta} && \tilde{B}^{A, bc}=\tilde{\chi}^A_{[ij], \gamma}\tilde{\eta}^{i,b}_{\alpha}\tilde{\eta}^{j,c}_{\beta}
\end{align}
that can be decomposed into a spin-$\frac{3}{2}$ baryon and two spin-$\frac{1}{2}$ baryons as in Eq. \eqref{eq:baryons101chieta}. The spin-$\frac{3}{2}$ baryon cannot be massless \cite{Weinberg:1980kq}, while the spin-$\frac{1}{2}$ baryons acquire a mass through the coupling with the mesons. This mass is not suppressed in the large-$N$ limit, as we now show. The large-$N$ scaling of the simplest meson and baryon correlators is
\begin{align}
    & \expval{\bar{B}U B} \sim N^2 && \expval{\bar{B}V B} \sim N^2  && \expval{B\bar{B}}\sim N^2\nonumber \\
    & \expval{UU}\sim N^2 && \expval{VV}\sim N \nonumber \\
    & \expval{U}\sim N^2 && \expval{V}\sim N
\end{align}
and similarly for the two- and three-point correlators involving $\tilde{B}$. It follows that the low-energy effective largangian for the mesons and light-baryons looks like
\begin{align}
    \mathcal{L}\sim& \frac{1}{N^2}\left(\bar{B}\slashed{\partial}B+\bar{\tilde{B}}\slashed{\partial}\tilde{B}\right)+\frac{1}{N^2}\bar{U}(-\partial^2)U+\frac{1}{N}\bar{V}(-\partial^2)V \nonumber \\
    +&\frac{g_{\chi}}{N^4}\left( \bar{B}UB+\bar{\tilde{B}}U\tilde{B}\right)+ \frac{g_{\eta}}{N^3}\left( \bar{B}VB+\bar{\tilde{B}}V\tilde{B}\right)
\end{align}
The Yukawa couplings $g_{\chi}$ and $g_{\eta} $ have been chosen to satisfy parity conservation, that in vector-like gauge theories cannot be spontaneously broken \cite{Vafa:1984xg}. Then, the mass $m_B$ of the light-baryons scales as
\begin{equation}
    m_B\sim N^2\left(\frac{g_{\chi}}{N^4}\expval{U}+\frac{g_{\eta}}{N^3}\expval{V}\right)\sim N^0
\end{equation}
The transformation laws of the UV and IR fields under the unbroken symmetry group are shown in table \ref{tab:chietavecunbrk}.

\begin{table*}
   \renewcommand*{\arraystretch}{1.3}
    \centering
    \begin{tabular}{|c|c|c|c|c|c|c|}
    \hline
        & & $SU(N)_{\rm c}$ & $SU(N_{\chi})_V$ & $SU(N_{\eta})_{V}$ & $U(1)_{\chi V}$ &  $U(1)_{\eta V}$\\
        \hline
        UV &$\chi$ & $N_{\chi}\cdot \yng(2)$ & $\frac{N(N-1)}{2}\cdot \yng(1)$ & $\frac{N(N-1)}{2}\cdot (\cdot)$ & $+1$ & $0$ \\
        &$\tilde{\chi}$ & $N_{\chi}\cdot \bar{\yng(2)}$ & $\frac{N(N-1)}{2}\cdot \bar{\yng(1)}$ & $\frac{N(N-1)}{2}\cdot (\cdot)$ & $-1$  & $0$ \\
        &$\eta$ & $N_{\eta}\cdot \bar{\yng(1)}$ & $N\cdot (\cdot)$ & $N\cdot \yng(1)$ & $0$    & $+1$ \\
        &$\tilde{\eta}$ & $N_{\eta}\cdot \yng(1)$ & $N\cdot (\cdot)$ & $N\cdot \bar{\yng(1)}$ & $0$  & $-1$  \\
        \hline
        IR$^*$ & $ B $ & $\frac{N_{\chi}N_{\eta}(N_{\eta}+1)}{2}\cdot (\cdot) $ & $ \frac{N_{\eta}(N_{\eta}+1)}{2}\cdot \yng(1)$  & $N_{\chi}\cdot \yng(2)$ & $+1$ & $+2$ \\
         & $\tilde{B}$ & $\frac{N_{\chi}N_{\eta}(N_{\eta}+1)}{2}\cdot (\cdot) $ & $ \frac{N_{\eta}(N_{\eta}+1)}{2}\cdot \bar{\yng(1)}$ & $N_{\chi}\cdot \bar{\yng(2)}$ & $-1$ & $-2$ \\
         & $\pi_1$ & $N_{\chi}^2\cdot (\cdot)$ & (Adj) & $N_{\chi}^2\cdot (\cdot)$ & $0$ & $0$ \\
         & $\pi_2$ & $N_{\eta}^2\cdot (\cdot)$ & $N_{\eta}^2\cdot (\cdot)$ & (Adj) & $0$ & $0$ \\
        \hline
    \end{tabular}
    \caption{\footnotesize Unbroken symmetries of the $\chi\tilde{\chi}\eta\tilde{\eta}$ vector-like model. }
    \label{tab:chietavecunbrk}
\end{table*}

\paragraph{Heavy baryons} We now turn to heavy baryons. We indicate a generic heavy baryon as
\begin{equation}
\label{eq:genericbaryon}
    {\cal B}_{(l,n,m,p,\bar{n},\bar{m},\bar{p})}
\end{equation}
where $l$ is the number of $\epsilon_{i_1\dots i_N}$ if $l>0$ (or $\epsilon^{i_1\dots i_N}$ if $l<0$ ). We have $N l$ indices than can be saturated with $n$ $\psi^{ij}$ or $\bar{n}$ $\bar{\tilde{\chi}}^{[ij]}$ , $m$ $\chi^{[i k]}\eta_{k}$ or $\bar{m}$ $\bar{\tilde{\chi}}^{[i k]}\bar{\tilde{\eta}}_{k}$, $p$ $\bar{\eta}^i $ or $\bar{p}$ ${\tilde{\eta}}^{i}$. $N l = 2n + m +p+2\bar{n} + \bar{m} +\bar{p}$.  For example the with $l=1$ $n=\bar{n}=0$, and limiting to only left-handed fermions  we have 
\be
  {\cal B}_{(1,0,m,p,0,0,0)}   = \epsilon_{i_1\dots i_N} \chi^{[i_1 k_1]}\eta_{k_1} \dots \chi^{[i_{m} k_{m}]}\eta_{k_{m}} \tilde{\eta}^{m+1} \dots \tilde{\eta}^{N}
\ee
An example of heavy baryon that is absent in the $\psi\tilde{\psi}\eta\tilde{\eta}$ models is \cite{Bolognesi:2006ws}
\begin{align}
    &\mathcal{B}_{(1, N/2,0,0,0,0,0)}=\epsilon_{i_1\dots i_N}\chi^{[i_1i_2]}\dots \chi^{[i_{N-1}i_N]} && N\ \text{even}
\end{align}
The baryons of QCD with the fundamental quarks are, in this notation
\begin{align}
  &{\cal B}_{(1,0,0,N,0,0,0)}   = \epsilon_{i_1\dots i_N} \  \tilde{\eta}^{1} \dots \tilde{\eta}^{N}\nonumber \\
  &{\cal B}_{(1,0,0,0,0,0,N)}   = \epsilon_{i_1\dots i_N} \  \bar{\eta}^{1} \dots \bar{\eta}^{N}
\end{align}
\sloppy The fully antisymmetric baryons with only $\chi$ or $\tilde{\chi}$ are ${\cal B}_{(N+1,\frac{N (N+1)}{2},0,0,0,0,0)}  $, ${\cal B}_{(N+1,0,0,0,0,0,\frac{N (N+1)}{2})}  $. Again, this classification does not encode the transformation properties of the heavy baryons under non-abelian symmetry subgroups. The $U(1)_{\eta V}$, $U(1)_{\chi V}$ charges and fermionic number of the baryons are given in table \ref{tablebaryonsvectorialchi}.

\begin{table}[H]
   \renewcommand*{\arraystretch}{1.3}
\centering 
\begin{tabular}{ |c| c|c|c| }
\hline
baryons  &      $U(1)_{\chi V}$  &    $U(1)_{\eta V}$   &  ${\mathbb{Z}}_{2}^F$   \\
\hline  
 $ B = \chi\eta\eta $      & $1$        & $2$   & $-1$ \\
 \hline  
 $ \tilde{B} = \tilde{\chi}\tilde{\eta}\tilde{\eta}  $      & $-1$        & $-2$   & $-1$ \\
   \hline  
 $  {\cal B}_{(1,0,0,N,0,0,0)} $    &$0$& $N$   & $(-)^{N}$  \\
  \hline  
 $ {\cal B}_{(N+1,\frac{N (N+1)}{2},0,0,0,0,0)}  $    &$\frac{N (N+1)}{2}$& $0$  & $(-)^{\frac{N (N+1)}{2}}$  \\
  \hline  
 ${\cal B}_{(l,n,m,p,\bar{n},\bar{m},\bar{p})} $    &$n-\bar{n}$& $m+p-\bar{m}-\bar{p}$   & $(-)^{n + \bar{n} +p+\bar{p}}$  \\
	\hline
	\end{tabular}
\caption{\footnotesize Baryons and the charges under the unbroken $U(1)$ symmetries in the $\chi\tilde{\chi}\eta\tilde{\eta} $ vector-like model. 
}\label{tablebaryonsvectorialchi}
\end{table}

The lattice of possible $U(1)_{V\chi}\times U(1)_{V\eta}$ charges is the subset of $\mathbb{Z}\times \mathbb{Z}$ satisfying
\begin{equation}
    2 q_{U(1)_{\eta V}}-q_{U(1)_{\chi V}}=0 \mod N
\end{equation}

\section{Skyrmions and WZW terms}
\label{sec:skwzw}

In this section, we study the topology of the coset space $\frac{G_{\rm f}}{H_{\rm f}\times H_{\rm cf}}$ in models introduced in section \ref{sec:baryons} and section \ref{sec:introvect} and show that in BY and GG models there are no Skyrmions and WZW terms. Some mathematical results used in this section are presented in appendix \ref{app:topology}. We recall that:
\begin{itemize}
    \item A necessary condition for the effective field theory for the NGBs to admit Skyrmion solutions is that $\pi_3\left(\frac{G_{\rm f}}{H_{\rm f}\times H_{\rm cf}}\right)\neq 0$.
    \item A necessary and sufficient condition for the existence of a nontrivial WZW term is that $\pi_4\left(\frac{G_{\rm f}}{H_{\rm f}\times H_{\rm cf}}\right)=0$ and $\pi_5\left(\frac{G_{\rm f}}{H_{\rm f}\times H_{\rm cf}}\right)\neq 0$.
\end{itemize}

\subsection{$\psi\eta$ and $\chi\eta$ models}
\label{sec:skyrmions}
The $\psi\eta$ model has the symmetry-breaking pattern with
\begin{align}
    &G_{\rm c} = SU(N)_{\rm c}\ , && G_{\rm f}=SU(N+4)_{\eta}\times U(1)_{\psi\eta} \nonumber \\
    &H_{\rm c} = \{1\}\ , && H_{\rm f} = SU(4)_{\eta} \times U(1)'\nonumber \\
    &H_{\rm cf} = SU(N)_{{\rm cf}_{\eta}}
\end{align}
where the symmetries are defined in subsection \ref{sec:intropsieta}. The interesting homotopy groups are then
\begin{align}
    &\pi_i\left(\frac{G_{\rm f}}{H_{\rm f}\times H_{\rm cf}}\right) &&i=3,4,5
\end{align}
where $SU(N)_{{\rm f}_{\eta}}\subset SU(N+4)_{\eta}$ is the part of $SU(N)_{{\rm cf}_{\eta}}$ acting only on the flavor indices of the fields. From Eqs. \eqref{eq:pih}, \eqref{eq:piequiv} in appendix \ref{app:topology} we know that
\begin{align}
    &\pi_i\left(\frac{G_{\rm f}}{H_{\rm f}\times H_{\rm cf}}\right)=\pi_i\left(\frac{U(N+4)}{U(N)\times U(4)}\right) && i=3,4,5
\end{align}
The manifold appearing in the homotopy group in the rhs is the Grassmann manifold $G_{N+4,4}$ (see appendix \ref{app:stiefel}). Since $\pi_i(G_{N+4,4})=0$ for every $N\geq 1$ we conclude that
\begin{align}
    &\pi_i\left(\frac{G_{\rm f}}{H_{\rm f}\times H_{\rm cf}}\right)=0 && i=3,4,5
\end{align}
We conclude that \emph{the low-energy effective field theory for the $\psi\eta$ model in the color-flavor locking phase does not admit Skyrmions and WZW terms}.
\par The $\chi\eta$ model has the symmetry-breaking pattern with
\begin{align}
    &G_{\rm c} = SU(N)_{\rm c}\ , && G_{\rm f}=SU(N-4)_{\eta}\times U(1)_{\chi\eta} \nonumber \\
    &H_{\rm c} = SU(4)_{\rm c} , && H_{\rm f} = U(1)'\nonumber \\
    &H_{\rm cf} = SU(N-4)_{\text{cf}_{\eta}}
\end{align}
By the arguments of appendix \ref{app:stiefel} we know that
\begin{align}
     &\pi_i\left(\frac{G_{\rm f}}{H_{\rm f}\times H_{\rm cf}}\right)=\pi_i(\{1\})=0 && i=3,4,5
\end{align}
where $\{1\}$ is the trivial group and where $SU(N-4)_{{\rm f}_{\eta}}= SU(N-4)_{\eta}$ is the part of $SU(N-4)_{{\rm cf}_{\eta}}$ acting only on the flavor indices of the fields. We conclude that \emph{the low-energy effective field theory for the $\chi\eta$ model in the color-flavor locking phase does not admit Skyrmions and WZW terms}.

\subsection{BY and GG models}
The arguments of subsection \ref{sec:skyrmions} can be generalized to the BY and to the GG models.
\par In the BY model, the symmetry-breaking has
\begin{widetext}
\begin{align}
    &G_{\rm c} = SU(N)_{\rm c}\ , && G_{\rm f}=SU(N+4+p)_{\eta}  \times  SU(p)_{\xi}  \times  U(1)_{\psi\eta}\times  U(1)_{\psi\xi}\times U(1)_{\chi\eta} \nonumber \\
    &H_{\rm c} = \{1\} , && H_{\rm f} = SU(4)_{\eta}  \times  SU(p)_{\eta\xi}  \times  U(1)_{\psi \eta}' \times  U(1)_{\psi \xi}' \nonumber \\
    & H_{\rm cf} = SU(N)_{{\rm cf}_{\eta}}
\end{align}
\end{widetext}
The interesting homotopy groups are then
\begin{align}
\label{eq:piby}
    &\pi_i\left(\frac{G_{\rm f}}{H_{\rm f}\times H_{\rm cf}}\right)=\pi_i\left(\frac{U(N+4+p)_{\eta}\times U(p)_\xi}{U(N)_{{\rm f}_{\eta}}\times U(4)_{\eta}\times U(p)_{\eta\xi}}\right) \nonumber \\
    & i = 3,4,5
\end{align}
where in passing from the first to the second line we used Eqs. \eqref{eq:pih}, \eqref{eq:piequiv}. We denoted as $SU(N)_{{\rm f}_{\eta}}\subset SU(N+4+p)_{\eta}$ the part of $SU(N)_{{\rm cf}_{\eta}}$ acting only on the flavor indices of the fields. To go further, we need to define the groups appearing in the argument of the $\pi_i$. Let us define the spinor
\begin{align}
    \begin{pmatrix}
        \eta_1 \\
        \eta_2 \\
        \eta_3 \\
        \xi
    \end{pmatrix}
\end{align}
The group in the numerator acts on this spinor as (see table \ref{tab:by})
{\footnotesize
\begin{align}
    U(N+4+p)_{\eta}\times U(p)_\xi:\  \begin{pmatrix}
        \eta_1 \\
        \eta_2 \\
        \eta_3 \\
        \xi
    \end{pmatrix} \longmapsto 
    \begin{pNiceArray}{ccc|c}
        \Block{3-3}{\scalebox{1.5}{$U$}} & \\
        & \\
        & \\
        \hline
        \phantom{ab} & \phantom{bc} & \phantom{cd} & V
    \end{pNiceArray}
    \begin{pmatrix}
        \eta_1 \\
        \eta_2 \\
        \eta_3 \\
        \xi
    \end{pmatrix}
\end{align}}
where $U\in U(N+4+p),\ V\in U(p)$. The group in the denominator acts on the spinor as (see table \ref{tab:bycfl})
{\footnotesize
\begin{align}
    U(N)_{{\rm f}_{\eta}}\times U(4)_{\eta}\times U(p)_{\eta\xi}\ :\ 
    \begin{pmatrix}
        \eta_1 \\
        \eta_2 \\
        \eta_3 \\
        \xi
    \end{pmatrix} 
    \longmapsto 
    \begin{pmatrix}
        u^{\dagger} & & & \\
        & v & & \\
        & & w^{\dagger} & \\
        & & & w
    \end{pmatrix}
    \begin{pmatrix}
        \eta_1 \\
        \eta_2 \\
        \eta_3 \\
        \xi
    \end{pmatrix} 
\end{align}}
where $u\in U(N),\ v \in U(4),\ w \in U(p)$. Hence, the coset in Eq. \eqref{eq:piby} is defined by the equivalence relations
\begin{align}
    \begin{pmatrix}
        & & \\
        & \scalebox{1.5}{$U$} & \\
        & & 
    \end{pmatrix}
    \sim
    \begin{pmatrix}
        & & \\
        & \scalebox{1.5}{$U$} & \\
        & & 
    \end{pmatrix}
    \begin{pmatrix}
        u^{\dagger} & & \\
        & v & \\
        & & w^{\dagger}
    \end{pmatrix}\ , \qquad V \sim Vw^{\dagger}
\end{align}
From the equivalence classes defined by this relation, we can always pick the representative with $V=1$, by choosing $w=V^{\dagger}$. In this way we have shown that
\begin{align}
    \frac{U(N+4+p)_{\eta}\times U(p)_\xi}{U(N)_{{\rm f}_{\eta}}\times U(4)_{\eta}\times U(p)_{\eta\xi}} \cong \frac{U(N+4+p)}{U(N)\times U(4)}
\end{align}
This manifold is a fibration
\begin{align}
    U(4) \longrightarrow G_{N+4+p, N} \longrightarrow \frac{U(N+4+p)}{U(N)\times U(4)}
\end{align}
where $G_{N+4+p, N}$ is a Grassmann manifold (see appendix \ref{app:stiefel}). Then the long exact sequence of fibrations of \eqref{eq:les} implies that
\begin{align}
    &\pi_3\left(\frac{G_{\rm f}}{H_{\rm f}\times H_{\rm cf}}\right)=\pi_5\left(\frac{G_{\rm f}}{H_{\rm f}\times H_{\rm cf}}\right)=0 \nonumber \\
    &\pi_4\left(\frac{G_{\rm f}}{H_{\rm f}\times H_{\rm cf}}\right)=\mathbb{Z}
\end{align}
This proves that \emph{the low-energy effective field theory for the BY models in the color-flavor locking phase does not admit Skyrmions and WZW terms}. The absence of Skyrmions in the low-energy EFT of the BY models can be seen as a consequence of the fact that heavy baryon states, that Skyrmions are thought to describe, seem to be always allowed to decay into lighter states by the the unbroken symmetry group, as stated in subsections \ref{sec:intropsieta} and \ref{sec:introby}.
\par We now turn to the GG model, whose symmetry-breaking pattern is
\begin{widetext}
\begin{align}
    &G_{\rm c} = SU(N)_{\rm c}\ , && G_{\rm f}=SU(N-4+p)_{\eta}  \times  SU(p)_{\xi}  \times  U(1)_{\chi\eta}\times  U(1)_{\chi\xi} \nonumber \\
    &H_{\rm c} = SU(4)_{{\rm  c}} , && H_{\rm f} = SU(p)_{\eta\xi}  \times  U(1)_{\chi \eta}' \times  U(1)_{\chi \xi}' \nonumber \\
    & H_{\rm cf} = SU(N-4)_{{\rm cf}_{\eta}}
\end{align}
\end{widetext}
The interesting homotopy groups are then
\begin{align}
    &\pi_i\left(\frac{G_{\rm f}}{H_{\rm f}\times H_{\rm cf}}\right) && i=3,4,5
\end{align}
where again $SU(N-4)_{{\rm f}_{\eta}}\subset SU(N-4+p)_{\eta}$ is the part of $SU(N-4)_{{\rm cf}_{\eta}}$ acting only on the flavor indices of the fields. Using the same arguments employed for the BY models we find that
\begin{align}
    &\pi_i\left(\frac{G_{\rm f}}{H_{\rm f}\times H_{\rm cf}}\right)=\pi_i\left(\frac{U(N-4+p)}{U(N-4)}\right) && 
    i = 3,4,5
\end{align}
and hence
\begin{align}
    &\pi_i\left(\frac{G_{\rm f}}{H_{\rm f}\times H_{\rm cf}}\right)=0 && i=3,4,5
\end{align}
This proves that \emph{the low-energy effective field theory for the GG models in the color-flavor locking phase does not admit Skyrmions and WZW terms}. The absence of Skyrmions in the low-energy EFT of the GG models (including the simpler $\chi\eta$ model) seems not to be a consequence of the instability of heavy baryon states, since, as it was stated in subsections \ref{sec:introchieta} and \ref{sec:ggintro}, the unbroken symmetry group generally does not forbid heavy baryon states to decay into lighter particles.

\subsection{Vector-like models}
We first consider the $\psi\tilde{\psi}\eta\tilde{\eta}$ modelss. Since the condensates are gauge-invariant, we have $H_{\rm cf}=\{1\}$ and $H_{\rm c}=G_{\rm c} = SU(N)_{\rm c} $. The flavor groups are then
\begin{widetext}
\begin{align}
    &G_{\rm f}=SU(N_{\psi})_L\times SU(N_{\eta})_L\times SU(N_{\psi})_R\times SU(N_{\eta})_R\times U(1)_{\psi V}\times U(1)_{\eta V}\times \widetilde{U}(1)_{\psi\eta} \nonumber \\
    &H_{\rm f} = SU(N_{\psi})_{V}  \times SU(N_{\eta})_{V} \times U(1)_{\psi V} \times  U(1)_{\eta V}
\end{align}
\end{widetext}
where the subgroups are defined in tables \ref{tab:psietavec}, \ref{tab:psietavecunbrk}. The interesting homotopy groups are
\begin{align}
    &\pi_i(G_{\rm f}/H_{\rm f}) && i=3,4,5
\end{align}
Since $i>1$ we can neglect the $U(1)$ subgroups by the arguments of appendix \ref{app:stiefel}. We can also use the identity
\begin{widetext}
\begin{align}
    &\pi_i\left(G_{\rm f}/H_{\rm f}\right)=\pi_i\left(\frac{SU(N_{\psi})_L\times SU(N_{\psi})_R}{SU(N_{\psi})_V}\right)\times \pi_i\left(\frac{SU(N_{\eta})_L\times SU(N_{\eta})_R}{SU(N_{\eta})_V}\right) && i=3,4,5
\end{align}
\end{widetext}
which shows that
\begin{align}
\label{eq:ztimesz}
    & \pi_3\left(G_{\rm f}/H_{\rm f}\right)=\pi_5\left(G_{\rm f}/H_{\rm f}\right)=\mathbb{Z}\times \mathbb{Z} \nonumber \\
    & \pi_4\left(G_{\rm f}/H_{\rm f}\right)=0 
\end{align}
This proves that \emph{the low-energy effective field theory for the $\psi\tilde{\psi}\eta\tilde{\eta}$ models admits both Skyrmions and WZW terms}.
\par We then consider the $\chi\tilde{\chi}\eta\tilde{\eta}$ models. Since the condensates are gauge-invariant we have $H_{\rm cf}=\{1\}$ and $H_{\rm c}=G_{\rm c} = SU(N)_{\rm c} $. The flavor groups are then
\begin{widetext}
\begin{align}
    &G_{\rm f}=SU(N_{\chi})_L\times SU(N_{\eta})_L\times SU(N_{\chi})_R\times SU(N_{\eta})_R\times U(1)_{\chi V}\times U(1)_{\eta V}\times \widetilde{U}_{\chi\eta}(1) \nonumber \\
    &H_{\rm f} = SU(N_{\psi})_{V}  \times SU(N_{\eta})_{V} \times U(1)_{\chi V} \times  U(1)_{\eta V}
\end{align}
\end{widetext}
where the subgroups are defined in tables \ref{tab:chietavec}, \ref{tab:chietavecunbrk}. The interesting homotopy groups are
\begin{align}
    &\pi_i(G_{\rm f}/H_{\rm f}) && i=3,4,5
\end{align}
Since $i>1$ we can neglect the $U(1)$ subgroups by the arguments of appendix \ref{app:stiefel}. We can also use the identity
\begin{widetext}
\begin{align}
    &\pi_i\left(G_{\rm f}/H_{\rm f}\right)=\pi_i\left(\frac{SU(N_{\chi})_L\times SU(N_{\chi})_R}{SU(N_{\chi})_V}\right)\times \pi_i\left(\frac{SU(N_{\eta})_L\times SU(N_{\eta})_R}{SU(N_{\eta})_V}\right) && i=3,4,5
\end{align}
\end{widetext}
which shows that
\begin{align}
    & \pi_3\left(G_{\rm f}/H_{\rm f}\right)=\pi_5\left(G_{\rm f}/H_{\rm f}\right)=\mathbb{Z}\times \mathbb{Z} \nonumber \\
    & \pi_4\left(G_{\rm f}/H_{\rm f}\right)=0 
\end{align}
This proves that \emph{the low-energy effective field theory for the $\chi\tilde{\chi}\eta\tilde{\eta}$ model admits both Skyrmions and WZW terms}.

\subsection{Baryons and WZW terms in vector-like models}
In this subsection we write the WZW terms in the vector-like models and identify the Skyrmions with the baryons introduced in section \ref{sec:introvect} by matching their quantum numbers and the large-$N$ scaling of their masses.
\par Let us first consider the $\psi\tilde{\psi}\eta\tilde{\eta}$ modelss. At low energies, the degrees of freedom in the effective lagrangian are:
\begin{itemize}
    \item The $\pi^a$ arising from the symmetry-breaking $SU(N_{\psi})_L\times SU(N_{\psi})_{R}\to SU(N_{\psi})_V $ and created by the operator $\tilde{\psi}\psi$.
    \item The $\tilde{\pi}^a$ arising from the symmetry-breaking $SU(N_{\eta})_L\times SU(N_{\eta})_{R}\to SU(N_{\eta})_V $ and created by the operator $\tilde{\eta}\eta$.
    \item The $\pi$ arising from the symmetry-breaking $U(1)_{\psi\eta}\to 1$ and created by both $\tilde{\psi}\psi $ and $\tilde{\eta}\eta$.
\end{itemize}
The fields corresponding to these operators can appear in the low-energy effective lagrangian only through the combinations
\begin{align}
\label{eq:uvmatrices}
    &U=\mathrm{exp}\left(i\frac{\pi^at^a}{f_{\pi}}-i\frac{N_{\eta}\pi}{f}\right) \nonumber \\
    &V=\mathrm{exp}\left(i\frac{\tilde{\pi}^a\tilde{t}^a}{\tilde{f}_{\pi}}+i\frac{N_{\psi}(N+2)\pi}{f}\right) 
\end{align}
The $f_{\pi}$, $\tilde{f}_{\pi}$ and $f$ are the decay constants of the particles $\pi^a$, $\tilde{\pi}^a$ and $\pi$ respectively. The matrices $t^a$ and $\tilde{t}^a$ are the generators of $SU(N_{\psi})$ and $SU(N_{\eta})$ respectively. By the usual large-$N$ counting rules we know that
\begin{align}
\label{eq:fpi}
    &f_{\pi}\sim N && \tilde{f}_{\pi} \sim \sqrt{N} && f\sim N
\end{align}
From the definitions \eqref{eq:uvmatrices} is follows that the WZW terms in the action are
\begin{align}
\label{eq:wzw}
\Gamma_{\rm WZW}[U, V]=&\Gamma_{\rm WZW}[U]+\Gamma_{\rm WZW}[V] \nonumber \\
=&\frac{N(N+1)}{2}\frac{1}{240\pi^2}\int_{\mathcal{M}_5}\mathrm{Tr}\left(U^{-1}dU\right)^5 \nonumber \\
&+\frac{N}{240\pi^2}\int_{\mathcal{M}_5}\mathrm{Tr}\left(V^{-1}dV\right)^5
\end{align}
From Eq. \eqref{eq:ztimesz} we know that each Skyrmion field configuration can be classified by a pair of integers $(n,m)\in \mathbb{Z}\times \mathbb{Z}$, where $n$ is the winding number of the field $U$ while $m$ is the winding number of the field $V$. Since in the Skyrme lagrangian the $U(1)$ NGBs are decoupled \cite{Skyrme:1961vq, Skyrme:1962vh, Adkins:1983ya} we can assume that $\pi=0$ everywhere. Let us consider the two simplest Skyrmions with winding numbers $(1,0)$ and $(0,1)$. From the large-$N$ scaling of $f_\pi$ and $\tilde{f}_\pi$ it follows that \cite{Bolognesi:2006ws}
\begin{align}
    &M_{(1,0)}\sim N^2 && M_{(0,1)}\sim N
\end{align}
By the arguments of Ref. \cite{Witten:1983tw} the statistics of the two Skyrmions under consideration can be read from the coefficients of the two terms in the rhs Eq. \eqref{eq:wzw}: the Skyrmion $(1,0)$ has statistics $(-1)^{\frac{N(N+1)}{2}}$ while the Skyrmion $(0,1)$ has statistics $(-1)^N$. The charge of the Skyrmions under the unbroken $U(1)_{\psi V}$ and $U(1)_{\eta V}$ symmetries (see table \ref{tab:psietavecunbrk}) can be found by applying the Goldstone-Wilczek procedure \cite{Goldstone:1981kk} to the currents
\begin{align}
    &J_{\psi V}^{\mu}=i\bar{\psi}\bar{\sigma}^{\mu}\psi-i\bar{\tilde{\psi}}\bar{\sigma}^{\mu}\tilde{\psi} \nonumber \\
    &J_{\eta V}^{\mu}=i\bar{\eta}\bar{\sigma}^{\mu}\eta-i\bar{\tilde{\eta}}\bar{\sigma}^{\mu}\tilde{\eta}
\end{align}
On a Skyrmion background the $U(1)_{\psi V}$ and $U(1)_{\eta V}$ currents correspond to the topological currents for $U$ and $V$
\begin{widetext}
\begin{align}
    &\expval{J_{\psi V}^{\mu}}_{\rm Skyrmion}=\frac{N(N+1)}{2}\frac{1}{24\pi^2}\varepsilon^{\mu\nu\rho\sigma}\mathrm{Tr}(U^{-1}\partial_{\nu}UU^{-1}\partial_{\rho}UU^{-1}\partial_{\sigma}U) \nonumber \\
    &\expval{J_{\eta V}^{\mu}}_{\rm Skyrmion}=\frac{N}{24\pi^2}\varepsilon^{\mu\nu\rho\sigma}\mathrm{Tr}(V^{-1}\partial_{\nu}VV^{-1}\partial_{\rho}VV^{-1}\partial_{\sigma}V)
\end{align}
\end{widetext}
It follows that for the Skyrmion with winding numbers $(n,m)$ the $U(1)_{\psi V}$-charge $q_{U(1)_{\psi V}}^{(n,m)}$ and the $U(1)_{\eta V}$-charge $q_{U(1)_{\eta V}}^{(n,m)}$ are
\begin{align}
    &q_{U(1)_{\psi V}}^{(n,m)}=\int d^3\mathbf{x}\ \expval{J_{\psi V}^0(\mathbf{x},0)}_{\rm Skyrmion}=\frac{N(N+1)}{2}n \nonumber \\
    &q_{U(1)_{\eta V}}^{(n,m)}=\int d^3\mathbf{x}\ \expval{J_{\eta V}^0(\mathbf{x},0)}_{\rm Skyrmion}=Nm
\end{align}
The $U(1)$ charges of the two Skyrmions $(1,0)$ and $(0,1)$ are then
\begin{align}
    &\left(q_{U(1)_{\psi V}}^{(1,0)}, q_{U(1)_{\eta V}}^{(1,0)}\right)=\left(\frac{N(N+1)}{2},0\right) \nonumber \\
    &\left(q_{U(1)_{\psi V}}^{(0,1)}, q_{U(1)_{\eta V}}^{(0,1)}\right)=\left(0,N\right)
\end{align}
All these facts suggest the identifications
{\small
\begin{align}
    & (1,0) \longleftrightarrow \mathcal{B}_{(N+1,n,0,0,\frac{N(N+1)}{2}-n,0,0)}\ , && n=0,\dots, \frac{N(N+1)}{2} \nonumber \\
    &(0,1) \longleftrightarrow \mathcal{B}_{(1,0,0,n,0,0,N-n)}\ , && n=0,1\dots, N
\end{align}}
where the baryons on the right are defined in Eq. \eqref{eq:genericbaryonpsi} and below.
\par The analysis for the $\chi\tilde{\chi}\eta\tilde{\eta}$ models is essentially the same.

\section{The $\theta$-periodicity anomaly}
\label{sec:abj}
In this section we show that any theory in the CFL phase with no residual color group (including the $\psi\eta$ and the BY models) have no $\theta$-periodicity anomaly, while the $\chi\eta$ theory and the vector-like gauge theories have a $\theta$-periodicity anomaly.

\subsection{$\theta$-periodicity anomaly}
The $\theta$-periodicity anomaly was first introduced in Refs. \cite{Cordova:2019jnf, Cordova:2019uob}, and later applied to constrain the low-energy structure of gauge theories in the confining phase in Refs. \cite{Anber:2019nze, Anber:2020gig, Kitano:2020evx, Nakajima:2022jxg}. In this subsection we show how to perform the $\theta$-periodicity anomaly matching for quantum gauge theories in the CFL phase with the symmetry-breaking pattern \eqref{eq:sb}.
\par The anomaly matching is performed between a UV theory and an IR theory, where the UV theory is a $SU(N)$ gauge theory with connection $a$ and a $\theta$-parameter. In both theories, we introduce a background connection $A$ for the global symmetries $H_{\rm f}\times H_{\rm cf}^{({\rm f})}$. Since we allow $a$ and $A$ to have fractional topological charges, it is convenient to express the UV and IR partition functions
\begin{align}
    &\mathcal{Z}_{UV}[\widetilde{A}, B, \theta] && \mathcal{Z}_{IR}[\widetilde{A}, B, \theta]
\end{align}
in terms of the ordinary connections $\widetilde{a}$, $\widetilde{A}$ and a $1$-form  connection $B$ defined \footnote{In the notation of subsection \ref{app:1formquant}, $\protect\widetilde{A}$ corresponds to the $\protect\widetilde{A}_i^G$, $\protect\widetilde{A}_i^{U(1)}$ collectively and $B$ corresponds to the $B_i$ and $b$ collectively} as in appendix \ref{app:1form}, transforming under an appropriate $1$-form  gauge symmetry as in Eq. \eqref{eq:reparametrization}. We normalize the UV $\theta$-term so that
\begin{align}
\label{eq:theta}
    \mathcal{Z}_{UV}[0, 0, \theta+2\pi]=\mathcal{Z}_{UV}[0, 0, \theta]\;.
\end{align}
The key to the anomaly matching condition is the equality
\begin{align}
\label{eq:zrg}
    \mathcal{Z}_{UV}[\widetilde{A}, B , \theta]=\mathcal{Z}_{IR}[\widetilde{A}, B, \theta]\;,
\end{align}
that arises from the renormalization group-invariance of the partition function. If we act on the UV and IR fields with a symmetry transformation, they both acquire a phase that depends on the background fields
\begin{align}
    &\mathcal{Z}_{UV}[\widetilde{A}, B , \theta] \longmapsto \mathcal{Z}_{UV}[\widetilde{A}, B , \theta]\ e^{i\gamma_{UV}[\widetilde{A}, B]}\;, \nonumber \\
    &\mathcal{Z}_{IR}[\widetilde{A}, B , \theta] \longmapsto \mathcal{Z}_{IR}[\widetilde{A}, B , \theta]\ e^{i\gamma_{IR}[\widetilde{A}, B]}\;,
\end{align}
where the phases $\gamma_{UV}[\widetilde{A}, B]$ and $\gamma_{IR}[\widetilde{A}, B]$ are renormalization group-invariant. The anomaly-matching condition is
\begin{align}
    e^{i\gamma_{UV}[\widetilde{A}, B]} = e^{i\gamma_{IR}[\widetilde{A}, B]}\;.
\end{align}
The symmetry we are interested in is the $\theta$-periodicity \eqref{eq:theta} that is explicitly broken by the background gauge fields $\widetilde{A}, B$. When the IR theory matches the $\theta$-periodicity anomaly of the UV theory we have
\begin{align}
\label{eq:conditions}
   \frac{\mathcal{Z}_{UV}[\widetilde{A}, B, \theta+2\pi]}{\mathcal{Z}_{UV}[\widetilde{A}, B, \theta]}=\frac{\mathcal{Z}_{IR}[\widetilde{A}, B, \theta+2\pi]}{\mathcal{Z}_{IR}[\widetilde{A}, B, \theta]}
\end{align}
This is a necessary but not sufficient condition for the relation \eqref{eq:zrg} to hold.
\par Since we will mostly consider theories in the CFL phase, we will take the UV and the IR partition functions to compare as
\begin{widetext}
\begin{align}
    &\mathcal{Z}_{UV}[\widetilde{A}, B, \theta]=\int\limits_{H\text{-bundles}} [d\chi_{UV}][d\widetilde{a}]\ \mathrm{exp}\Big(-S_{UV}[\chi_{UV}, \widetilde{a}, A, B]+i\theta F_{UV}[\widetilde{a}, \widetilde{A}, B]\Big) \nonumber \\
    & \mathcal{Z}_{IR}[\widetilde{A}, B, \theta] = \int[d\chi_{IR}]\ \mathrm{exp}\Big(-S_{IR}[\chi_{IR}, \widetilde{A}, B, \theta]+i\theta F_{IR}[\widetilde{A}, B]\Big)
\end{align}
\end{widetext}
We denoted the UV and IR matter fields in the $\chi_{UV}$ and $\chi_{IR}$ respectively. The actions $S_{UV}$, $S_{IR}$ and the functionals $F_{UV}$, $F_{IR}$ must be invariant under the $1$-form  gauge symmetry \eqref{eq:reparametrization} in order for the theory to make sense.
\par The unconventional element in this definition is the restriction of the path-integral for $\mathcal{Z}_{UV}$ to $H$-bundles. This restriction is performed because the sectors in which the transition functions are $H$-valued are the only ones where the CFL can occur (see subsection \ref{sec:background} for a justification of this statement). 
\par We can now state that \emph{the IR theory matches the $\theta$-periodicity anomaly of the UV theory only is it is possible to construct a gauge-invariant and $1$-form  gauge-invariant functional $F_{IR}[\widetilde{A}, B]$ such that the condition \eqref{eq:conditions} is satisfied}. This criterion does not change for theories that possess additional free parameters, including fermion masses.
\par We now describe the mechanism that may cause the $\theta$-periodicity anomaly matching to fail. If we shift $\theta$ by $2\pi$, the \emph{integrand} of the UV partition function will acquire a phase factor $e^{i2\pi F_{IR}[\widetilde{a}, \widetilde{A}, B]}$. Thanks to the fact that the topological charge of $\widetilde{a}$ is integer-valued by definition and thanks to the normalization condition \eqref{eq:theta}, the dependence from $\widetilde{a}$ in the exponential always drops and we can always write
\begin{align}
    \mathcal{Z}_{UV}[\widetilde{A}, B, \theta+2\pi]= \mathcal{Z}_{UV}[\widetilde{A}, B, \theta] e^{i2\pi F_{UV}[0, \widetilde{A}, B]}
\end{align}
The $\theta$-periodicity anomaly is matched by the IR theory if it is possible to construct a functional $F_{IR}[\widetilde{A}, B]$ that is both gauge-invariant and $1$-form  gauge invariant satisfying
\begin{align}
    e^{i2\pi F_{UV}[0, \widetilde{A}, B]} = e^{i2\pi F_{IR}[\widetilde{A}, B]}
\end{align}
This task is highly nontrivial because, in general, $F_{UV}[0, \widetilde{A}, B]$ is not $1$-form  gauge-invariant.
\par \sloppy We conclude this section by recalling the rule to compute $F_{UV}$. Suppose that the matter content of the UV theory consists of the massless left-handed fermions $\psi_1,\dots, \psi_n$ (counted without multiplicities) with integer charges $q_1,\dots, q_n$ under the anomalous $U(1)_{\rm an}$. We call $\mathcal{D}_i[\widetilde{a}, \widetilde{A}, B]=\bar{\sigma}^{\mu}D_{i \mu}[\widetilde{a}, \widetilde{A}, B]$ the covariant derivatives of the $i$-th fermion field $\psi_i$. Then, $F_{UV}$ is given by
\begin{align}
\label{eq:fuv}
    F_{UV}[\widetilde{a}, \widetilde{A}, B]=\frac{\sum_{i=1}^n q_i\mathcal{I}\left(\mathcal{D}_i[\widetilde{a}, \widetilde{A}, B]\right)}{\sum_{i=1}^n q_i}
\end{align}
where $\mathcal{I}$ denotes the index of its argument. Note that since both $F_{UV}$ and $\mathcal{D}_i$ are $1$-form  gauge-invariant they can always be reexpressed in terms of the original connections
\begin{align}
    & F_{UV}[\widetilde{a}, \widetilde{A}, B]=F_{UV}[a, A] && \mathcal{D}_i[\widetilde{a}, \widetilde{A}, B]=\mathcal{D}_i[a, A]
\end{align}
We will express the $F_{UV}$ and $\mathcal{D}_i$ in this way in subsection \ref{sec:complete}. For future use, we introduce also the functional densities
\begin{align}
\label{eq:densities}
    & F_{UV}[\widetilde{a}, \widetilde{A}, B] = \int \mathcal{F}_{UV}[\widetilde{a}, \widetilde{A}, B] \nonumber \\
    & F_{IR}[\widetilde{A}, B] = \int \mathcal{F}_{IR}[\widetilde{A}, B] 
\end{align}

\subsection{Complete CFL}
\label{sec:complete}
In this subsection, we prove that the $\theta$-periodicity anomaly is always absent in the complete CFL phase. We consider a quantum field theory with a dynamical $SU(N)_{\rm c}$ gauge field $a$ and a flavor symmetry group $G_{\rm f}$. We consider a scenario of complete CFL
\begin{align}
    G \longrightarrow H\;,
\end{align}
with
\begin{align}
   &G=\frac{SU(N)_{\rm c}\times G_{\rm f}}{\Gamma_G}\;, && H=\frac{H_{\rm cf}\times H_{\rm f}}{\Gamma_H}\;,
\end{align}
where $\Gamma_G$ and $\Gamma_H$ are discrete abelian subgroups of $SU(N)_{\rm c}\times G_{\rm f}$ and $H_{\rm cf}\times H_{\rm f}$. We define the (sub)algebras $\mathfrak{g}_{\rm f}$, $\mathfrak{h}_{\rm f}$, $\mathfrak{h}_{\rm cf}^{(\rm c)}$ and $\mathfrak{h}_{\rm cf}^{(\rm f)}$ of the (sub)groups $G_{\rm f}$, $H_{\rm f}$, $H_{\rm cf}^{(\rm c)}$ and $H_{\rm cf}^{(\rm f)}$ respectively (see subsection \ref{sec:cfldef}). We couple the $G_{\rm f}$-Noether currents to a background gauge field $A$, and decompose $A$ and $a$ as
\begin{widetext}
\begin{align}
    &A=A^{\mathfrak{h}_{\rm cf}}+A^{\mathfrak{h}_{\rm f}}+A_{\perp} && A^{\mathfrak{h}_{\rm cf}}\in \mathfrak{h}_{\rm cf}^{({\rm f})}\ , \quad A^{\mathfrak{h}_{\rm f}} \in \mathfrak{h}_{\rm f}\ , \quad A_{\perp}\in \mathfrak{g}_{\rm f}\ominus(\mathfrak{h}_{\rm cf}^{({\rm f})}\oplus \mathfrak{h}_{\rm f}) \nonumber \\
    & a = a^{\mathfrak{h}_{\rm cf}}+a_{\perp} && a^{\mathfrak{h}_{\rm cf}} \in \mathfrak{h}_{\rm cf}^{({\rm c})}\ , \quad a_{\perp} \in \mathfrak{su}(N)_{\rm c}\ominus \mathfrak{h}_{\rm cf}^{({\rm c})}\;.
\end{align}
\end{widetext}
Consider the functional defined in Eq. \eqref{eq:fuv}. In the topological sectors where the CFL phase occurs, the color and flavor transition functions $(g_{\rm c})_{ij}$ and $(g_{\rm f})_{ij}$ must have the form
\begin{align}
    &(g_{\rm c})_{ij}=(h_{\rm cf})_{ij}  && (h_{\rm cf})_{ij}\in H_{\rm cf}^{({\rm c})} \nonumber  \\
    &(g_{\rm f})_{ij}=(h_{\rm cf})_{ij}(h_{\rm f})_{ij} && (h_{\rm cf})_{ij}\in H_{\rm cf}^{({\rm f})}\ , \quad (h_{\rm f})_{ij}\in H_{\rm f}\;.
\end{align}
These transition functions satisfy some generalized cocycle conditions. Since the indices depend only on the transition functions (i.e., on the principal bundle) and not on the values of the fields in the bulk (i.e., on the connection), we have
\begin{align}
    &\mathcal{I}(\mathcal{D}_i[a, A])= \mathcal{I}(\mathcal{D}_i[A^{\mathfrak{h}_{\rm cf}}, A^{\mathfrak{h}_{\rm cf}}+A^{\mathfrak{h}_{\rm f}}]) && i=1,\dots, n
\end{align}
on $H$-bundles. These are exactly the indices that one would obtain by coupling a background gauge field $A_H$ to the $(H_{\rm cf}^{({\rm c})}\times H_{\rm cf}^{({\rm f})} \times H_{\rm f})$-Noether currents depending on $A^{\mathfrak{h}_{\rm cf}}$ and $A^{\mathfrak{h}_{\rm f}}$ according to the identity
\begin{align}
     &F_{UV}[a, A]\rvert_{H\text{-bundles}}= \nonumber \\
     &\frac{\sum_{i=1}^n q_i\  \mathcal{I}(\mathcal{D}_i[A^{\mathfrak{h}_{\rm cf}}, A^{\mathfrak{h}_{\rm cf}}+A^{\mathfrak{h}_{\rm f}}])}{\sum_{i=1}^n q_i} \equiv F_{UV}[A_{H}]
\end{align}
Hence, we can always match the $\theta$-periodicity anomaly by choosing
\begin{align}
    F_{IR}[A_{H}]=F_{UV}[A_{H}]\;.
\end{align}
This choice is automatically invariant under the $1$-form  gauge symmetry because it depends on the transition functions only through the $1$-form  gauge-invariant quantities $A^{\mathfrak{h}_{\rm cf}}$, $A^{\mathfrak{h}_{\rm f}}$. Hence, we conclude that \emph{the $\theta$-periodicity anomaly in the complete CFL phase can always be matched by the background gauge connections alone}. This result applies to all theories with complete CFL, including the $\psi\eta$ model (see subsection \ref{sec:intropsieta}) and the BY models (see subsection \ref{sec:introby}).

\subsection{Partial CFL: the case of the $\chi\eta$ model}
Theories with partial CFL are more interesting. In this subsection we consider the $\chi\eta$ model (see subsection \ref{sec:introchieta}), whose unbroken symmetry group is shown in Eqs. \eqref{eq:chietaunbroken}, \eqref{eq:chietacenter} where the factors are defined in table \ref{tab:chietacfl}.
\par We introduce the following background connections
\begin{itemize}
    \item $A_{\rm cf}$ for $SU(N-4)_{\rm cf}$
    \item $A$ for $U(1)'$
    \item $A_2$ for $\mathbb{Z}^F_2$
\end{itemize}
The connection $A_2$ is a flat connection satisfying the constraints
\begin{align}
    &\oint_{\gamma} \frac{A_2}{2\pi} \in \frac{\mathbb{Z}}{2} && \qquad \int_{\Sigma_2} \frac{dA_2}{2\pi}\in \mathbb{Z} 
\end{align}
where $\gamma$ is any closed loop and $\Sigma_2$ is any closed surface. These constraints can be satisfied simultaneously only if $A_2$ has a singularity \cite{Bolognesi:2020mpe, Bolognesi:2023xxv}. The dynamical gauge connection $a$ is associated to the color group $SU(N)_{\rm c}$. However, due to the CFL inside the restricted path integral its transition functions $(g_{\rm c})_{ij}$ must decompose into
\begin{align}
    (g_{\rm c})_{ij}=\begin{pmatrix}
        (g_4)_{ij} & 0 \\
        0 & (g_{\rm cf})_{ij}
    \end{pmatrix}\ , \qquad \begin{matrix}(g_4)_{ij}\in SU(4)\\ (g_{\rm cf})_{ij}\in SU(N-4)\end{matrix}
\end{align}
where the $(g_{\rm cf})_{ij}$ are transition functions of $A_{\rm cf}$ and $(g_4)_{ij}$ is a transition function associated to $SU(4)_{\rm c}$ whose topology is not fixed by the external gauge connections. We introduce a new \emph{fictitious} gauge connection 
\begin{itemize}
    \item $a_4$ for $SU(4)_{\rm c}$
\end{itemize}
whose transition functions are $(g_4)_{ij}$. In the restricted path integral for this theory, one has to sum over the topologies of the $(g_4)_{ij}$.
\par To deal with the fractional fluxes due to the $1$-form  center symmetries we introduce the $U(N-4)$ connection $\widetilde{A}_{\rm cf}$, the $U(1)$ connections $\widetilde{A}$, $\widetilde{A}_2$, $B_{\rm cf}$, $b$ and the $U(4)$ connection $\widetilde{a}_4$ that are related to the original connections by
\begin{align}
\label{eq:chietadecomp}
    & A_{\rm cf}=\widetilde{A}_{\rm cf}-\frac{B_{\rm cf}}{N-4} && A=\widetilde{A}-\frac{B_{\rm cf}}{N-4}+\frac{b}{4} \nonumber \\
    & a_{4}=\widetilde{a}_{4}-\frac{b}{4} && A_2 = \widetilde{A}_2 -\frac{b}{2}
\end{align}
with the 1-form gauge symmetry
\begin{align}
\label{eq:repchieta}
    &\widetilde{A}_{\rm cf}\longmapsto \widetilde{A}_{\rm cf}+(N-4)\lambda_{\rm cf} && B_{\rm cf}\longmapsto B_{\rm cf}+\lambda_{\rm cf} \nonumber \\
    &\widetilde{a}_4 \longmapsto \widetilde{a}_4+4\lambda_4 && b\longmapsto b+4\lambda_4 \nonumber \\
    &\widetilde{A}\longmapsto \widetilde{A}+\lambda_{\rm cf}-\lambda_4 && \widetilde{A}_2 \longmapsto \widetilde{A}_2+2\lambda_4
\end{align}
We also define the 1-form gauge-invariant topological charges
\begin{align}
    &Q_4 = \frac{1}{8\pi^2}\int \mathrm{Tr}\left(f_4\wedge f_4\right) \nonumber \\
    &Q_{\rm cf}= \frac{1}{8\pi^2}\int \mathrm{Tr}\left(F_{\rm cf}\wedge F_{\rm cf}\right)  \nonumber \\
    &Q_2= \frac{1}{8\pi^2} \int dA_2\wedge dA_2  \nonumber \\
    &Q=\frac{1}{8\pi^2} \int dA\wedge dA\nonumber \\
    & Q'=\frac{1}{8\pi^2} \int (2dA+dA_2)\wedge (2dA+dA_2)
\end{align}
where $f_4$ is the curvature of $a_4$ and $F_{\rm cf}$ is the curvature of $A_{\rm cf}$. By using the decomposition of Eq. \eqref{eq:chietadecomp} it is possible to compute the corresponding fluxes
\begin{align}
\label{eq:fluxes}
    & Q_4 = n-\frac{mm'}{4} \nonumber \\
    & Q_{\rm cf}=n_{\rm cf}-\frac{m_{\rm cf}m_{\rm cf}'}{N-4} \nonumber \\
    & Q_2 = \left(n_2-\frac{m}{2}\right)\left(n_2'-\frac{m'}{2}\right) \nonumber \\
    & Q = \left(n_1-\frac{m_{\rm cf}}{N-4}+\frac{m}{4}\right)\left(n_1'-\frac{m_{\rm cf}'}{N-4}+\frac{m'}{4}\right) \nonumber \\
    & Q' = \left(2n_1+n_2-\frac{2m_{\rm cf}}{N-4}\right)\left(2n_1'+n_2'-\frac{2m_{\rm cf}'}{N-4}\right)
\end{align}
We stress again the fact that the quantity $n$ is dynamical (even in the restricted path integral there is a sum over $n$) while all the other integers are fixed.
\par The functional $F_{UV}$ can be computed as in the previous subsection by coupling the UV fermions in table \ref{tab:chietacfl} to the external gauge connections introduced above, through the covariant derivatives
\begin{align}
    &D\chi_1=\left(d-i\bar{\mathcal{R}}_A(A_{\rm cf})-2iA-iA_2\right)\chi_1 \nonumber \\
    &D\chi_2=\left(d--i\bar{\mathcal{R}}_F(A_{\rm cf})-i\bar{\mathcal{R}}_{F}(a_4)-iA-iA_2\right)\chi_2 \nonumber \\
    &D\chi_3=\left(d-i\bar{\mathcal{R}}_A(a_4)-iA_2\right)\chi_3 \nonumber \\
    &D\eta_1=\left(d-i\mathcal{R}_{S\oplus A}(A_{\rm cf})-2iA-iA_2\right)\eta_1 \nonumber \\
    &D\eta_2=\left(d-i\mathcal{R}_F(A_{\rm cf})-i\mathcal{R}_F(a_4)-iA-iA_2\right)\eta_2
\end{align}
The functional $F_{UV}$ can be computed by the Fujikawa method. One finds
{\small
\begin{align}
    F_{UV}=& Q_4-\frac{(N-2)(N+1)}{2}Q_{\rm cf}+(N-1)Q_2 \nonumber \\
    +&4(N-4)Q-\frac{(N-4)(N^2-2N+15)}{4}Q'
\end{align}}
We now need to find a functional $F_{IR}$ that is independent of the dynamical degree of freedom $\widetilde{a}_4$ (and hence on its second Chern class $n$) and that satisfies
\begin{align}
    e^{i2\pi F_{UV}}=e^{i2\pi F_{IR}}
\end{align}
We can split $F_{IR}$ in two terms $F_{IR}^{(1)}$, $F_{IR}^{(2)}$ where
{\small
\begin{align}
\label{eq:ftilde1}
    F_{IR}^{(1)}=&-\frac{(N-2)(N+1)}{2}Q_{\rm cf}+(N-1)Q_2 \nonumber \\
    &+4(N-4)Q-\frac{(N-4)(N^2-2N+15)}{4}Q'
\end{align}}
and $F_{IR}^{(2)}$ must satisfy
\begin{align}
    e^{2\pi F_{IR}^{(2)}}=e^{-2\pi i \frac{mm'}{4}}
\end{align}
where the integers $m$, $m'$ are defined in the first line of Eq. \eqref{eq:fluxes}. We now show that for $N$ even it is impossible to construct $F_{IR}^{(2)}$ from the background fields a quantity that is invariant under the $1$-form  gauge symmetry \eqref{eq:repchieta} and satisfies
\begin{align}
    F_{IR}^{(2)}\equiv -\frac{mm'}{4}\ \mathrm{mod}\ 1
\end{align}
The most general candidate for $F_{IR}^{(2)}$ is the linear combination
\begin{align}
    F_{IR}^{(2)}=s_{\rm cf} Q_{\rm cf}+s_2 Q_2+ s Q+s'Q'
\end{align}
In order for $ F_{IR}^{(2)}$ to be independent of $n, n_{\rm cf}, n_1, n_2$, the coefficients must satisfy
\begin{align}
    s_{\rm cf} = \tilde{s}_{\rm cf}\ , \qquad & s = \frac{4(N-4)}{\mathrm{gcd}(N,4)}\tilde{s} && \nonumber \\
    s_2 = 2\tilde{s}_2\ , \qquad & s' =\frac{N-4}{\mathrm{gcd}(N,2)}\tilde{s}' && \tilde{s}_{\rm cf},\tilde{s}_2,\tilde{s},\tilde{s}'\in \mathbb{Z}
\end{align}
With these conditions, the functional becomes
\begin{align}
     F_{IR}^{(2)}\equiv &  \Bigg\{\frac{m_{\rm cf}m_{\rm cf}'}{N-4}\left[-\tilde{s}_{\rm cf}+\frac{\tilde{s}}{\mathrm{gcd}(N,4)}+\frac{4\tilde{s}'}{\mathrm{gcd}(N,4)}\right]  \nonumber \\
    &+mm'\left[\frac{\tilde{s}_2}{2}+\frac{N-4}{4\ \mathrm{gcd}(N,4)\tilde{s}}\right] \nonumber \\
    &+ \left(mm_{\rm cf}'+m'm_{\rm cf}\right)\left[-\frac{\tilde{s}}{\mathrm{gcd}(N,4)}\right] \Bigg\} \mathrm{mod}\ 1 
\end{align}
The third line cancels when
\begin{align}
     \tilde{s} = \mathrm{gcd}(N,4) \tilde{\tilde{s}}\ , \qquad \tilde{\tilde{s}}\in \mathbb{Z}
\end{align}
The second line cancels when
\begin{equation}
\label{eq:secondline}
    \frac{\tilde{s}_2}{2}+\frac{N}{4}\tilde{\tilde{s}} \equiv \frac{1}{4} \mathrm{mod}\ 1
\end{equation}
The first line does not yield any interesting constraint thanks to the freedom of choosing $s'$. The condition \eqref{eq:secondline} can be satisfied if and only if
\begin{equation}
    \mathrm{gcd}(N, 4) =1 \Leftrightarrow N\ \text{is odd}
\end{equation}
We conclude this subsection by stating that \emph{in the $\chi\eta$ model the $\theta$-periodicity anomaly cannot be matched by background gauge connections alone when $N$ is even.}

\subsection{Vector-like models}
We first study the $\psi\tilde{\psi}\eta\tilde{\eta}$ models, whose unbroken symmetry group is shown in Eqs. \eqref{eq:psietavecunbrk}, \eqref{eq:psietavecunbrk} where the factors are defined in table \ref{tab:psietavecunbrk}. We introduce the following background connections
\begin{itemize}
    \item $A_{\psi}$ for $SU(N_{\psi})_V$
    \item $A_{\eta}$ for $SU(N_{\eta})_V$
    \item $A_1$ for $U(1)_{\psi V}$
    \item $A_2$ for $U(1)_{\eta V}$
\end{itemize}
As usual, we call $a$ the dynamical $SU(N)_{\rm c}$ connection. To deal with the fractional fluxes due to the $1$-form  center symmetries we introduce a $U(N_{\psi})$ connection $\widetilde{A}_{\psi}$, a $U(N_{\eta})$ connection $\widetilde{A}_{\eta}$ and the $U(1)$ connections $\widetilde{A}_1, \widetilde{A}_2, b, B_{\psi}, B_{\eta}$ that are related to the original connections by
\begin{align}
\label{eq:psietavecdecomp}
    & A_{\psi}=\widetilde{A}_{\psi}-\frac{B_{\psi}}{N_{\psi}} && A_1=\widetilde{A}_1+\frac{2b}{N}+\frac{B_{\psi}}{N_{\psi}} \nonumber \\
    & A_{\eta}=\widetilde{A}_{\eta}-\frac{B_{\eta}}{N_{\psi}} && A_2=\widetilde{A}_2-\frac{b}{N}+\frac{B_{\eta}}{N_{\eta}} \nonumber \\
    & a = \widetilde{a}-\frac{b}{N} &&
\end{align}
with the 1-form gauge symmetry
\begin{align}
\label{eq:reppsietavec}
    & \widetilde{a} \longmapsto \widetilde{a} +\lambda_{\rm c}  && b \longmapsto b+N\lambda_{\rm c} \nonumber \\
    & \widetilde{A}_{\psi} \longmapsto \widetilde{A}_{\psi}+\lambda_{\psi} && B_{\psi} \longmapsto B_{\psi} + N_{\psi}\lambda_{\psi} \nonumber \\
    & \widetilde{A}_{\eta} \longmapsto \widetilde{A}_{\eta}+\lambda_{\eta} && B_{\eta} \longmapsto B_{\eta} + N_{\eta}\lambda_{\eta} \nonumber \\
    & \widetilde{A}_1 \longmapsto  \widetilde{A}_1 -\lambda_{\psi}-2\lambda_{\rm c} && \widetilde{A}_2 \longmapsto \widetilde{A}_2 -\lambda_{\eta}+\lambda_{\rm c}
\end{align}
We also define the 1-form gauge-invariant topological charges
\begin{align}
    & Q=\frac{1}{8\pi^2}\int \mathrm{Tr}(f\wedge f)  && Q_1 = \frac{1}{8\pi^2}\int dA_1\wedge dA_1 \nonumber \\
    & Q_{\psi}= \frac{1}{8\pi^2}\int \mathrm{Tr}(F_{\psi}\wedge F_{\psi}) && Q_2 = \frac{1}{8\pi^2}\int dA_2\wedge dA_2 \nonumber \\
    & Q_{\eta}= \frac{1}{8\pi^2}\int \mathrm{Tr}(F_{\eta}\wedge F_{\eta}) && Q_{12} = \frac{1}{8\pi^2}\int dA_1\wedge dA_2
\end{align}
where $f$, $F_{\psi}$, $F_{\eta}$ are the curvatures for $a$, $A_{\psi}$, $A_{\eta}$ respectively. By using the decomposition \eqref{eq:psietavecdecomp} it is possible to compute
\begin{align}
\label{eq:vectfluxes}
    & Q=n-\frac{mm'}{N}  \nonumber \\
    & Q_{\psi}= n_{\psi}-\frac{m_{\psi}m_{\psi}}{N_{\psi}} \nonumber \\
    & Q_{\eta}= n_{\eta}-\frac{m_{\eta}m_{\eta}}{N_{\eta}}  \nonumber \\ 
    & Q_1 = \left(n_1+\frac{2m}{N}+\frac{m_{\psi}}{N_{\psi}}\right)\left(n_1'+\frac{2m'}{N}+\frac{m_{\psi}'}{N_{\psi}}\right) \nonumber \\
    & Q_2 = \left(n_2-\frac{m}{N}+\frac{m_{\eta}}{N_{\eta}}\right)\left(n_2'-\frac{m'}{N}+\frac{m_{\eta}'}{N_{\eta}}\right) \nonumber \\
    & Q_{12} = \left(n_1+\frac{2m}{N}+\frac{m_{\psi}}{N_{\psi}}\right) \left(n_2'-\frac{m'}{N}+\frac{m_{\eta}'}{N_{\eta}}\right) \nonumber \\
    & \phantom{Q_{12}} + \left(n_1'+\frac{2m'}{N}+\frac{m_{\psi}'}{N_{\psi}}\right)\left(n_2-\frac{m}{N}+\frac{m_{\eta}}{N_{\eta}}\right) \nonumber \\
\end{align}
Again, the only dynamical quantity (i.e. summed over in the path-integral) is $n$.
\par The functional $F_{UV}$ can be computed as in the previous subsection by coupling the UV fermions in table \ref{tab:psietavecunbrk} to the external gauge connections introduced above, through the covariant derivatives
\begin{align}
    &D\begin{pmatrix}
        \psi \\
        \bar{\tilde{\psi}}
    \end{pmatrix}
    = \left[d-i\mathcal{R}_S(a)-i\mathcal{R}_{F}(A_{\psi})-iA_1\right]\begin{pmatrix}
        \psi \\
        \bar{\tilde{\psi}}
    \end{pmatrix} \nonumber \\
    &D\begin{pmatrix}
        \eta \\
        \bar{\tilde{\eta}}
    \end{pmatrix}
    = \left[d-i\mathcal{R}_F(a)-i\mathcal{R}_{F}(A_{\eta})+iA_2\right]\begin{pmatrix}
        \eta \\
        \bar{\tilde{\eta}}
    \end{pmatrix}
\end{align}
Using the Fujikawa method, we find
\begin{align}
    F_{UV}=&\frac{1}{N_{\eta}+N_{\psi}(N+2)}\Bigg\{\left[N_{\eta}+N_{\psi}(N+2)\right]Q \nonumber \\
    &+\frac{N(N+1)}{2}Q_{\psi}+NQ_{\eta} & \nonumber \\
    &+N_{\psi}\frac{N(N+1)}{2}Q_1+N_{\eta}N Q_2 \Bigg\}&
\end{align}
We now need to find a functional $F_{IR}$ that is independent of the dynamical degree of freedom $\widetilde{a}$ (and hence on its second Chern class $n$) and that satisfies
\begin{align}
    e^{i2\pi F_{UV}}=e^{i2\pi F_{IR}}
\end{align}
We can split $F_{IR}$ in two terms $F_{IR}^{(1)}$, $F_{IR}^{(2)}$ where
\begin{align}
\label{eq:fir1}
    F_{IR}^{(1)}=&\frac{1}{N_{\eta}+N_{\psi}(N+2)}\Bigg\{\frac{N(N+1)}{2}Q_{\psi}+NQ_{\eta} \nonumber \\
    &+N_{\psi}\frac{N(N+1)}{2}Q_1+N_{\eta}N Q_2 \Bigg\}
\end{align}
and $F_{IR}^{(2)}$ must satisfy
\begin{align}
    e^{2\pi F_{IR}^{(2)}}=e^{-2\pi i \frac{mm'}{N}}
\end{align}
or, in other words
\begin{equation}
    F_{IR}^{(2)}\equiv -\frac{mm'}{N}\ \mathrm{mod}\ 1
\end{equation}
where the integers $m$, $m'$ are defined in the first line of Eq. \eqref{eq:vectfluxes}. The most general candidate for $F_{IR}^{(2)}$ is the combination
\begin{align}
    F_{IR}^{(2)}=s_{\psi}Q_{\psi}+s_{\eta}Q_{\eta}+s_1Q_1+s_2Q_2+s_{12}Q_{12}
\end{align}
By inspection of the topological charges in Eq. \eqref{eq:vectfluxes} we find that a necessary condition for the dependence of $F_{IR}^{(2)}\ \mathrm{mod}\ 1$ from $n, n_{\psi}, n_{\eta}, n_1, n_2 $ to drop is that
\begin{align}
\label{eq:sconditions}
    & s_1 = \frac{N_{\psi}N}{\mathrm{gcd}(N,2)\mathrm{gcd}(N_{\psi}, \widetilde{N})}\tilde{s}_1 && \nonumber \\
    & s_2 = \frac{N_{\eta}N}{\mathrm{gcd}(N_{\psi}, N)}\tilde{s}_2 && \nonumber \\
    & s_{12} = \frac{N_{\psi}N_{\eta}N}{\mathrm{gcd}(N_{\psi}, N_{\eta}, N)}\tilde{s}_{12} && \tilde{s}_1, \tilde{s}_2, \tilde{s}_{12}\in \mathbb{Z} 
\end{align}
where and $\widetilde{N} = \frac{N}{\mathrm{gcd}(N,2)}$. When these conditions are satisfied, $F_{IR}^{(2)}$ reduces to
{\footnotesize
\begin{align}
    F_{IR}^{(2)}=\Bigg\{ & \frac{mm'}{N}\left(\frac{4s_1}{N}+\frac{s_2}{N}-\frac{4s_{12}}{N}\right)+\frac{m_{\psi}m_{\psi}'}{N_{\psi}}\left(-s_{\psi}+\frac{s_1}{N_{\psi}}\right) \nonumber \\
    &+\frac{m_{\eta}m_{\eta}'}{N_{\eta}}\left(-s_{\eta}+\frac{s_2}{N_{\eta}}\right) \nonumber \\
    &+\frac{1}{NN_{\psi}}(2s_1-s_{12})(mm_{\psi}'-m'm_{\psi}) \nonumber \\
    &+\frac{1}{NN_{\eta}}(2s_{12}-s_2)(mm_{\eta}'-m'm_{\eta}) \nonumber \\
    &+\frac{s_{12}}{N_{\psi}N_{\eta}}(m_{\psi}m_{\eta}'+m_{\psi}'m_{\eta})\Bigg\}\mathrm{mod}\ 1
\end{align}}
The term in the third line disappears thanks to the third condition in Eq. \eqref{eq:sconditions}. The terms in the second line disappear if
{\footnotesize
\begin{align}
    & 2s_1-s_{12} = N_{\psi}N t_1  \implies s_1 =\frac{N_{\psi}N}{2} \left(t_1+\frac{N_{\eta}}{\mathrm{gcd}(N, N_{\psi}, N_{\eta})}\tilde{s}_{12}\right) \\
    & 2s_{12}-s_2 = N_{\eta}N t_2 \implies s_2= N_{\eta}N\left(\frac{2N_{\psi}}{\mathrm{gcd}(N, N_{\psi}, N_{\eta})}\tilde{s}_{12}-t_2\right)
\end{align}}
where $t_1,t_2\in\mathbb{Z}$ and we used the third condition in Eq. \eqref{eq:sconditions}. It is always possible to choose $\tilde{s}_{12}, s_{\psi}, s_{\eta}$ so that the second and third terms in the first line disappear. The first term of the first line cancels if and only if
\begin{align}
    4s_1+s_2-4s_{12}=2N_{\psi}t_1-N_{\eta}t_2 = N\ \mathrm{mod}\ 1
\end{align}
By the Bézout identity, we know that there are some $t_1,t_2\in \mathbb{Z}$ that satisfy this relation if only if $N$ is a multiple of $\mathrm{gcd}(2N_{\psi}, N_{\eta})$. We conclude that \emph{in the vector-like $\psi\tilde{\psi}\eta\tilde{\eta}$ models the $\theta$-periodicity anomaly can be matched by background gauge connections alone unless $N$ is a multiple of $\mathrm{gcd}(2N_{\psi}, N_{\eta})$}.
\par A similar calculation shows that \emph{in the vector-like $\chi\tilde{\chi}\eta\tilde{\eta}$ models the $\theta$-periodicity anomaly can be matched by background gauge connections alone unless $N$ is a multiple of $\mathrm{gcd}(2N_{\chi}, N_{\eta})$}.

\section{Effective field theories on domain walls}
\label{sec:dw}
In this section we define a class of classical field configurations called \emph{domain walls} that may appear in field theories with massive compact scalar fields. We show that in gauge theories with a $\theta$-periodicity anomaly, the domain walls made of the $U(1)_{\rm an}$ pseudo-NGB $\varphi$ must support a Chern-Simons effective field theory.

\subsection{Domain walls}
Let us consider a gauge theory with left-handed spinors $\psi_1,\dots, \psi_n$ with $U(1)_{\rm an}$ charges $q_i,\dots, q_n$. We assume there is a subgroup  $\mathbb{Z}_{\ell} $ unbroken by the anomaly. Under a $U(1)_{\rm an}$ transformation, the UV partition function in the presence of the background fields $\widetilde{A}, B$ transforms as
\begin{align}
    U(1)_{\rm an}:\ & \psi_i \longmapsto e^{iq_i\alpha}\psi_i \nonumber \\
    & \mathcal{Z}_{UV}[\widetilde{A}, B, \theta] \longmapsto \mathcal{Z}_{UV}\textstyle
    \left[\widetilde{A}, B, \theta+\left(\sum_{i=0}^n q_i\right)\alpha\right]
\end{align}
The anomaly matching requires that the transformation law of $\mathcal{Z}_{IR}[\widetilde{A}, B, \theta]$ under $U(1)_{\rm an}$ is also
\begin{align}
\label{eq:u1anir}
    U(1)_{\rm an}:\ & \mathcal{Z}_{IR}[\widetilde{A}, B, \theta] \longmapsto \mathcal{Z}_{IR}\textstyle
    \left[\widetilde{A}, B, \theta+\left(\sum_{i=0}^n q_i\right)\alpha\right]
\end{align}
At sufficiently low energies, the only field that transforms nontrivially under $U(1)_{\rm an}$ is the anomalous pseudo-NGB $\varphi$, whose transformation law is, in general,
\begin{align}
\label{eq:transformations}
    U(1)_{\rm an}:\ \varphi \longmapsto \varphi+k\alpha
\end{align}
for some integer $k$ such that
\begin{align}
    &\sum_{i=0}^n q_i=k \ell && k, \ell \in \mathbb{Z}
\end{align}
It follows that in the low-energy effective action $\varphi$ and $\theta$ can appear only through the combination $(\ell\varphi+\theta)$. The low-energy effective action is invariant under the unbroken subgroup $\mathbb{Z}_{\ell}$, that acts on $\varphi$ as
\begin{align}
    \mathbb{Z}_{\ell}:\ & \varphi \longmapsto \varphi+\frac{2\pi}{\ell}m && m=0,1,\dots, \ell-1
\end{align}
This property is consistent with the $2\pi$-periodicity of $\theta$ in the absence of background fields. It follows that in the presence of background fields the low-energy effective action can be split into
\begin{align}
\label{eq:sir}
    S_{IR}=S_{\varphi}[\varphi, \theta]+S_{\text{other}}[\chi_{IR}, \widetilde{A}, B, \theta]-i\Delta S
\end{align}
where the normalization of the third term has been chosen for convenience. The first term $S_{\varphi}$ depends only on $\varphi$
\begin{widetext}
\begin{align}
\label{eq:sphi}
    S_{\varphi}[\varphi, \theta]=&Z\int\left[\frac{1}{2}\partial_{\mu}\varphi\partial^{\mu}\varphi+\frac{m_{\varphi}^2}{2\ell^2} \min\limits_{n\in \mathbb{Z}}\left(\ell\varphi+\theta +2\pi n\right)^2+\mathcal{L}(\partial \varphi, \partial^2\varphi,\dots)\right]
\end{align}
\end{widetext}
The mass term for the $\varphi$ field comes from the 't Hooft's vertex. The normalization factor $Z$ has been introduced because in general the normalization in which Eq. \eqref{eq:transformations} is valid is not canonical, while $\mathcal{L}(\partial \varphi, \partial^2\varphi,\dots)$ contains only derivative interaction terms. By construction, $S_{\varphi}$ transforms under $U(1)_{\rm an}$ as
\begin{align}
    U(1)_{\rm an}:\ S_{\varphi}[\varphi, \theta] \longmapsto S_{\varphi}\left[\varphi, \textstyle\theta+\left(\sum_{i=0}^n q_i\right)\alpha\right]
\end{align}
We stop for a moment to note that in the chiral models of section \ref{sec:baryons} and the vector-like models in section \ref{sec:introvect}, the mass $m_{\varphi}$ scales as
\begin{align}
    m_{\varphi}\sim \begin{cases}
        \frac{1}{N} & \text{when\ } N_{\psi}=N_{\chi}=0 \\
        N^0 & \text{otherwise}
    \end{cases}
\end{align}
due to the presence of two-index quarks. When $\varphi $ is canonical normalized the interaction lagrangian \emph{does} vanish in the large-$N$ limit due to the usual large-$N$ counting rules. Let $\varphi_c=\sqrt{Z}\varphi$ the canonically normalized field. We have
\begin{align}
    &\lim\limits_{N\to\infty}\mathcal{L}_c(\partial\varphi_c,\partial^2\varphi_c,\dots) \nonumber \\
    =&\lim\limits_{N\to\infty}Z\mathcal{L}(\partial\varphi/\sqrt{Z}, \partial^2\varphi/\sqrt{Z},\dots)=0
\end{align}
So when $N\to\infty$ the field $\varphi$ can be treated as a free massive field\footnote{This is true also in the chiral gauge theories of section \ref{sec:baryons} where the number of flavors is constrained to grow linearly with $N$ and the $N\to\infty$ limit is a Veneziano limit. The key observation is that in all these cases the $S$-matrix becomes trivial when $N\to\infty$.}. We do not know if at large-$N$ the theory has other states of mass $m\lesssim m_\varphi$. Were this the case, they should be included in the low-energy effective action to preserve unitarity. We assume that neglecting these states (if they are present) leads to no inconsistency because the mass $m_{\varphi}$ naturally arises when one includes the 't Hooft instanton vertex in the low-energy effective action for the massless degrees of freedom \cite{Bolognesi:2021hmg}.
\par The second term $S_{\rm other}$ depends on all the other IR fields $\chi_{IR}$ and is inert under $U(1)_{\rm an}$. The third term $\Delta S$ transforms nontrivially under $U(1)_{\rm an}$. To determine its form we write the IR partition function as
\begin{align}
    \mathcal{Z}_{IR}[\widetilde{A}, B, \theta] = \int \mathrm{exp}\Big(-S_{\varphi}-S_{\text{other}}+i\Delta S\Big)
\end{align}
where we omitted the integration measure for brevity. If we want the transformation law \eqref{eq:u1anir} and the $\theta$-periodicity property \eqref{eq:conditions} to be satisfied, the term $\Delta S$ must necessarily be of the form
\begin{align}
    \Delta S = \int(\theta+\ell\varphi)\ C_4
\end{align}
where $C_4$ is a gauge-invariant and $1$-form  invariant $4$-form satisfying the normalization condition
\begin{align}
    \int C_4 \equiv F_{UV}[0,\widetilde{A},B]\ \mathrm{mod}\ 1
\end{align}
on $H$-bundles. If the $\theta$-periodicity anomaly cannot be matched via a functional depending on the background connections $\widetilde{A}, B$ alone, the $4$-form $C_4$ must be constructed out of new degrees of freedom that must be included in the IR EFT. Although this is difficult to do in general, we can find a closed form for $\Delta S$ on a particular $\varphi$-background called \emph{domain wall}.
\par Suppose the theory lives on a four-manifold $\mathcal{M}$ and let $\gamma$ be a noncontractible loop. Thanks to the $\mathbb{Z}_{\ell}$ symmetry, $\varphi$ is allowed to have a nontrivial monodromy around $\gamma$
\begin{align}
    &\oint_{\gamma}d\varphi = \frac{2\pi}{\ell}m && m=0,1,\dots, \ell-1
\end{align}
For $m \neq 0$ these field configurations are stable, while for $m=0$ they are generally unstable or metastable. The simplest field configuration of this kind is the domain wall \cite{Dvali:1996xe, Gabadadze:1999pp, Witten:1997ep,Armoni:2005sp,Armoni:2003ji}. A domain wall is a time-independent field configuration of $\varphi$ depending only on one of the spacetime coordinates, say $z$, satisfying the boundary conditions
\begin{align}
    &\lim\limits_{z\to+\infty}\varphi(z)=\lim\limits_{z\to-\infty}\varphi(z)+\frac{2\pi}{\ell}m && m=0,\dots, \ell-1
\end{align}
If $\varphi$ has a mass $m_{\varphi}$, the domain wall usually has a profile that is almost flat for $z\to\pm \infty$ and jumps rapidly by $\frac{2\pi}{\ell}m$ in a region of size $\sim 1/m_{\varphi}$ near $z=0$. At extremely low energies a domain wall can be approximated by a step function
\begin{equation}
    \varphi(z)=\begin{cases}
        0 & z<0 \\
        \frac{2\pi}{\ell}m & z>0
    \end{cases}
\end{equation}
By dimensional analysis, the tension of the domain wall in the large-$N$ limit can be estimated as

\begin{align}
    \sigma\propto Zm_{\varphi}
\end{align}
The exact expression is derived in appendix \ref{app:dw} under the approximation where $\varphi$ is a free field. The decay rate of the metastable domain wall with $m=0$ has been estimated semiclassically in Ref. \cite{Preskill:1992ck} in the thin-wall approximation. The decay process is described by a bounce solution in Euclidean space in which a hole bounded by a string nucleates. This leads to the estimate
\begin{align}
    \Gamma \propto \mathrm{exp}\left(-\frac{16\pi\mu^2}{3\sigma^2}\right)
\end{align}
where $\mu$ is the tension of the string and $\sigma$ is the tension of the domain wall. The value of $\mu$ is highly model-dependent \cite{Eto:2013bxa, Eto:2022lhu}.

\subsection{$\chi\eta$ model}
We consider the $\chi\eta$ when $N$ is even and the background gauge fields are not sufficient to match the $\theta$-periodicity anomaly.
\par In this case, the anomaly breaks the subgroup $U(1)_{\rm an}$ (see subsection \ref{sec:introchieta}) to the discrete subgroup
\begin{align}
    \mathbb{Z}_{2}^F: \quad & \psi \longmapsto e^{i\pi k}\psi && \eta \longmapsto e^{i \pi k}\eta && k=0,1
\end{align}
because $N^*=\mathrm{gcd}(N,2)=2$ when $N$ is even. The anomalous pseudo-NGB $\varphi$ is created by the  't Hooft instanton-vertex
\begin{align}
    &\mathrm{det}_{i, A}\left(\chi^{[ij]}\eta_j^A\right)\mathrm{det}_{\ell,\ell'}\left(\chi^{[\ell\ell']}\right)\sim e^{iN^*\varphi} \nonumber \\
    &i,A=1,\dots,N-4\ , \quad \ell,\ell'=N-3, ..., N
\end{align}
$U(1)_{\rm an}$ transformations act on $\varphi$ as
\begin{align}
    U(1)_{\rm an}:\ &\varphi \longmapsto \varphi + \alpha \nonumber \\
    &\theta \longmapsto \theta+2\alpha
\end{align}
The only allowed domain walls have the boundary conditions with boundary conditions
\begin{align}
    &\lim\limits_{z\to+\infty}\varphi(z)=\lim\limits_{z\to-\infty}\varphi(z)+\pi m  && m=0,1
\end{align}
and in the deep IR is approximated by the step functions
\begin{align}
   & \varphi(z)=\begin{cases}
        0 & z<0 \\
        \pi m & z>0 \\
    \end{cases} && m=0,1
\end{align}
The domain wall with $m=1$ is stable, while the domain wall with $m=2$ is metastable.
\par We now put $\theta=0$ through a $U(1)_{\rm an}$ transformation. Then, on a domain wall background $\Delta S$ takes the form
\begin{align}
    \Delta S = \pi m \int_{z>0}\mathcal{F}_{IR}^{(1)}-\frac{m}{4\pi}\int_{z>0}\frac{db\wedge db}{4}+\int_{z=0} c_3
\end{align}
where $\mathcal{F}^{(1)}_{IR}$ is the functional density for the $F_{IR}^{(1)}$ defined in Eq. \eqref{eq:ftilde1}, while $c_3$ is a $3$-form that contains the new degrees of freedom of the domain wall. $1$-form gauge invariance of $\Delta S$ constrains $c_3$ to transform as
\begin{align}
    c_3 \longmapsto c_3+\frac{m}{2\pi}\int_{z=0}\lambda_{\rm c}\wedge db
\end{align}
\par There are several choices of $c_3$ that satisfy all these requirements. The simplest example is the following. We introduce a new $\mathfrak{u}(1)$-valued $1$-form field $c$ transforming as
\begin{align}
    c\longmapsto c-\lambda_{\rm c}
\end{align}
and take
\begin{align}
    \int_{z=0} c_3 = -\frac{m}{4\pi}\int_{z=0} \left(4c\wedge dc+2c\wedge db\right)
\end{align}
Therefore, the domain wall can be described by a $U(1)_{-4}$ Chern-Simons theory. Alternatively, one can introduce a $\mathfrak{u}(4)$ $1$-form  field $c$ with the same $1$-form  transformation law as above and take 
\begin{align}
    \int_{z=0} c_3 = -\frac{m}{4\pi}\int_{z=0} \mathrm{Tr}\left(c\wedge dc+\frac{2}{3}c\wedge c \wedge c+\frac{1}{2}c\wedge db\right)
\end{align}
which is a $U(4)_{-1}$ Chern-Simons theory. As a matter of fact, the $U(1)_{-4}$ and the $U(4)_{-1}$ Chern-Simons theories are equivalent by level-rank duality \cite{Hsin:2016blu}. Another possibility is to take a $U(2)_{-1}\times U(2)_{-1}$ Chern-Simons theory for two $\mathfrak{u}(2)$ $1$-form s $c_1, c_2$, with action
\begin{widetext}
\begin{align}
    \int_{z=0} c_3 =-\frac{m}{4\pi}\int_{z=0} \mathrm{Tr}\left[c_1\wedge dc_1+\frac{2}{3}c_1\wedge c_1 \wedge c_1+\frac{1}{2}c_1\wedge db\right]-\frac{m}{4\pi}\int_{z=0} \mathrm{Tr}\left[c_2\wedge dc_2+\frac{2}{3}c_2\wedge c_2 \wedge c_2+\frac{1}{2}c_2\wedge db\right]
\end{align}
\end{widetext}
It is possible to construct other theories that match the $\theta$-periodicity anomaly with the same procedure.
\par We stress the fact that even when the background gauge fields alone are sufficient to match the $\theta$-periodicity anomaly the domain wall may still support an effective field theory. An example of this sort is provided in Ref. \cite{Kitano:2020evx}, where by a large-$N$ argument it is shown that in $SU(N)$ QCD with $N_{\rm f}$ flavors and $\mathrm{gcd}(N, N_{\rm f})=1$, the $\eta'$ domain wall supports a nontrivial effective field theory even though they are not necessary to match any $\theta$-periodicity anomaly.

\subsection{Vector-like models}
We now consider the vector-like $\psi\tilde{\psi}\eta\tilde{\eta}$  models when $N$ is \emph{not} a multiple of $\mathrm{gcd}(2N_{\psi}, N_{\eta})$. 
\par
\begin{figure*}
	\begin{center}
	\includegraphics[scale=0.8]{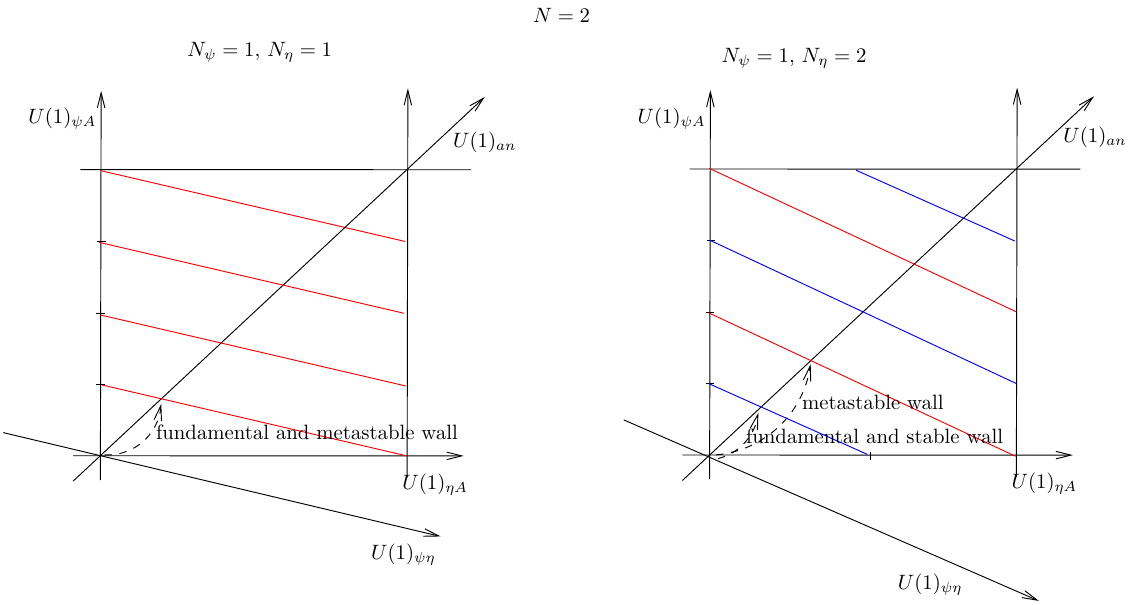} 
		\caption{Examples of the braking of the torus $U(1)_{\psi A} \times U(1)_{\eta A} $.}
		\label{twou}
	\end{center}
\end{figure*}

 In the $\psi\tilde{\psi}\eta\tilde{\eta}$, the axial anomaly breaks the subgroup $U(1)_{\rm an}$ (see table \ref{tab:psietavec}) to the discrete subgroup 
\begin{align}
    \mathbb{Z}_{\widetilde{N}}: \quad & \psi \longmapsto e^{\frac{2\pi i k}{\widetilde{N}}}\psi && \tilde{\psi} \longmapsto e^{\frac{2\pi i k}{\widetilde{N}}}\tilde{\psi} &&  \nonumber \\
     & \eta \longmapsto e^{\frac{2\pi i k}{\widetilde{N}}}\eta && \tilde{\eta} \longmapsto e^{\frac{2\pi i k}{\widetilde{N}}}\tilde{\eta} && k=0,\dots \widetilde{N}-1
\end{align}
where $\widetilde{N}=\mathrm{gcd}(2N_{\eta},2N_{\psi}(N+2))$. The anomalous pseudo-NGB $\varphi$ is created by the 't Hooft vertex
\begin{align}
   \mathrm{det}\left(\tilde{\psi}^A\psi^B\right)^{N+2}\mathrm{det}\left(\tilde{\eta}^a\eta^b\right)\sim e^{i\widetilde{N}\varphi}
\end{align}
We now explain this normalization of $\varphi$. The $U(1)_{\rm an}$ transformations act as
\begin{align}
    U(1)_{\rm an}:\ &\varphi \longmapsto \varphi + \frac{2[N_{\eta}+N_{\psi}(N+2)]}{\widetilde{N}}\alpha \nonumber \\
    &\theta \longmapsto \theta+2[N_{\eta}+N_{\psi}(N+2)]\alpha
\end{align}
so $\ell = \widetilde{N}$ as expected. The domain walls have boundary conditions
\begin{align}
    &\lim\limits_{z\to+\infty}\varphi(z)=\lim\limits_{z\to-\infty}\varphi(z)+\frac{2\pi m}{\widetilde{N}}\ , && m=0,1,\dots,\widetilde{N}-1
\end{align}
and in the deep IR can be approximated by
\begin{equation}
    \varphi(z)=\begin{cases}
        0 & z<0 \\
        \frac{2\pi}{\widetilde{N}}m & z>0 \qquad \qquad m=0,1,\dots,\widetilde{N}-1
    \end{cases}
\end{equation}
The structure of the domain walls in two simple examples is graphically represented in Figure \ref{twou}. The procedure is completely analogous to that of the $\chi\eta$ model, so we show the results without proof. After putting $\theta=0$ by a $U(1)_{\rm an}$ transformation
\begin{align}
\label{eq:c2c3vec}
    \Delta S=2\pi m\int_{z> 0} \mathcal{F}^{(1)}_{IR}-\frac{m}{4\pi}\int_{z>0}\frac{db\wedge db}{N}+\int_{z=0} c_3
\end{align}
where $\mathcal{F}^{(1)}_{IR}$ is the functional density for the $F_{IR}^{(1)}$ defined in Eq. \eqref{eq:fir1}. Again, there are several choices for $c_3$. In the simplest case, the only degree of freedom consists of a $U(1)$ $1$-form  $c$ transforming under \eqref{eq:reppsietavec} as
\begin{align}
    c \longmapsto c-\lambda_{\rm c}
\end{align}
and with action
\begin{align}
    \int_{z=0}c_3=-\frac{m}{4\pi}\int_{z=0}\left(Nc\wedge dc+2db\wedge c\right)
\end{align}
which is the action of a $U(1)_{-mN}$ Chern-Simons theory. An interesting possibility arises for the metastable domain walls
\begin{equation}
    \varphi(z)=\begin{cases}
        0 & z<0 \\
        2\pi & z>0
    \end{cases}
\end{equation}
i.e. with $m=\widetilde{N}$. In this case, we introduce a $\mathfrak{u}(N_{\psi})$ $1$-form  field $c_{\psi}$ and a $\mathfrak{u}(N_{\eta})$ $1$-form  field $c_{\eta}$. The simplest $3$-form $c_3$ that we can construct out of these fields such that $\Delta S$ is $1$-form  gauge-invariant is
\begin{widetext}
\begin{align}
    \int_{z=0}c_3=&-\frac{N+2}{4\pi}\int_{z=0}\mathrm{Tr}\left[N\left(c_{\psi}\wedge dc_{\psi}+\frac{2}{3}c_{\psi}\wedge c_{\psi}\wedge c_{\psi}\right)+2c_{\psi}\wedge db\right] \nonumber \\
    &-\frac{1}{4\pi}\int_{z=0}\mathrm{Tr}\left[N\left(c_{\eta}\wedge dc_{\eta}+\frac{2}{3}c_{\eta}\wedge c_{\eta}\wedge c_{\eta}\right)+2c_{\eta}\wedge db\right]
\end{align}
\end{widetext}

For the vector-like $\chi\tilde{\chi}\eta\tilde{\eta}$ models when $N$ is \emph{not} a multiple of $\mathrm{gcd}(2N_{\chi}, N_{\eta})$ the procedure and results are identical except for the fact that $\tilde{N}=\mathrm{gcd}(2N_{\eta}, 2N_{\chi}(N-2)) $

\section{Conclusion}
\label{sec:conclusion}

In this paper, we study the solitonic sector of some chiral $SU(N)$ gauge theories and vector-like theories with fermions in mixed fundamental and two-index representations. As the paper presents various results, we find it necessary to summarize the main findings here.

From the structure of the symmetries and large-$N$ considerations, we know that in all the chiral models (specifically the $\psi\eta$, $\chi\eta$, and their generalization) there are stable light baryons composed of three elementary fermions with no $\epsilon$-tensor. The story is different for heavy baryons. In the $\psi\eta$ and BY models, we verified numerically that the unbroken symmetry group allows for the decay of heavy baryons into lighter particles e.g. light baryons and NGBs. On the contrary, in the $\chi\eta$ and GG models these decays are not always allowed: some heavy baryons are forbidden to decay into lighter particles by the unbroken symmetry group.

By assuming that the CFL phase is realized in the IR, we computed the homotopy groups of the coset where the NGBs of the theories live. We find that the $\pi_3$ is trivial; therefore, it is impossible to find topologically stable Skyrmions in such a phase. For the $\psi\eta$ and BY models, this result could be thought of as an unrigorous consistency check: it would be difficult to reconcile the presence of topologically stable Skyrmions if the corresponding baryons are supposed to decay\footnote{The check is not rigorous because it is not possible to determine if such solitons are solutions of finite energy, and within the regime of validity of the IR EFT.}. In the $\chi\eta$ and GG models, however, stable heavy baryons seem to exist. We consider this fact rather surprising: even at finite $N$, it is legitimate to expect the Skyrme model to predict \textit{qualitatively} the baryon observables, as it happens in QCD with $N=3$. Hence, we think that the dynamical mechanism that leads to the absence of topologically stable Skyrmions deserves an explanation.

In vector-like models, we have two ``baryon numbers'', and two heavy baryons are stable. Consistently, by assuming the chiral condensates, the relevant coset has $\pi_3 =\mathbb{Z} \times \mathbb{Z}$, leading to two independently conserved topological charges. By computing the WZW terms, we can extract the quantum numbers of such Skyrmions, which can be identified with the baryons.

Then we study the domain walls of such theories. In many of these theories, there are stable and metastable domain walls. To obtain some insight into the degrees of freedom of the domain wall, computed the $\theta$-periodicity anomaly: the anomalies of the theory on the wall must reproduce the $\theta$-periodicity anomaly of the theory.

Interestingly, such an anomaly is always trivialized in the case of complete CFL. The reason is that, in the complete CFL case, the topology of the gauge field is locked to reproduce the external $H_{\rm cfl}$ principal bundle topology. As the anomaly therefore depends only on such data, it can be written in a gauge invariant (and, crucially, 1-form gauge invariant) way in terms of the external fields only, and thus, can be reproduced by a local counterterm in the IR EFT.

In case of partial CFL, conversely, generically, there is a nontrivial $\theta$-periodicity anomaly, which cannot be reproduced in the IR by a 1-form gauge invariant counterterm constructed from the background gauge fields alone. For example, we find that the $\theta$-periodicity anomaly of the $\chi\eta$ is nontrivial when $N$ is even, while the $\psi\tilde{\psi}\eta\tilde{\eta}$ is nontrivial when ${\rm gcd}(2N_\psi, N_\eta)$ divides $N$.

These results, by themselves, do not imply that the IR EFT is not consistent with the UV data. They mean, however, that on the domain walls of the theory, there are massless or topological degrees of freedom. We propose some alternatives for this worldsheet dynamics in terms of Chen-Simons theories, with the caveat that there are inequivalent possibilities that cannot be singled out with these considerations alone.

This result is required to study the possibility of having ``pancakes'?, i.e., solitons consisting of a metastable domain wall, bounded by a string-vortex. In the context of $N_{\rm f}=1$ QCD, it has been shown that such states are stable, carry baryon number, and spin $\sim N/2$. As such, they are candidates to correspond to the baryons.

The same mechanism can be at work in our cases. Unfortunately, we do not have enough control over the theory to discuss the stability. Nevertheless, it would be interesting to compute the charges of such states and determine if they can play any role in the picture above. For example, in orbifold QCD or QCD with $N_\psi=N_\eta=1$, we expect that the mechanism could be analogous to the case of $N_{\rm f}=1$ QCD: heavy baryons are to be expected because of conservation of the vector-like symmetries $U(1)_{\psi V}$ and $U(1)_{\eta V}$, but the coset topology does not allow Skyrmions.

\appendix

\section{Gauging $1$-form symmetries}
\label{app:1form}

\subsection{Fractional fluxes}
Consider a manifold $\mathcal{M}$ covered by the open patches $U_i$. Let $\psi$ be a fermion field and $A$ a connection defined locally in each patch such that
\begin{align}
    \psi(x) = \psi_i(x)\ , \qquad A(x)=A_i(x)\ , \qquad x \in U_i
\end{align}
These local definition are glued together in the overlap between the patches by a set of transition functions with values in $SU(N)$. We call $g_{ij}$ the transition function associated to the intersection $U_i\cap U_j$. The relation between $\psi_i$ and $\psi_j$ is then
\begin{widetext}
\begin{align}
    & \psi_i(x)=\mathcal{R}(g_{ij}(x))\psi_j(x) \nonumber \\
    & A_i(x) = g_{ij}(x)A_i(x)g_{ij}^{-1}(x)-idg_{ij}(x) g_{ij}^{-1}(x)\ , \qquad x\in U_i\cap U_j \nonumber
\end{align}
\end{widetext}
where $\mathcal{R}$ is a representation of $SU(N)$. Suppose that the kernel of the representation $\mathcal{R}$ is a subgroup $\mathbb{Z}_p \subset SU(N)$. In order for $\psi$ and $A$ to be single valued on $\mathcal{M}$, it is sufficient that the transition functions satisfy the generalized cocycle condition
\begin{equation}
\label{eq:cocycle}
    g_{ij}g_{jk}g_{ki}=z_{ijk}\in \mathbb{Z}_p
\end{equation}
The transition functions $g_{ij}$ satisfying the condition \eqref{eq:cocycle} define a $SU(N)/\mathbb{Z}_p$ bundle. We now show how to describe a $SU(N)/\mathbb{Z}_p$ in terms of a $U(N)$ and a $U(1)$ bundles with suitable structure functions.
\par We introduce $U(1)$-connection $B$ with transition functions $e^{i\alpha_{ij}}$ satisfying the relation
\begin{equation}
\label{eq:u1coycle}
    e^{i\alpha_{ij}/p}e^{i\alpha_{jk}/p}e^{i\alpha_{ki}/p}=z_{ijk}^{-1}
\end{equation}
Taking the $p$-th power of this equality, one sees that the transition functions for $B$ satisfy the ordinary cocycle conditions because $z_{ijk}^p=1$. We introduce also the $U(N)$-connection
\begin{align}
\label{eq:widetilde}
    \widetilde{A}=A+\frac{B}{p}
\end{align}
Inserting \eqref{eq:u1coycle} into \eqref{eq:cocycle}, one sees that the transition functions $\widetilde{g}_{ij}=g_{ij}e^{i\alpha_{ij}/p}$ for $\widetilde{A}$ satisfy the ordinary cocycle conditions $\widetilde{g}_{ij}\widetilde{g}_{jk}\widetilde{g}_{ki}=1$. Note also that the definition \eqref{eq:widetilde} is invariant under the reparametrizations
\begin{equation}
\label{eq:reparametrization0}
    \widetilde{A}\longmapsto \widetilde{A}+\lambda\ , \qquad B\longmapsto B+p\lambda
\end{equation}
where $\lambda$ is some $1$-form . Thanks to Eq. \eqref{eq:widetilde}, it is possible to define the twisted connection $A$ in terms of two ordinary connections $\widetilde{A}$ and $B$ at the price of introducing the reparametrization invariance \eqref{eq:reparametrization}. We can use this trick to easily compute the second Chern class of $A$
\begin{align}
\label{eq:fractional}
    \int \frac{1}{8\pi^2}\mathrm{Tr}\ F\wedge F=\int \frac{1}{8\pi^2}\mathrm{Tr}\ F\wedge F-\frac{N}{p^2}\int \frac{1}{8\pi^2}dB\wedge dB
\end{align}
Since both $\widetilde{A}$ and $B$ are ordinary connections we have
\begin{align}
    &\int \frac{1}{8\pi^2}\mathrm{Tr}\ F\wedge F \in \mathbb{Z} \nonumber \\
    &\int\frac{1}{8\pi^2}dB\wedge dB \in \mathbb{Z}
\end{align}
Hence, we computed the second Chern class of a $SU(N)/\mathbb{Z}_p$-connection in terms of two ordinary connections $\widetilde{A}$, $B$.
\par We can generalize this procedure to bundles with structure group
\begin{align}
    \frac{G^{(1)}\times \dots \times G^{(n)}\times U(1)^{(1)}\times \dots \times U(1)^{(m)}}{\mathbb{Z}_{p_1}\times \mathbb{Z}_{p_2}\times \dots\times \mathbb{Z}_{p_n}}
\end{align}
where the groups $G^{(i)}$ are compact, simple and semi-simple and the $\mathbb{Z}_{p_i}$ are defined as
\begin{align}
    \Big(\underbrace{e^{\frac{2\pi i k_1}{p_1}},\dots,e^{\frac{2\pi i k_n}{p_n}}}_{G^{(1)}\times \dots \times G^{(n)}}, \underbrace{e^{2\pi i \sum_{i=1}^n\frac{q_{1i}k_i}{p_i}},\dots, e^{2\pi i \sum_{i=1}^n\frac{q_{mi}k_i}{p_i}}}_{U(1)^{(1)}\times \dots \times U(1)^{(m)}}\Big)
\end{align}
where $q_{ij}\in \mathbb{Z}$. Let $A^{G}_i$ be the connection for the group $G^{(i)}$ and $A^{U(1)}_i$ be the connection for the group $U(1)^{(i)}$. We can redefine these connections as
\begin{align}
    & A^{G}_i = \widetilde{A}^G_i-\frac{B_i}{p_i} && i=1,\dots, n   \nonumber \\
    & A_i^{U(1)} = \widetilde{A}_i^{U(1)}-\sum_{j=1}^n\frac{q_{ij}}{p_j}B_j && i=1,\dots, m
\end{align}
where now $\widetilde{A}_i^{G}$ is an ordinary $G^{(i)}\times U(1)$-connection and $\widetilde{A}_i^{U(1)}$ is an ordinary $U(1)$-connection. This definition is invariant under the $1$-form  gauge symmmetry
\begin{align}
\label{eq:reparametrization}
    & \widetilde{A}^G_i \longmapsto  \widetilde{A}^G_i+\lambda_i && i=1,\dots, n \nonumber \\
    &B_i \longmapsto B_i+p_i\lambda_i && i=1,\dots, n \nonumber \\
    &\widetilde{A}_i^{U(1)}\longmapsto \widetilde{A}_i^{U(1)}+\sum_{j=1}^nq_{ij}\lambda_j  && i = 1,\dots, m 
\end{align}
that arises from the freedom to redefine the $\widetilde{A}^G_i$, $\widetilde{A}_i^{U(1)}$, $B_i$ while keeping $A^{G}_i$, $A_i^{U(1)}$ constant.

\subsection{Quantization}
\label{app:1formquant}
Let us consider a quantum field theory with color group $SU(N)_{\rm c}$, a dynamical gauge connection $a$ and some matter fields $\psi$. The universal cover of the group of global symmetries is
\begin{align}
    \overline{G}=SU(N)_{\rm c}\times G^{(1)}\times \dots \times G^{(n)}\times U(1)^{(1)}\times \dots \times U(1)^{(m)}
\end{align}
The matter fields $\psi$ transform under a representation $\mathcal{R}$ of $\overline{G}$ with kernel
\begin{align}
    \mathrm{ker}(\mathcal{R})=\mathbb{Z}_N\times\mathbb{Z}_{p_1}\times \mathbb{Z}_{p_2}\times \dots\times \mathbb{Z}_{p_n} \subset \overline{G}
\end{align}
The true global symmetry group of the theory is then $G=\overline{G}/\mathrm{ker}(R)$. In the absence of background gauge fields, the Euclidean partition function of the theory is
\begin{align}
    \mathcal{Z}=\int\limits_{SU(N)_{\rm_c}\text{-bundles}} [da][d\psi]\ e^{-S[\psi,a]}
\end{align}
and the correlators between local operators $\mathcal{O}_i(x_i)$ are defined as
\begin{widetext}
\begin{align}
    \expval{\mathcal{O}_1(x_1)\dots \mathcal{O}_1(x_1)}=\frac{1}{\mathcal{Z}}\int\limits_{SU(N)_{\rm_c}\text{-bundles}} [da][d\psi]\mathcal{O}_1(x_1)\dots \mathcal{O}_1(x_1)\ e^{-S[\psi, a]}
\end{align}
\end{widetext}
We put the theory on a four-manifold $\mathcal{M}_4$ with nontrivial topology and turn on some background gauge connections for the symmetry group
\begin{align}
    \frac{SU(N)_{\rm c}\times G^{(1)}\times \dots \times G^{(n)}\times U(1)^{(1)}\times \dots \times U(1)^{(m)}}{\mathbb{Z}_N\times\mathbb{Z}_{p_1}\times \mathbb{Z}_{p_2}\times \dots\times \mathbb{Z}_{p_n}}
\end{align}
As shown above, the connections that appear in the partition function are
\begin{itemize}
    \item The $U(N)_{\rm c}$-connection $\widetilde{a}$
    \item The $G^{(i)}\times U(1)$-connections $\widetilde{A}_i^{G}$
    \item The $U(1)^{(i)}$ connections $\widetilde{A}_i^{U(1)}$
    \item The $U(1)$-connection $b$ for the $1$-form  symmetry $\mathbb{Z}_N$
    \item The $U(1)$-connections $B_i$ for the $1$-form  symmetries $\mathbb{Z}_{p_i}$
\end{itemize}
Physical quantities, including the action $S[\psi, \widetilde{a}, \widetilde{A}, b, B]$, must be invariant under the $1$-form  gauge symmetry
\begin{align}
    & \widetilde{a}\longmapsto \widetilde{a}+\lambda_{\rm c} && \nonumber \\
    & \widetilde{A}^G_i \longmapsto  \widetilde{A}^G_i+\lambda_i && i=1,\dots, n \nonumber \\
    &\widetilde{A}_i^{U(1)}\longmapsto \widetilde{A}_i^{U(1)}+q_i\lambda_{\rm c}+\sum_{j=1}^nq_{ij}\lambda_j  && i = 1,\dots, m \nonumber \\
    & b \longmapsto b+N\lambda_{\rm c}  && \nonumber \\
    &B_i \longmapsto B_i+p_i\lambda_i && i=1,\dots, n
\end{align}
In this work, the partition function for the theory coupled to these external background fields is \emph{defined} as
\begin{align}
\label{eq:fractionalz}
    \mathcal{Z}=\int\limits_{G\text{-bundles}}[d\widetilde{a}][d\psi]\ e^{-S[\psi, \widetilde{a}, \widetilde{A}, b, B]}
\end{align}
while the correlators are
\begin{widetext}
\begin{align}
    \expval{\mathcal{O}_1(x_1)\dots \mathcal{O}_1(x_1)}=\frac{1}{\mathcal{Z}}\int\limits_{G\text{-bundles}} [d\widetilde{a}][d\psi]\mathcal{O}_1(x_1)\dots \mathcal{O}_1(x_1)\ e^{-S[\psi, \widetilde{a}, \widetilde{A}, b, B]}
\end{align}
\end{widetext}
Note that we summed over the connection $\widetilde{a}$, and not both $\widetilde{a}$ and $b$. In other words, we are summing over the bundles by keeping the $z_{ijk}$ in Eq. \eqref{eq:cocycle} constant.

\section{Some algebraic topology}
\label{app:topology}
The material presented in this appendix is taken from Refs. \cite{MR1867354, Steenrod1951TopologyOF}.

\subsection{Fundamental results}
A fibration
\begin{equation}
    F\longrightarrow E \longrightarrow B
\end{equation}
admits a long exact sequence of homotopy groups
\begin{equation}
\label{eq:les}
    \dots\longrightarrow \pi_i(F)\longrightarrow \pi_i(E) \longrightarrow \pi_i (B) \longrightarrow \pi_{i-1}(F)\longrightarrow \dots
\end{equation}
In the following sections, $F$ and $E$ will be compact, connected Lie groups, and $B$ will be a coset space.
\par In particular, we will be interested in the homotopy groups $\pi_i$ with $i\geq 2$. In this case, a convenient simplification occurs. A coset
\begin{equation}
    \frac{G \times H}{G' \times H'}
\end{equation}
is part of the fibrations
\begin{align}
\label{eq:fibration}
     &H \longrightarrow \frac{G \times H}{G'\times H'} \longrightarrow \frac{G}{G'\times H'} \nonumber \\
     &H' \longrightarrow \frac{G\times H}{G'} \longrightarrow \frac{G\times H}{G'\times H'}
\end{align}
Suppose that
\begin{equation}
\label{eq:pih}
    \pi_i(H)=\pi_i(H')=0\ , \qquad \text{for} \qquad i\geq 2
\end{equation}
Using the long exact sequence \eqref{eq:les} of homotopy groups for the fibrations \eqref{eq:fibration}, we see that
\begin{equation}
\label{eq:piequiv}
    \pi_i\left(\frac{G\times H}{G'\times H'}\right)=\pi_i(G/G')\ , \qquad \text{for} \qquad i\geq 2
\end{equation}
Since the derivations of the following sections involve the segment of long exact sequences containing only $\pi_i$ for $i\geq 2$, this result allows us to discard any discrete or $U(1)$ factors appearing in the coset spaces of interest.
\par The spaces we are interested in are made of products and cosets of $SU(n)$. We summarise here the topological properties of this group that we need.
\begin{subequations}
\label{eq:properties}
    \begin{equation}
        \pi_2(SU(n))=0\ , \qquad n\geq 2
    \end{equation}
    \begin{equation}
        \pi_3(SU(n))=\mathbb{Z}\ , \qquad n\geq 2
    \end{equation}
    \begin{equation}
        \pi_4(SU(n))=\begin{cases}
            \mathbb{Z}\ , \quad & n=2\\
            0\ , \quad & n\geq 3
        \end{cases}
    \end{equation}
    \begin{equation}
        \pi_i(SU(n))=\begin{cases}
            0 \, \quad & i\leq 2n-1\  \text{odd} \\
            \mathbb{Z} \, \quad & i\leq 2n-1\  \text{even}
        \end{cases}
    \end{equation}
    \begin{equation}
        U(n)\simeq \frac{SU(n)\times U(1)}{\mathbb{Z}_n}
\end{equation}
\end{subequations}

\subsection{Stiefel and Grassmann manifolds}
\label{app:stiefel}
A \emph{Stiefel manifold} is the coset space
\begin{equation}
    V_{n+m,m}=\frac{U(n+m)}{U(m)}
\end{equation}
To parametrize this space, we decompose a matrix $g\in U(n+m)$ as
\begin{equation}
    \begin{pmatrix}
       X & Y 
    \end{pmatrix}
\end{equation}
where $X$ is a $(n+m)\times n$ matrix and $X$ is a $(n+m)\times m$ matrix. The unitarity conditions $gg^{\dagger}=g^{\dagger}g=1$ imply that $X$ and $Y$ satisfy
\begin{align}
    & X^{\dagger}X=1 \qquad && X^{\dagger}Y=Y^{\dagger}X=0 \nonumber \\
    & Y^{\dagger}Y=1 \qquad && XX^{\dagger}+YY^{\dagger}=1
\end{align}
$X$ and $Y$ can be interpreted as $n$- and $m$-plets of orthonormal vectors in $\mathbb{C}^{n+m}$. At this point $V_{n+m-m}$ can be defined by the equivalence relation
\begin{equation}
    \begin{pmatrix}
       X & Y 
    \end{pmatrix}\sim
    \begin{pmatrix}
       X & Y 
    \end{pmatrix}\begin{pmatrix}
        1 & 0 \\
        0 & u
    \end{pmatrix}=\begin{pmatrix}
       X & Yu 
    \end{pmatrix}\ , \qquad u\in U(m)
\end{equation}
It is known that Stiefel manifolds satisfy
\begin{equation}
\label{eq:stiefelpi}
    \pi_i(V_{n+m,m})=0\ , \quad i\leq 2m
\end{equation}
One can define also the \emph{Grassmann manifolds} as
\begin{equation}
    G_{n+m,m}=\frac{U(n+m)}{U(n)\times U(m)}
\end{equation}
which by \eqref{eq:fibration} define the fibrations
\begin{equation}
\label{eq:grfib}
\begin{gathered}
    U(n) \longrightarrow V_{n+m,m} \longrightarrow G_{n+m,m} \\
    U(m) \longrightarrow V_{n+m,n} \longrightarrow G_{n+m,m}
\end{gathered}
\end{equation}
Of course, in our notation $G_{N,k}=G_{N,N-k}$. In analogy to the Stiefel manifolds, Grassmann manifolds are defined by the equivalence relation
\begin{equation}
    \begin{pmatrix}
       X & Y 
    \end{pmatrix}\sim
    \begin{pmatrix}
       X & Y 
    \end{pmatrix}\begin{pmatrix}
        u & 0 \\
        0 & v
    \end{pmatrix}=\begin{pmatrix}
       Xu & Yv 
    \end{pmatrix}\ , \qquad u\in U(n),\ v\in U(m)
\end{equation}
The application to the long exact sequence \eqref{eq:les} to the fibration \eqref{eq:grfib} leads to the relation
\begin{equation}
\label{eq:grassmannpi}
    \pi_i(G_{n+m,m})=\pi_{i-1}(U(k))\ , \qquad i\leq 2k\ ,\ k=\mathrm{max}(m,n)
\end{equation}

\section{Domain walls at large $N$}
\label{app:dw}
The effective action for the canonically normalized anomalous pseudo-NGB $\varphi_{c}$ in the large-$N$ limit, in the absence of background fields and $\theta$-parameter and in Minkowski space can be derived from Eq. \eqref{eq:sphi}
\begin{align}
    S_{\varphi}^{M}[\varphi_c]=\int d^4x\left[\frac{1}{2}\partial_{\mu}\varphi_c\partial^{\mu}\varphi_c-\frac{1}{2}m_{\varphi}^2\min\limits_{n\in\mathbb{Z}}\left(\varphi_c+2\pi \sqrt{Z}n/\ell\right)^2\right]
\end{align}
Let us consider the simplest domain wall, satisfying the boundary conditions
\begin{align}
   &\lim\limits_{z\to-\infty}\varphi_c(z)=0 \nonumber \\
   &  \lim\limits_{z\to+\infty}\varphi_c(z)=\frac{2\pi\sqrt{Z}}{\ell}
\end{align}
Parity and translational symmetry allow us to impose the further condition
\begin{align}
    \varphi_c(0)=\frac{\pi\sqrt{Z}}{\ell}
\end{align}
The equations of motion for this field configurations are
\begin{widetext}
\begin{align}
    &\partial_z^2\varphi_c(z)=m_{\varphi}^2\left(\varphi_c(z)+2\pi\sqrt{Z}n/\ell\right) &&\textstyle \text{for}\quad \frac{2\pi\sqrt{Z}}{\ell}\left(n-\frac{1}{2}\right)\leq \varphi_c(z)\leq \frac{2\pi\sqrt{Z}}{\ell}\left(n+\frac{1}{2}\right)
\end{align}
\end{widetext}
The only solution satisfying all the boundary conditions is
\begin{align}
    \varphi_c(z)=\begin{cases}
        \varphi_c^{(-)}(z)=\frac{\pi\sqrt{Z}}{\ell}e^{m_{\varphi}z} & z<0 \\
        \varphi_c^{(+)}(z)=\frac{\pi\sqrt{Z}}{\ell}\left(2-e^{-m_{\varphi}z}\right) & z>0
    \end{cases}
\end{align}
The energy per unit area i.e. the tension of the domain wall is given by
{\small
\begin{align}
    \sigma=&\int_{-\infty}^{0}dz\left[\frac{1}{2}(\partial_{z}\varphi_c^{(-)}(z))^2+\frac{1}{2}m_{\varphi}^2(\varphi_c^{(-)}(z))^2\right] \nonumber \\
    +&\int_{0}^{+\infty}dz\left[\frac{1}{2}(\partial_{z}\varphi_c^{(+)}(z))^2+\frac{1}{2}m_{\varphi}^2(\varphi_c^{(+)}(z))^2\right] = \frac{\pi^2 Z m_{\varphi}}{\ell^2}
\end{align}}

\section*{Acknowledgments} 

We thank M. Bochicchio, K. Konishi,  and F. Scardino  for useful discussions.  This work is supported by the INFN special 
research initiative grants, "GAST" (Gauge and String Theories) and "GAGRA" (Gauge and Gravity).

\bibliography{references}

\begin{thebibliography}{67}%
\makeatletter
\providecommand \@ifxundefined [1]{%
 \@ifx{#1\undefined}
}%
\providecommand \@ifnum [1]{%
 \ifnum #1\expandafter \@firstoftwo
 \else \expandafter \@secondoftwo
 \fi
}%
\providecommand \@ifx [1]{%
 \ifx #1\expandafter \@firstoftwo
 \else \expandafter \@secondoftwo
 \fi
}%
\providecommand \natexlab [1]{#1}%
\providecommand \enquote  [1]{``#1''}%
\providecommand \bibnamefont  [1]{#1}%
\providecommand \bibfnamefont [1]{#1}%
\providecommand \citenamefont [1]{#1}%
\providecommand \href@noop [0]{\@secondoftwo}%
\providecommand \href [0]{\begingroup \@sanitize@url \@href}%
\providecommand \@href[1]{\@@startlink{#1}\@@href}%
\providecommand \@@href[1]{\endgroup#1\@@endlink}%
\providecommand \@sanitize@url [0]{\catcode `\\12\catcode `\$12\catcode `\&12\catcode `\#12\catcode `\^12\catcode `\_12\catcode `\%12\relax}%
\providecommand \@@startlink[1]{}%
\providecommand \@@endlink[0]{}%
\providecommand \url  [0]{\begingroup\@sanitize@url \@url }%
\providecommand \@url [1]{\endgroup\@href {#1}{\urlprefix }}%
\providecommand \urlprefix  [0]{URL }%
\providecommand \Eprint [0]{\href }%
\providecommand \doibase [0]{http://dx.doi.org/}%
\providecommand \selectlanguage [0]{\@gobble}%
\providecommand \bibinfo  [0]{\@secondoftwo}%
\providecommand \bibfield  [0]{\@secondoftwo}%
\providecommand \translation [1]{[#1]}%
\providecommand \BibitemOpen [0]{}%
\providecommand \bibitemStop [0]{}%
\providecommand \bibitemNoStop [0]{.\EOS\space}%
\providecommand \EOS [0]{\spacefactor3000\relax}%
\providecommand \BibitemShut  [1]{\csname bibitem#1\endcsname}%
\let\auto@bib@innerbib\@empty
\bibitem [{\citenamefont {Bolognesi}\ \emph {et~al.}(2020)\citenamefont {Bolognesi}, \citenamefont {Konishi},\ and\ \citenamefont {Luzio}}]{Bolognesi:2020mpe}%
  \BibitemOpen
  \bibfield  {author} {\bibinfo {author} {\bibfnamefont {S.}~\bibnamefont {Bolognesi}}, \bibinfo {author} {\bibfnamefont {K.}~\bibnamefont {Konishi}}, \ and\ \bibinfo {author} {\bibfnamefont {A.}~\bibnamefont {Luzio}},\ }\href {\doibase 10.1007/JHEP09(2020)001} {\bibfield  {journal} {\bibinfo  {journal} {JHEP}\ }\textbf {\bibinfo {volume} {09}},\ \bibinfo {pages} {001} (\bibinfo {year} {2020})},\ \Eprint {http://arxiv.org/abs/2004.06639} {arXiv:2004.06639 [hep-th]} \BibitemShut {NoStop}%
\bibitem [{\citenamefont {Bolognesi}\ \emph {et~al.}(2021{\natexlab{a}})\citenamefont {Bolognesi}, \citenamefont {Konishi},\ and\ \citenamefont {Luzio}}]{Bolognesi:2021yni}%
  \BibitemOpen
  \bibfield  {author} {\bibinfo {author} {\bibfnamefont {S.}~\bibnamefont {Bolognesi}}, \bibinfo {author} {\bibfnamefont {K.}~\bibnamefont {Konishi}}, \ and\ \bibinfo {author} {\bibfnamefont {A.}~\bibnamefont {Luzio}},\ }\href {\doibase 10.1103/PhysRevD.103.094016} {\bibfield  {journal} {\bibinfo  {journal} {Phys. Rev. D}\ }\textbf {\bibinfo {volume} {103}},\ \bibinfo {pages} {094016} (\bibinfo {year} {2021}{\natexlab{a}})},\ \Eprint {http://arxiv.org/abs/2101.02601} {arXiv:2101.02601 [hep-th]} \BibitemShut {NoStop}%
\bibitem [{Note1()}]{Note1}%
  \BibitemOpen
  \bibinfo {note} {Being the quark bilinear charged both under color and flavor symmetries, its condensation breaks the two groups to a diagonal subgroup. Such a mechanism is known as Color-flavor locking.}\BibitemShut {Stop}%
\bibitem [{\citenamefont {Bolognesi}\ \emph {et~al.}(2024)\citenamefont {Bolognesi}, \citenamefont {Konishi}, \citenamefont {Luzio},\ and\ \citenamefont {Orso}}]{Bolognesi:2024bnm}%
  \BibitemOpen
  \bibfield  {author} {\bibinfo {author} {\bibfnamefont {S.}~\bibnamefont {Bolognesi}}, \bibinfo {author} {\bibfnamefont {K.}~\bibnamefont {Konishi}}, \bibinfo {author} {\bibfnamefont {A.}~\bibnamefont {Luzio}}, \ and\ \bibinfo {author} {\bibfnamefont {M.}~\bibnamefont {Orso}},\ }\href {\doibase 10.1103/PhysRevD.110.114037} {\bibfield  {journal} {\bibinfo  {journal} {Phys. Rev. D}\ }\textbf {\bibinfo {volume} {110}},\ \bibinfo {pages} {114037} (\bibinfo {year} {2024})},\ \Eprint {http://arxiv.org/abs/2410.01315} {arXiv:2410.01315 [hep-th]} \BibitemShut {NoStop}%
\bibitem [{Note2()}]{Note2}%
  \BibitemOpen
  \bibinfo {note} {The hadronic spectrum in chiral gauge theories has recently been considered in \cite {Kristensen:2024vmi,Girmohanta:2019cth}.}\BibitemShut {Stop}%
\bibitem [{\citenamefont {Skyrme}(1962)}]{Skyrme:1962vh}%
  \BibitemOpen
  \bibfield  {author} {\bibinfo {author} {\bibfnamefont {T.~H.~R.}\ \bibnamefont {Skyrme}},\ }\href {\doibase 10.1016/0029-5582(62)90775-7} {\bibfield  {journal} {\bibinfo  {journal} {Nucl. Phys.}\ }\textbf {\bibinfo {volume} {31}},\ \bibinfo {pages} {556} (\bibinfo {year} {1962})}\BibitemShut {NoStop}%
\bibitem [{\citenamefont {Witten}(1979)}]{Witten:1979kh}%
  \BibitemOpen
  \bibfield  {author} {\bibinfo {author} {\bibfnamefont {E.}~\bibnamefont {Witten}},\ }\href {\doibase 10.1016/0550-3213(79)90232-3} {\bibfield  {journal} {\bibinfo  {journal} {Nucl. Phys. B}\ }\textbf {\bibinfo {volume} {160}},\ \bibinfo {pages} {57} (\bibinfo {year} {1979})}\BibitemShut {NoStop}%
\bibitem [{\citenamefont {Adkins}\ \emph {et~al.}(1983)\citenamefont {Adkins}, \citenamefont {Nappi},\ and\ \citenamefont {Witten}}]{Adkins:1983ya}%
  \BibitemOpen
  \bibfield  {author} {\bibinfo {author} {\bibfnamefont {G.~S.}\ \bibnamefont {Adkins}}, \bibinfo {author} {\bibfnamefont {C.~R.}\ \bibnamefont {Nappi}}, \ and\ \bibinfo {author} {\bibfnamefont {E.}~\bibnamefont {Witten}},\ }\href {\doibase 10.1016/0550-3213(83)90559-X} {\bibfield  {journal} {\bibinfo  {journal} {Nucl. Phys. B}\ }\textbf {\bibinfo {volume} {228}},\ \bibinfo {pages} {552} (\bibinfo {year} {1983})}\BibitemShut {NoStop}%
\bibitem [{\citenamefont {Bolognesi}(2007)}]{Bolognesi:2006ws}%
  \BibitemOpen
  \bibfield  {author} {\bibinfo {author} {\bibfnamefont {S.}~\bibnamefont {Bolognesi}},\ }\href {\doibase 10.1103/PhysRevD.75.065030} {\bibfield  {journal} {\bibinfo  {journal} {Phys. Rev. D}\ }\textbf {\bibinfo {volume} {75}},\ \bibinfo {pages} {065030} (\bibinfo {year} {2007})},\ \Eprint {http://arxiv.org/abs/hep-th/0605065} {arXiv:hep-th/0605065} \BibitemShut {NoStop}%
\bibitem [{\citenamefont {Cherman}\ and\ \citenamefont {Cohen}(2006)}]{Cherman:2006iy}%
  \BibitemOpen
  \bibfield  {author} {\bibinfo {author} {\bibfnamefont {A.}~\bibnamefont {Cherman}}\ and\ \bibinfo {author} {\bibfnamefont {T.~D.}\ \bibnamefont {Cohen}},\ }\href {\doibase 10.1088/1126-6708/2006/12/035} {\bibfield  {journal} {\bibinfo  {journal} {JHEP}\ }\textbf {\bibinfo {volume} {12}},\ \bibinfo {pages} {035} (\bibinfo {year} {2006})},\ \Eprint {http://arxiv.org/abs/hep-th/0607028} {arXiv:hep-th/0607028} \BibitemShut {NoStop}%
\bibitem [{\citenamefont {Komargodski}(2018)}]{Komargodski:2018odf}%
  \BibitemOpen
  \bibfield  {author} {\bibinfo {author} {\bibfnamefont {Z.}~\bibnamefont {Komargodski}},\ }\href@noop {} {\  (\bibinfo {year} {2018})},\ \Eprint {http://arxiv.org/abs/1812.09253} {arXiv:1812.09253 [hep-th]} \BibitemShut {NoStop}%
\bibitem [{\citenamefont {Bigazzi}\ \emph {et~al.}(2023)\citenamefont {Bigazzi}, \citenamefont {Cotrone},\ and\ \citenamefont {Olzi}}]{Bigazzi:2022luo}%
  \BibitemOpen
  \bibfield  {author} {\bibinfo {author} {\bibfnamefont {F.}~\bibnamefont {Bigazzi}}, \bibinfo {author} {\bibfnamefont {A.~L.}\ \bibnamefont {Cotrone}}, \ and\ \bibinfo {author} {\bibfnamefont {A.}~\bibnamefont {Olzi}},\ }\href {\doibase 10.1007/JHEP02(2023)194} {\bibfield  {journal} {\bibinfo  {journal} {JHEP}\ }\textbf {\bibinfo {volume} {02}},\ \bibinfo {pages} {194} (\bibinfo {year} {2023})},\ \Eprint {http://arxiv.org/abs/2211.05147} {arXiv:2211.05147 [hep-th]} \BibitemShut {NoStop}%
\bibitem [{\citenamefont {Karasik}(2022)}]{Karasik:2022tmd}%
  \BibitemOpen
  \bibfield  {author} {\bibinfo {author} {\bibfnamefont {A.}~\bibnamefont {Karasik}},\ }\href {\doibase 10.3390/sym14112347} {\bibfield  {journal} {\bibinfo  {journal} {Symmetry}\ }\textbf {\bibinfo {volume} {14}},\ \bibinfo {pages} {2347} (\bibinfo {year} {2022})}\BibitemShut {NoStop}%
\bibitem [{\citenamefont {Lin}\ and\ \citenamefont {Ma}(2024)}]{Lin:2023qya}%
  \BibitemOpen
  \bibfield  {author} {\bibinfo {author} {\bibfnamefont {F.}~\bibnamefont {Lin}}\ and\ \bibinfo {author} {\bibfnamefont {Y.-L.}\ \bibnamefont {Ma}},\ }\href {\doibase 10.1007/JHEP05(2024)270} {\bibfield  {journal} {\bibinfo  {journal} {JHEP}\ }\textbf {\bibinfo {volume} {05}},\ \bibinfo {pages} {270} (\bibinfo {year} {2024})},\ \Eprint {http://arxiv.org/abs/2310.16438} {arXiv:2310.16438 [hep-th]} \BibitemShut {NoStop}%
\bibitem [{Note3()}]{Note3}%
  \BibitemOpen
  \bibinfo {note} {Such a TQFT provides degrees of freedom to the edge of the disk.}\BibitemShut {Stop}%
\bibitem [{\citenamefont {C\'ordova}\ \emph {et~al.}(2020{\natexlab{a}})\citenamefont {C\'ordova}, \citenamefont {Freed}, \citenamefont {Lam},\ and\ \citenamefont {Seiberg}}]{Cordova:2019jnf}%
  \BibitemOpen
  \bibfield  {author} {\bibinfo {author} {\bibfnamefont {C.}~\bibnamefont {C\'ordova}}, \bibinfo {author} {\bibfnamefont {D.~S.}\ \bibnamefont {Freed}}, \bibinfo {author} {\bibfnamefont {H.~T.}\ \bibnamefont {Lam}}, \ and\ \bibinfo {author} {\bibfnamefont {N.}~\bibnamefont {Seiberg}},\ }\href {\doibase 10.21468/SciPostPhys.8.1.001} {\bibfield  {journal} {\bibinfo  {journal} {SciPost Phys.}\ }\textbf {\bibinfo {volume} {8}},\ \bibinfo {pages} {001} (\bibinfo {year} {2020}{\natexlab{a}})},\ \Eprint {http://arxiv.org/abs/1905.09315} {arXiv:1905.09315 [hep-th]} \BibitemShut {NoStop}%
\bibitem [{\citenamefont {C\'ordova}\ \emph {et~al.}(2020{\natexlab{b}})\citenamefont {C\'ordova}, \citenamefont {Freed}, \citenamefont {Lam},\ and\ \citenamefont {Seiberg}}]{Cordova:2019uob}%
  \BibitemOpen
  \bibfield  {author} {\bibinfo {author} {\bibfnamefont {C.}~\bibnamefont {C\'ordova}}, \bibinfo {author} {\bibfnamefont {D.~S.}\ \bibnamefont {Freed}}, \bibinfo {author} {\bibfnamefont {H.~T.}\ \bibnamefont {Lam}}, \ and\ \bibinfo {author} {\bibfnamefont {N.}~\bibnamefont {Seiberg}},\ }\href {\doibase 10.21468/SciPostPhys.8.1.002} {\bibfield  {journal} {\bibinfo  {journal} {SciPost Phys.}\ }\textbf {\bibinfo {volume} {8}},\ \bibinfo {pages} {002} (\bibinfo {year} {2020}{\natexlab{b}})},\ \Eprint {http://arxiv.org/abs/1905.13361} {arXiv:1905.13361 [hep-th]} \BibitemShut {NoStop}%
\bibitem [{\citenamefont {Anber}\ and\ \citenamefont {Poppitz}(2019)}]{Anber:2019nze}%
  \BibitemOpen
  \bibfield  {author} {\bibinfo {author} {\bibfnamefont {M.~M.}\ \bibnamefont {Anber}}\ and\ \bibinfo {author} {\bibfnamefont {E.}~\bibnamefont {Poppitz}},\ }\href {\doibase 10.1007/JHEP11(2019)063} {\bibfield  {journal} {\bibinfo  {journal} {JHEP}\ }\textbf {\bibinfo {volume} {11}},\ \bibinfo {pages} {063} (\bibinfo {year} {2019})},\ \Eprint {http://arxiv.org/abs/1909.09027} {arXiv:1909.09027 [hep-th]} \BibitemShut {NoStop}%
\bibitem [{\citenamefont {Kitano}\ and\ \citenamefont {Matsudo}(2021)}]{Kitano:2020evx}%
  \BibitemOpen
  \bibfield  {author} {\bibinfo {author} {\bibfnamefont {R.}~\bibnamefont {Kitano}}\ and\ \bibinfo {author} {\bibfnamefont {R.}~\bibnamefont {Matsudo}},\ }\href {\doibase 10.1007/JHEP03(2021)023} {\bibfield  {journal} {\bibinfo  {journal} {JHEP}\ }\textbf {\bibinfo {volume} {03}},\ \bibinfo {pages} {023} (\bibinfo {year} {2021})},\ \Eprint {http://arxiv.org/abs/2011.14637} {arXiv:2011.14637 [hep-th]} \BibitemShut {NoStop}%
\bibitem [{\citenamefont {Frohlich}\ \emph {et~al.}(1980)\citenamefont {Frohlich}, \citenamefont {Morchio},\ and\ \citenamefont {Strocchi}}]{Frohlich:1980gj}%
  \BibitemOpen
  \bibfield  {author} {\bibinfo {author} {\bibfnamefont {J.}~\bibnamefont {Frohlich}}, \bibinfo {author} {\bibfnamefont {G.}~\bibnamefont {Morchio}}, \ and\ \bibinfo {author} {\bibfnamefont {F.}~\bibnamefont {Strocchi}},\ }\href {\doibase 10.1016/0370-2693(80)90594-8} {\bibfield  {journal} {\bibinfo  {journal} {Phys. Lett. B}\ }\textbf {\bibinfo {volume} {97}},\ \bibinfo {pages} {249} (\bibinfo {year} {1980})}\BibitemShut {NoStop}%
\bibitem [{\citenamefont {Frohlich}\ \emph {et~al.}(1981)\citenamefont {Frohlich}, \citenamefont {Morchio},\ and\ \citenamefont {Strocchi}}]{Frohlich:1981yi}%
  \BibitemOpen
  \bibfield  {author} {\bibinfo {author} {\bibfnamefont {J.}~\bibnamefont {Frohlich}}, \bibinfo {author} {\bibfnamefont {G.}~\bibnamefont {Morchio}}, \ and\ \bibinfo {author} {\bibfnamefont {F.}~\bibnamefont {Strocchi}},\ }\href {\doibase 10.1016/0550-3213(81)90448-X} {\bibfield  {journal} {\bibinfo  {journal} {Nucl. Phys. B}\ }\textbf {\bibinfo {volume} {190}},\ \bibinfo {pages} {553} (\bibinfo {year} {1981})}\BibitemShut {NoStop}%
\bibitem [{\citenamefont {Elitzur}(1975)}]{Elitzur:1975im}%
  \BibitemOpen
  \bibfield  {author} {\bibinfo {author} {\bibfnamefont {S.}~\bibnamefont {Elitzur}},\ }\href {\doibase 10.1103/PhysRevD.12.3978} {\bibfield  {journal} {\bibinfo  {journal} {Phys. Rev. D}\ }\textbf {\bibinfo {volume} {12}},\ \bibinfo {pages} {3978} (\bibinfo {year} {1975})}\BibitemShut {NoStop}%
\bibitem [{\citenamefont {Coleman}\ \emph {et~al.}(1969)\citenamefont {Coleman}, \citenamefont {Wess},\ and\ \citenamefont {Zumino}}]{Coleman:1969sm}%
  \BibitemOpen
  \bibfield  {author} {\bibinfo {author} {\bibfnamefont {S.~R.}\ \bibnamefont {Coleman}}, \bibinfo {author} {\bibfnamefont {J.}~\bibnamefont {Wess}}, \ and\ \bibinfo {author} {\bibfnamefont {B.}~\bibnamefont {Zumino}},\ }\href {\doibase 10.1103/PhysRev.177.2239} {\bibfield  {journal} {\bibinfo  {journal} {Phys. Rev.}\ }\textbf {\bibinfo {volume} {177}},\ \bibinfo {pages} {2239} (\bibinfo {year} {1969})}\BibitemShut {NoStop}%
\bibitem [{\citenamefont {Callan}\ \emph {et~al.}(1969)\citenamefont {Callan}, \citenamefont {Coleman}, \citenamefont {Wess},\ and\ \citenamefont {Zumino}}]{Callan:1969sn}%
  \BibitemOpen
  \bibfield  {author} {\bibinfo {author} {\bibfnamefont {C.~G.}\ \bibnamefont {Callan}, \bibfnamefont {Jr.}}, \bibinfo {author} {\bibfnamefont {S.~R.}\ \bibnamefont {Coleman}}, \bibinfo {author} {\bibfnamefont {J.}~\bibnamefont {Wess}}, \ and\ \bibinfo {author} {\bibfnamefont {B.}~\bibnamefont {Zumino}},\ }\href {\doibase 10.1103/PhysRev.177.2247} {\bibfield  {journal} {\bibinfo  {journal} {Phys. Rev.}\ }\textbf {\bibinfo {volume} {177}},\ \bibinfo {pages} {2247} (\bibinfo {year} {1969})}\BibitemShut {NoStop}%
\bibitem [{\citenamefont {Weinberg}(1976)}]{Weinberg:1975gm}%
  \BibitemOpen
  \bibfield  {author} {\bibinfo {author} {\bibfnamefont {S.}~\bibnamefont {Weinberg}},\ }\href {\doibase 10.1103/PhysRevD.19.1277} {\bibfield  {journal} {\bibinfo  {journal} {Phys. Rev. D}\ }\textbf {\bibinfo {volume} {13}},\ \bibinfo {pages} {974} (\bibinfo {year} {1976})},\ \bibinfo {note} {[Addendum: Phys.Rev.D 19, 1277--1280 (1979)]}\BibitemShut {NoStop}%
\bibitem [{\citenamefont {Leutwyler}(1994)}]{Leutwyler:1993iq}%
  \BibitemOpen
  \bibfield  {author} {\bibinfo {author} {\bibfnamefont {H.}~\bibnamefont {Leutwyler}},\ }\href {\doibase 10.1006/aphy.1994.1094} {\bibfield  {journal} {\bibinfo  {journal} {Annals Phys.}\ }\textbf {\bibinfo {volume} {235}},\ \bibinfo {pages} {165} (\bibinfo {year} {1994})},\ \Eprint {http://arxiv.org/abs/hep-ph/9311274} {arXiv:hep-ph/9311274} \BibitemShut {NoStop}%
\bibitem [{Note4()}]{Note4}%
  \BibitemOpen
  \bibinfo {note} {A familiar setting where there is continuity between the CFL regime and a strongly coupled regime is finite density QCD with three flavors. In that case, depending on the chemical potential, it is more convenient to use the CFL description \cite {Alford:1998mk} or a chiral Lagrangian description. However, the two regimes are believed to be connected, and the qualitative features match \cite {Schafer:1998ef}.}\BibitemShut {Stop}%
\bibitem [{Note5()}]{Note5}%
  \BibitemOpen
  \bibinfo {note} {In the notation of subsection \ref {app:1formquant}, $\protect \widetilde {A}$ corresponds to the $\protect \widetilde {A}_i^G$, $\protect \widetilde {A}_i^{U(1)}$ collectively and $B$ corresponds to the $B_i$ and $b$ collectively.}\BibitemShut {Stop}%
\bibitem [{\citenamefont {Yonekura}(2021)}]{Yonekura:2020upo}%
  \BibitemOpen
  \bibfield  {author} {\bibinfo {author} {\bibfnamefont {K.}~\bibnamefont {Yonekura}},\ }\href {\doibase 10.1007/JHEP03(2021)057} {\bibfield  {journal} {\bibinfo  {journal} {JHEP}\ }\textbf {\bibinfo {volume} {03}},\ \bibinfo {pages} {057} (\bibinfo {year} {2021})},\ \Eprint {http://arxiv.org/abs/2009.04692} {arXiv:2009.04692 [hep-th]} \BibitemShut {NoStop}%
\bibitem [{\citenamefont {Brower}\ \emph {et~al.}(2003)\citenamefont {Brower}, \citenamefont {Chandrasekharan}, \citenamefont {Negele},\ and\ \citenamefont {Wiese}}]{Brower:2003yx}%
  \BibitemOpen
  \bibfield  {author} {\bibinfo {author} {\bibfnamefont {R.}~\bibnamefont {Brower}}, \bibinfo {author} {\bibfnamefont {S.}~\bibnamefont {Chandrasekharan}}, \bibinfo {author} {\bibfnamefont {J.~W.}\ \bibnamefont {Negele}}, \ and\ \bibinfo {author} {\bibfnamefont {U.~J.}\ \bibnamefont {Wiese}},\ }\href {\doibase 10.1016/S0370-2693(03)00369-1} {\bibfield  {journal} {\bibinfo  {journal} {Phys. Lett. B}\ }\textbf {\bibinfo {volume} {560}},\ \bibinfo {pages} {64} (\bibinfo {year} {2003})},\ \Eprint {http://arxiv.org/abs/hep-lat/0302005} {arXiv:hep-lat/0302005} \BibitemShut {NoStop}%
\bibitem [{\citenamefont {Aoki}\ \emph {et~al.}(2007)\citenamefont {Aoki}, \citenamefont {Fukaya}, \citenamefont {Hashimoto},\ and\ \citenamefont {Onogi}}]{Aoki:2007ka}%
  \BibitemOpen
  \bibfield  {author} {\bibinfo {author} {\bibfnamefont {S.}~\bibnamefont {Aoki}}, \bibinfo {author} {\bibfnamefont {H.}~\bibnamefont {Fukaya}}, \bibinfo {author} {\bibfnamefont {S.}~\bibnamefont {Hashimoto}}, \ and\ \bibinfo {author} {\bibfnamefont {T.}~\bibnamefont {Onogi}},\ }\href {\doibase 10.1103/PhysRevD.76.054508} {\bibfield  {journal} {\bibinfo  {journal} {Phys. Rev. D}\ }\textbf {\bibinfo {volume} {76}},\ \bibinfo {pages} {054508} (\bibinfo {year} {2007})},\ \Eprint {http://arxiv.org/abs/0707.0396} {arXiv:0707.0396 [hep-lat]} \BibitemShut {NoStop}%
\bibitem [{\citenamefont {Bolognesi}\ \emph {et~al.}(2023)\citenamefont {Bolognesi}, \citenamefont {Konishi},\ and\ \citenamefont {Luzio}}]{Bolognesi:2023xxv}%
  \BibitemOpen
  \bibfield  {author} {\bibinfo {author} {\bibfnamefont {S.}~\bibnamefont {Bolognesi}}, \bibinfo {author} {\bibfnamefont {K.}~\bibnamefont {Konishi}}, \ and\ \bibinfo {author} {\bibfnamefont {A.}~\bibnamefont {Luzio}},\ }\href {\doibase 10.1007/JHEP08(2023)125} {\bibfield  {journal} {\bibinfo  {journal} {JHEP}\ }\textbf {\bibinfo {volume} {08}},\ \bibinfo {pages} {125} (\bibinfo {year} {2023})},\ \Eprint {http://arxiv.org/abs/2307.03822} {arXiv:2307.03822 [hep-th]} \BibitemShut {NoStop}%
\bibitem [{Note6()}]{Note6}%
  \BibitemOpen
  \bibinfo {note} {To prove this point, it is necessary to consider the entire global structure of the symmetry group. In other words, to correctly background gauge the $\protect \mathbb {Z}^{(1)}_N$ symmetry.}\BibitemShut {Stop}%
\bibitem [{Note7()}]{Note7}%
  \BibitemOpen
  \bibinfo {note} {This is one of the simplest realizations of the NAM paradigm \cite {Bolognesi:2024bnm}.}\BibitemShut {Stop}%
\bibitem [{Note8()}]{Note8}%
  \BibitemOpen
  \bibinfo {note} {The two-dimensional $\epsilon $-tensor $\epsilon ^{\alpha \beta }$ acting on spinor indices is defined through the conventions $\epsilon ^{12}=-\epsilon ^{21}=-\epsilon _{12}=\epsilon _{21}=1$.}\BibitemShut {Stop}%
\bibitem [{\citenamefont {Weinberg}\ and\ \citenamefont {Witten}(1980)}]{Weinberg:1980kq}%
  \BibitemOpen
  \bibfield  {author} {\bibinfo {author} {\bibfnamefont {S.}~\bibnamefont {Weinberg}}\ and\ \bibinfo {author} {\bibfnamefont {E.}~\bibnamefont {Witten}},\ }\href {\doibase 10.1016/0370-2693(80)90212-9} {\bibfield  {journal} {\bibinfo  {journal} {Phys. Lett. B}\ }\textbf {\bibinfo {volume} {96}},\ \bibinfo {pages} {59} (\bibinfo {year} {1980})}\BibitemShut {NoStop}%
\bibitem [{Note9()}]{Note9}%
  \BibitemOpen
  \bibinfo {note} {Which cannot be weakly coupled, as shown in \cite {Li:2025tvu} using exact functional RG methods.}\BibitemShut {Stop}%
\bibitem [{Note10()}]{Note10}%
  \BibitemOpen
  \bibinfo {note} {It is cubersome to write the global form of the faithful symmetry group for such models \cite {Bolognesi:2021yni}.}\BibitemShut {Stop}%
\bibitem [{\citenamefont {Bolognesi}\ \emph {et~al.}(2021{\natexlab{b}})\citenamefont {Bolognesi}, \citenamefont {Konishi},\ and\ \citenamefont {Luzio}}]{Bolognesi:2021hmg}%
  \BibitemOpen
  \bibfield  {author} {\bibinfo {author} {\bibfnamefont {S.}~\bibnamefont {Bolognesi}}, \bibinfo {author} {\bibfnamefont {K.}~\bibnamefont {Konishi}}, \ and\ \bibinfo {author} {\bibfnamefont {A.}~\bibnamefont {Luzio}},\ }\href {\doibase 10.1007/JHEP08(2021)028} {\bibfield  {journal} {\bibinfo  {journal} {JHEP}\ }\textbf {\bibinfo {volume} {08}},\ \bibinfo {pages} {028} (\bibinfo {year} {2021}{\natexlab{b}})},\ \Eprint {http://arxiv.org/abs/2105.03921} {arXiv:2105.03921 [hep-th]} \BibitemShut {NoStop}%
\bibitem [{\citenamefont {Vafa}\ and\ \citenamefont {Witten}(1984)}]{Vafa:1984xg}%
  \BibitemOpen
  \bibfield  {author} {\bibinfo {author} {\bibfnamefont {C.}~\bibnamefont {Vafa}}\ and\ \bibinfo {author} {\bibfnamefont {E.}~\bibnamefont {Witten}},\ }\href {\doibase 10.1103/PhysRevLett.53.535} {\bibfield  {journal} {\bibinfo  {journal} {Phys. Rev. Lett.}\ }\textbf {\bibinfo {volume} {53}},\ \bibinfo {pages} {535} (\bibinfo {year} {1984})}\BibitemShut {NoStop}%
\bibitem [{\citenamefont {Armoni}\ \emph {et~al.}(2006)\citenamefont {Armoni}, \citenamefont {Shore},\ and\ \citenamefont {Veneziano}}]{Armoni:2005wt}%
  \BibitemOpen
  \bibfield  {author} {\bibinfo {author} {\bibfnamefont {A.}~\bibnamefont {Armoni}}, \bibinfo {author} {\bibfnamefont {G.}~\bibnamefont {Shore}}, \ and\ \bibinfo {author} {\bibfnamefont {G.}~\bibnamefont {Veneziano}},\ }\href {\doibase 10.1016/j.nuclphysb.2006.01.044} {\bibfield  {journal} {\bibinfo  {journal} {Nucl. Phys. B}\ }\textbf {\bibinfo {volume} {740}},\ \bibinfo {pages} {23} (\bibinfo {year} {2006})},\ \Eprint {http://arxiv.org/abs/hep-ph/0511143} {arXiv:hep-ph/0511143} \BibitemShut {NoStop}%
\bibitem [{Note11()}]{Note11}%
  \BibitemOpen
  \bibinfo {note} {With $N=3$ and $N_{\chi }=1$, the main purpose of \cite {Armoni:2005wt,Armoni:2014ywa} was to have a large $N$ planar equivalence of multi-flavor QCD with supersymmetric YM.}\BibitemShut {Stop}%
\bibitem [{\citenamefont {Skyrme}(1961)}]{Skyrme:1961vq}%
  \BibitemOpen
  \bibfield  {author} {\bibinfo {author} {\bibfnamefont {T.~H.~R.}\ \bibnamefont {Skyrme}},\ }\href {\doibase 10.1098/rspa.1961.0018} {\bibfield  {journal} {\bibinfo  {journal} {Proc. Roy. Soc. Lond. A}\ }\textbf {\bibinfo {volume} {260}},\ \bibinfo {pages} {127} (\bibinfo {year} {1961})}\BibitemShut {NoStop}%
\bibitem [{\citenamefont {Witten}(1983)}]{Witten:1983tw}%
  \BibitemOpen
  \bibfield  {author} {\bibinfo {author} {\bibfnamefont {E.}~\bibnamefont {Witten}},\ }\href {\doibase 10.1016/0550-3213(83)90063-9} {\bibfield  {journal} {\bibinfo  {journal} {Nucl. Phys. B}\ }\textbf {\bibinfo {volume} {223}},\ \bibinfo {pages} {422} (\bibinfo {year} {1983})}\BibitemShut {NoStop}%
\bibitem [{\citenamefont {Goldstone}\ and\ \citenamefont {Wilczek}(1981)}]{Goldstone:1981kk}%
  \BibitemOpen
  \bibfield  {author} {\bibinfo {author} {\bibfnamefont {J.}~\bibnamefont {Goldstone}}\ and\ \bibinfo {author} {\bibfnamefont {F.}~\bibnamefont {Wilczek}},\ }\href {\doibase 10.1103/PhysRevLett.47.986} {\bibfield  {journal} {\bibinfo  {journal} {Phys. Rev. Lett.}\ }\textbf {\bibinfo {volume} {47}},\ \bibinfo {pages} {986} (\bibinfo {year} {1981})}\BibitemShut {NoStop}%
\bibitem [{\citenamefont {Anber}\ and\ \citenamefont {Poppitz}(2020)}]{Anber:2020gig}%
  \BibitemOpen
  \bibfield  {author} {\bibinfo {author} {\bibfnamefont {M.~M.}\ \bibnamefont {Anber}}\ and\ \bibinfo {author} {\bibfnamefont {E.}~\bibnamefont {Poppitz}},\ }\href {\doibase 10.1007/JHEP04(2020)097} {\bibfield  {journal} {\bibinfo  {journal} {JHEP}\ }\textbf {\bibinfo {volume} {04}},\ \bibinfo {pages} {097} (\bibinfo {year} {2020})},\ \Eprint {http://arxiv.org/abs/2002.02037} {arXiv:2002.02037 [hep-th]} \BibitemShut {NoStop}%
\bibitem [{\citenamefont {Nakajima}\ \emph {et~al.}(2023)\citenamefont {Nakajima}, \citenamefont {Sakai},\ and\ \citenamefont {Yokokura}}]{Nakajima:2022jxg}%
  \BibitemOpen
  \bibfield  {author} {\bibinfo {author} {\bibfnamefont {T.}~\bibnamefont {Nakajima}}, \bibinfo {author} {\bibfnamefont {T.}~\bibnamefont {Sakai}}, \ and\ \bibinfo {author} {\bibfnamefont {R.}~\bibnamefont {Yokokura}},\ }\href {\doibase 10.1007/JHEP01(2023)175} {\bibfield  {journal} {\bibinfo  {journal} {JHEP}\ }\textbf {\bibinfo {volume} {01}},\ \bibinfo {pages} {175} (\bibinfo {year} {2023})},\ \Eprint {http://arxiv.org/abs/2212.12987} {arXiv:2212.12987 [hep-th]} \BibitemShut {NoStop}%
\bibitem [{Note12()}]{Note12}%
  \BibitemOpen
  \bibinfo {note} {In the notation of subsection \ref {app:1formquant}, $\protect \widetilde {A}$ corresponds to the $\protect \widetilde {A}_i^G$, $\protect \widetilde {A}_i^{U(1)}$ collectively and $B$ corresponds to the $B_i$ and $b$ collectively}\BibitemShut {NoStop}%
\bibitem [{Note13()}]{Note13}%
  \BibitemOpen
  \bibinfo {note} {This is true also in the chiral gauge theories of section \ref {sec:baryons} where the number of flavors is constrained to grow linearly with $N$ and the $N\to \infty $ limit is a Veneziano limit. The key observation is that in all these cases the $S$-matrix becomes trivial when $N\to \infty $.}\BibitemShut {Stop}%
\bibitem [{\citenamefont {Dvali}\ and\ \citenamefont {Shifman}(1997)}]{Dvali:1996xe}%
  \BibitemOpen
  \bibfield  {author} {\bibinfo {author} {\bibfnamefont {G.~R.}\ \bibnamefont {Dvali}}\ and\ \bibinfo {author} {\bibfnamefont {M.~A.}\ \bibnamefont {Shifman}},\ }\href {\doibase 10.1016/S0370-2693(97)00131-7} {\bibfield  {journal} {\bibinfo  {journal} {Phys. Lett. B}\ }\textbf {\bibinfo {volume} {396}},\ \bibinfo {pages} {64} (\bibinfo {year} {1997})},\ \bibinfo {note} {[Erratum: Phys.Lett.B 407, 452 (1997)]},\ \Eprint {http://arxiv.org/abs/hep-th/9612128} {arXiv:hep-th/9612128} \BibitemShut {NoStop}%
\bibitem [{\citenamefont {Gabadadze}\ and\ \citenamefont {Shifman}(2000)}]{Gabadadze:1999pp}%
  \BibitemOpen
  \bibfield  {author} {\bibinfo {author} {\bibfnamefont {G.}~\bibnamefont {Gabadadze}}\ and\ \bibinfo {author} {\bibfnamefont {M.~A.}\ \bibnamefont {Shifman}},\ }\href {\doibase 10.1103/PhysRevD.61.075014} {\bibfield  {journal} {\bibinfo  {journal} {Phys. Rev. D}\ }\textbf {\bibinfo {volume} {61}},\ \bibinfo {pages} {075014} (\bibinfo {year} {2000})},\ \Eprint {http://arxiv.org/abs/hep-th/9910050} {arXiv:hep-th/9910050} \BibitemShut {NoStop}%
\bibitem [{\citenamefont {Witten}(1997)}]{Witten:1997ep}%
  \BibitemOpen
  \bibfield  {author} {\bibinfo {author} {\bibfnamefont {E.}~\bibnamefont {Witten}},\ }\href {\doibase 10.1016/S0550-3213(97)00648-2} {\bibfield  {journal} {\bibinfo  {journal} {Nucl. Phys. B}\ }\textbf {\bibinfo {volume} {507}},\ \bibinfo {pages} {658} (\bibinfo {year} {1997})},\ \Eprint {http://arxiv.org/abs/hep-th/9706109} {arXiv:hep-th/9706109} \BibitemShut {NoStop}%
\bibitem [{\citenamefont {Armoni}\ and\ \citenamefont {Hollowood}(2005)}]{Armoni:2005sp}%
  \BibitemOpen
  \bibfield  {author} {\bibinfo {author} {\bibfnamefont {A.}~\bibnamefont {Armoni}}\ and\ \bibinfo {author} {\bibfnamefont {T.~J.}\ \bibnamefont {Hollowood}},\ }\href {\doibase 10.1088/1126-6708/2005/07/043} {\bibfield  {journal} {\bibinfo  {journal} {JHEP}\ }\textbf {\bibinfo {volume} {07}},\ \bibinfo {pages} {043} (\bibinfo {year} {2005})},\ \Eprint {http://arxiv.org/abs/hep-th/0505213} {arXiv:hep-th/0505213} \BibitemShut {NoStop}%
\bibitem [{\citenamefont {Armoni}\ and\ \citenamefont {Shifman}(2003)}]{Armoni:2003ji}%
  \BibitemOpen
  \bibfield  {author} {\bibinfo {author} {\bibfnamefont {A.}~\bibnamefont {Armoni}}\ and\ \bibinfo {author} {\bibfnamefont {M.}~\bibnamefont {Shifman}},\ }\href {\doibase 10.1016/S0550-3213(03)00409-7} {\bibfield  {journal} {\bibinfo  {journal} {Nucl. Phys. B}\ }\textbf {\bibinfo {volume} {664}},\ \bibinfo {pages} {233} (\bibinfo {year} {2003})},\ \Eprint {http://arxiv.org/abs/hep-th/0304127} {arXiv:hep-th/0304127} \BibitemShut {NoStop}%
\bibitem [{\citenamefont {Preskill}\ and\ \citenamefont {Vilenkin}(1993)}]{Preskill:1992ck}%
  \BibitemOpen
  \bibfield  {author} {\bibinfo {author} {\bibfnamefont {J.}~\bibnamefont {Preskill}}\ and\ \bibinfo {author} {\bibfnamefont {A.}~\bibnamefont {Vilenkin}},\ }\href {\doibase 10.1103/PhysRevD.47.2324} {\bibfield  {journal} {\bibinfo  {journal} {Phys. Rev. D}\ }\textbf {\bibinfo {volume} {47}},\ \bibinfo {pages} {2324} (\bibinfo {year} {1993})},\ \Eprint {http://arxiv.org/abs/hep-ph/9209210} {arXiv:hep-ph/9209210} \BibitemShut {NoStop}%
\bibitem [{\citenamefont {Eto}\ \emph {et~al.}(2014)\citenamefont {Eto}, \citenamefont {Hirono},\ and\ \citenamefont {Nitta}}]{Eto:2013bxa}%
  \BibitemOpen
  \bibfield  {author} {\bibinfo {author} {\bibfnamefont {M.}~\bibnamefont {Eto}}, \bibinfo {author} {\bibfnamefont {Y.}~\bibnamefont {Hirono}}, \ and\ \bibinfo {author} {\bibfnamefont {M.}~\bibnamefont {Nitta}},\ }\href {\doibase 10.1093/ptep/ptu013} {\bibfield  {journal} {\bibinfo  {journal} {PTEP}\ }\textbf {\bibinfo {volume} {2014}},\ \bibinfo {pages} {033B01} (\bibinfo {year} {2014})},\ \Eprint {http://arxiv.org/abs/1309.4559} {arXiv:1309.4559 [hep-ph]} \BibitemShut {NoStop}%
\bibitem [{\citenamefont {Eto}\ and\ \citenamefont {Nitta}(2022)}]{Eto:2022lhu}%
  \BibitemOpen
  \bibfield  {author} {\bibinfo {author} {\bibfnamefont {M.}~\bibnamefont {Eto}}\ and\ \bibinfo {author} {\bibfnamefont {M.}~\bibnamefont {Nitta}},\ }\href {\doibase 10.1007/JHEP09(2022)077} {\bibfield  {journal} {\bibinfo  {journal} {JHEP}\ }\textbf {\bibinfo {volume} {09}},\ \bibinfo {pages} {077} (\bibinfo {year} {2022})},\ \Eprint {http://arxiv.org/abs/2207.00211} {arXiv:2207.00211 [hep-th]} \BibitemShut {NoStop}%
\bibitem [{\citenamefont {Hsin}\ and\ \citenamefont {Seiberg}(2016)}]{Hsin:2016blu}%
  \BibitemOpen
  \bibfield  {author} {\bibinfo {author} {\bibfnamefont {P.-S.}\ \bibnamefont {Hsin}}\ and\ \bibinfo {author} {\bibfnamefont {N.}~\bibnamefont {Seiberg}},\ }\href {\doibase 10.1007/JHEP09(2016)095} {\bibfield  {journal} {\bibinfo  {journal} {JHEP}\ }\textbf {\bibinfo {volume} {09}},\ \bibinfo {pages} {095} (\bibinfo {year} {2016})},\ \Eprint {http://arxiv.org/abs/1607.07457} {arXiv:1607.07457 [hep-th]} \BibitemShut {NoStop}%
\bibitem [{Note14()}]{Note14}%
  \BibitemOpen
  \bibinfo {note} {The check is not rigorous because it is not possible to determine if such solitons are solutions of finite energy, and within the regime of validity of the IR EFT.}\BibitemShut {Stop}%
\bibitem [{\citenamefont {Hatcher}(2002)}]{MR1867354}%
  \BibitemOpen
  \bibfield  {author} {\bibinfo {author} {\bibfnamefont {A.}~\bibnamefont {Hatcher}},\ }\href@noop {} {\emph {\bibinfo {title} {Algebraic topology}}}\ (\bibinfo  {publisher} {Cambridge University Press},\ \bibinfo {address} {Cambridge},\ \bibinfo {year} {2002})\ pp.\ \bibinfo {pages} {xii+544}\BibitemShut {NoStop}%
\bibitem [{\citenamefont {Steenrod}(1951)}]{Steenrod1951TopologyOF}%
  \BibitemOpen
  \bibfield  {author} {\bibinfo {author} {\bibfnamefont {N.~E.}\ \bibnamefont {Steenrod}}\ }(\bibinfo {year} {1951})\BibitemShut {NoStop}%
\bibitem [{\citenamefont {Kristensen}\ and\ \citenamefont {Ryttov}(2024)}]{Kristensen:2024vmi}%
  \BibitemOpen
  \bibfield  {author} {\bibinfo {author} {\bibfnamefont {A.~H.}\ \bibnamefont {Kristensen}}\ and\ \bibinfo {author} {\bibfnamefont {T.~A.}\ \bibnamefont {Ryttov}},\ }\href {\doibase 10.1103/PhysRevD.110.014012} {\bibfield  {journal} {\bibinfo  {journal} {Phys. Rev. D}\ }\textbf {\bibinfo {volume} {110}},\ \bibinfo {pages} {014012} (\bibinfo {year} {2024})},\ \Eprint {http://arxiv.org/abs/2404.12947} {arXiv:2404.12947 [hep-ph]} \BibitemShut {NoStop}%
\bibitem [{\citenamefont {Girmohanta}\ \emph {et~al.}(2019)\citenamefont {Girmohanta}, \citenamefont {Ryttov},\ and\ \citenamefont {Shrock}}]{Girmohanta:2019cth}%
  \BibitemOpen
  \bibfield  {author} {\bibinfo {author} {\bibfnamefont {S.}~\bibnamefont {Girmohanta}}, \bibinfo {author} {\bibfnamefont {T.~A.}\ \bibnamefont {Ryttov}}, \ and\ \bibinfo {author} {\bibfnamefont {R.}~\bibnamefont {Shrock}},\ }\href {\doibase 10.1103/PhysRevD.99.116022} {\bibfield  {journal} {\bibinfo  {journal} {Phys. Rev. D}\ }\textbf {\bibinfo {volume} {99}},\ \bibinfo {pages} {116022} (\bibinfo {year} {2019})},\ \Eprint {http://arxiv.org/abs/1903.09672} {arXiv:1903.09672 [hep-th]} \BibitemShut {NoStop}%
\bibitem [{\citenamefont {Alford}\ \emph {et~al.}(1999)\citenamefont {Alford}, \citenamefont {Rajagopal},\ and\ \citenamefont {Wilczek}}]{Alford:1998mk}%
  \BibitemOpen
  \bibfield  {author} {\bibinfo {author} {\bibfnamefont {M.~G.}\ \bibnamefont {Alford}}, \bibinfo {author} {\bibfnamefont {K.}~\bibnamefont {Rajagopal}}, \ and\ \bibinfo {author} {\bibfnamefont {F.}~\bibnamefont {Wilczek}},\ }\href {\doibase 10.1016/S0550-3213(98)00668-3} {\bibfield  {journal} {\bibinfo  {journal} {Nucl. Phys. B}\ }\textbf {\bibinfo {volume} {537}},\ \bibinfo {pages} {443} (\bibinfo {year} {1999})},\ \Eprint {http://arxiv.org/abs/hep-ph/9804403} {arXiv:hep-ph/9804403} \BibitemShut {NoStop}%
\bibitem [{\citenamefont {Sch{\"a}fer}\ and\ \citenamefont {Wilczek}(1999)}]{Schafer:1998ef}%
  \BibitemOpen
  \bibfield  {author} {\bibinfo {author} {\bibfnamefont {T.}~\bibnamefont {Sch{\"a}fer}}\ and\ \bibinfo {author} {\bibfnamefont {F.}~\bibnamefont {Wilczek}},\ }\href {\doibase 10.1103/PhysRevLett.82.3956} {\bibfield  {journal} {\bibinfo  {journal} {Phys. Rev. Lett.}\ }\textbf {\bibinfo {volume} {82}},\ \bibinfo {pages} {3956} (\bibinfo {year} {1999})},\ \Eprint {http://arxiv.org/abs/hep-ph/9811473} {arXiv:hep-ph/9811473} \BibitemShut {NoStop}%
\bibitem [{\citenamefont {Li}\ \emph {et~al.}(2025)\citenamefont {Li}, \citenamefont {Pastor-Guti{\'e}rrez}, \citenamefont {Vatani},\ and\ \citenamefont {Xu}}]{Li:2025tvu}%
  \BibitemOpen
  \bibfield  {author} {\bibinfo {author} {\bibfnamefont {H.-L.}\ \bibnamefont {Li}}, \bibinfo {author} {\bibfnamefont {{\'A}.}~\bibnamefont {Pastor-Guti{\'e}rrez}}, \bibinfo {author} {\bibfnamefont {S.}~\bibnamefont {Vatani}}, \ and\ \bibinfo {author} {\bibfnamefont {L.-X.}\ \bibnamefont {Xu}},\ }\href@noop {} {\  (\bibinfo {year} {2025})},\ \Eprint {http://arxiv.org/abs/2507.21208} {arXiv:2507.21208 [hep-th]} \BibitemShut {NoStop}%
\bibitem [{\citenamefont {Armoni}\ \emph {et~al.}(2015)\citenamefont {Armoni}, \citenamefont {Shifman}, \citenamefont {Shore},\ and\ \citenamefont {Veneziano}}]{Armoni:2014ywa}%
  \BibitemOpen
  \bibfield  {author} {\bibinfo {author} {\bibfnamefont {A.}~\bibnamefont {Armoni}}, \bibinfo {author} {\bibfnamefont {M.}~\bibnamefont {Shifman}}, \bibinfo {author} {\bibfnamefont {G.}~\bibnamefont {Shore}}, \ and\ \bibinfo {author} {\bibfnamefont {G.}~\bibnamefont {Veneziano}},\ }\href {\doibase 10.1016/j.physletb.2014.12.035} {\bibfield  {journal} {\bibinfo  {journal} {Phys. Lett. B}\ }\textbf {\bibinfo {volume} {741}},\ \bibinfo {pages} {184} (\bibinfo {year} {2015})},\ \Eprint {http://arxiv.org/abs/1412.3389} {arXiv:1412.3389 [hep-th]} \BibitemShut {NoStop}%
\end{thebibliography}%
\bibliographystyle{apsrev4-1}

\end{document}